\documentclass[prd,showpacs,superscriptaddress,twocolumn,nofootinbib,floatfix,10pt]{revtex4-2}
\usepackage[utf8]{inputenc}
\usepackage{amsmath,amssymb}
\usepackage{slashed}
\usepackage{subfigure}
\usepackage[colorlinks=true,
breaklinks=true,
urlcolor=magenta,
citecolor=blue]{hyperref}
\usepackage[usenames,dvipsnames]{color}
\usepackage{appendix}
\usepackage{braket,bm}
\usepackage{multirow}
\usepackage{enumitem}
\usepackage{color}

\usepackage{orcidlink}
\usepackage{graphicx}
\usepackage{tikz}
\usepackage{booktabs}
\usepackage{array}
\usepackage{overpic}
\usepackage{float}
\usepackage{threeparttable}
\allowdisplaybreaks[4]

\newcommand{\itp}{\affiliation{CAS Key Laboratory of Theoretical Physics, Institute of Theoretical Physics, \\Chinese Academy of Sciences, Beijing 100190, China}}

\newcommand{\ucas}{\affiliation{School of Physical Sciences, University of Chinese Academy of Sciences, Beijing 100049, China}}

\newcommand{\peng}{\affiliation{Peng Huanwu Collaborative Center for Research and Education, Beihang University, Beijing 100191, China}}
\newcommand{\scnt}{\affiliation{Southern Center for Nuclear-Science Theory (SCNT), Institute of Modern Physics,\\ Chinese Academy of Sciences, Huizhou 516000, China}}

\newcommand{\thu}{\affiliation{Department of Physics, Tsinghua University, Beijing 100084, China}}

\begin{document}
\title{Possible  $\Sigma_c^* \bar{\Sigma}$ molecular states}

\author{Bing Wu\orcidlink{0009-0004-8178-3015}}\email{wubing@itp.ac.cn}
\itp\ucas
\author{Xiang-Kun Dong\orcidlink{0000-0001-6392-7143}}\email{xiangkun@hiskp.uni-bonn.de}
\affiliation{Helmholtz-Institut~f\"{u}r~Strahlen-~und~Kernphysik~and~Bethe~Center~for
Theoretical~Physics, \\Universit\"{a}t~Bonn,  D-53115~Bonn,~Germany}
\author{Feng-Kun Guo\orcidlink{0000-0002-2919-2064}}\email{fkguo@itp.ac.cn}
\itp\ucas\peng\scnt
\author{Bing-Song Zou\orcidlink{0000-0002-3000-7540}} \email{zoubs@itp.ac.cn}
\thu \itp \ucas \peng \scnt  

\begin{abstract}
We investigate the possibility of deuteron-like $\Sigma_c^*\bar{\Sigma}$ bound states within the one-boson-exchange model and systematically analyze the effects of the contact-range $\delta^{3}(\vec{r}\,)$ potential, the tensor term from the vector-meson exchange, and nonlocal potentials due to the dependence on the sum of the initial and final state center-of-mass momenta. We find that the pion-exchange potential including the $\delta^{3}(\vec{r}\,)$ term and the tensor term of the $\rho$-exchange potential exhibit comparable magnitudes but opposite signs for any $S$-wave baryon-antibaryon systems. For the $\Sigma_c^*\bar{\Sigma}$ system, it is most likely to form bound states with mass around 3.7~GeV in the $I(J^P)=0(2^-)$ and $1(2^-)$ channels.
\end{abstract}

\maketitle
\section{Introduction}
Since the landmark discovery of $X(3872)$ in 2003~\cite{Belle:2003nnu}, there has been a significant surge in both experimental and theoretical investigations into exotic states. Up to now, dozens of exotic states or their candidates have been observed in experiments, and theoretical frameworks explaining the underlying structures of these exotic states, such as molecular states, multiquark states, hybrids, or glueballs, are continuously evolving and being refined. We refer to Refs.~\cite{Chen:2016qju,Hosaka:2016pey,Richard:2016eis,Lebed:2016hpi,Esposito:2016noz,Guo:2017jvc,Ali:2017jda,Olsen:2017bmm,Kou:2018nap,Kalashnikova:2018vkv,Cerri:2018ypt,Liu:2019zoy,Brambilla:2019esw,Guo:2019twa,Yang:2020atz,Ortega:2020tng,Dong:2021juy,Dong:2021bvy} for reviews of the experimental and theoretical studies.
Intriguingly, many of the observed exotic states are located in close proximity to the thresholds of a pair of hadrons that they can couple to, including the following famous examples, $X(3872)$~\cite{Belle:2003nnu} and $Z_c(3900)^\pm$~\cite{BESIII:2013ris,Belle:2013yex,BESIII:2013qmu} around the $D\bar D^*$ threshold, the $T_{cc}(3875)$~\cite{LHCb:2021vvq,LHCb:2021auc} near the $DD^*$ threshold, the $Z_c(4020)^\pm$~\cite{BESIII:2013ouc,BESIII:2013mhi} near the $D^*\bar D^*$ threshold, the $Z_b(10610)^\pm$ and $Z_b(10650)^\pm$~\cite{Belle:2011aa,Belle:2015upu} near the $B\bar B^*$ and $B^*\bar B^*$ thresholds, the $Z_{cs}(3985)$~\cite{BESIII:2020qkh,BESIII:2022qzr,LHCb:2021uow,LHCb:2023hxg} near the $\bar D_s D^*$ and $\bar D_s^* D$ thresholds, the $P_c$ states~\cite{LHCb:2019kea} near the $\bar D^{(*)}\Sigma_c$ thresholds, the $P_{cs}$ states~\cite{LHCb:2020jpq,LHCb:2022ogu} near the $\bar D^{(*)}\Xi_c$ threshold and so on. It is natural to explain them as hadronic molecules composed of the corresponding hadron pairs~\cite{Guo:2017jvc,Dong:2020hxe}. 

The hadronic molecule picture has undergone a process of ongoing refinement and evolution. The first proposal of a hadronic molecule composed of a pair of charmed and anticharmed mesons was advanced in 1976~\cite{Voloshin:1976ap}. Merely a year later, the $\psi(4040)$ peak observed in $e^+e^-$ annihilation, which was ultimately interpreted as a charmonium state, was speculated to be a result of the production of a $D^*\bar{D}^*$ molecule based on preliminary analysis~\cite{DeRujula:1976zlg}. Given the notable success of the one-pion-exchange (OPE) potential model in describing the deuteron and nucleon-nucleon scattering, it was widely conceived that the pions play a significant role in the formation of hadronic molecules. In 1980s, an accurate description of the nuclear force was achieved with the one-boson-exchange (OBE) model~\cite{Nagels:1975fb,Machleidt:1987hj,Machleidt:1989tm,Stoks:1994wp}. 
In 1991 and 1994, T\"ornqvist carried out a comprehensive analysis of the potential existence of deuteron-like meson-meson bound states using the OPE, employing both qualitative and quantitative methods~\cite{Tornqvist:1991ks,Tornqvist:1993ng}.

The theoretical analyses mentioned thus far can be considered as preliminary attempts to model two-body hadronic molecular states, in the absence of definitive experimental results apart from the deuteron. 
Nevertheless, with the discovery of the $X(3872)$ by Belle Collaboration, which lies beyond the conventional charmonium spectrum~\cite{Eichten:1979ms,Godfrey:1985xj}, these initial attempts have been extended to study possible hadronic molecules in various hadron systems. Numerous studies suggest that the $X(3872)$ may be a $D\bar D^*$ molecule~\cite{Tornqvist:2003na,Close:2003sg,Pakvasa:2003ea,Braaten:2003he,Wong:2003xk,Tornqvist:2004qy}, based on its distinct characteristics near the $D\bar{D}^*$ threshold, and the observed ratio of its isospin breaking decays $\Gamma(X \to J/\psi \pi^+ \pi^-)$ and $\Gamma(X \to J/\psi \pi^+ \pi^- \pi^0)$, which can be easily explained within the molecular picture~\cite{Gamermann:2009fv,Gamermann:2009uq}. 
In 2008, Thomas and Close undertook a comprehensive analysis, examining and verifying the calculations of the molecular state model in the literature thus far. They scrutinized several pivotal aspects, including different conventions for charge conjugation eigenstates, the $\delta^{3}(\vec{r}\,)$ term and the tensor force~\cite{Thomas:2008ja}. Their research suggested that the $X(3872)$ could potentially be a bound state within the OPE model. However, these results demonstrated a significant sensitivity to  the cutoff in the form factor. For an in-depth discussion on the form factor and renormalization related to the short-distance interactions, we refer to Refs.~\cite{CalleCordon:2009pit,Reinert:2017usi}.
Furthermore, in Ref.~\cite{Ding:2009vj}, the authors elaborated on the OPE model in a constituent quark model by integrating additional contributions from mid- and short-range interactions. These interactions were linked to exchanges of the $\eta$, $\sigma$, $\rho$ and $\omega$ mesons.

In this study, we will investigate the potential existence of $\Sigma_c^*\bar{\Sigma}$ hadronic molecules with quark components $c\bar{s}qq\bar{q}\bar{q}$. If such states exist, they would significantly enrich the excited $D_s$ state spectrum in a higher energy region beyond the scope of conventional $c\bar s$ mesons and their mixture of $c\bar{s}q\bar{q}$ configurations~\cite{Hao:2022vwt}. We will explore various issues associated with the OBE model, including the effects of $\delta^{3}(\vec{r}\,)$ which has been repeatedly discussed, the contribution of the tensor term in the vector-meson exchange, and the impact of nonlocal terms due to the dependence on the sum of the initial and final state center-of-mass (c.m.) momenta (denoted as $\vec k$), which has not been thoroughly investigated in the hadronic molecular context. It is important to clarify that this work is not aiming at precisely predicting the masses of possible $\Sigma_c^*\bar{\Sigma}$ bound states, but rather at exploring the potential existence of such hadronic molecules and attempting to formalize the calculation process of the OBE model after considering various factors.

This paper is organized as follows. After the Introduction, we begin by presenting the effective potential of $\Sigma_c^*\bar{\Sigma}$ in Sec.~\ref{sec:potential}. We then proceed to discuss various factors, including the effects of momentum $\vec{k}$, the $\delta^{3}(\vec{r}\,)$ term and the tensor potential in the OBE model in Sec.~\ref{Sec:III}. Subsequently, we  present the numerical outcomes of the OBE model in Sec.~\ref{the general OBE}. In Sec.~\ref{general relation from quark model}, we show that cancellations generally exists between the pion and $\rho$-meson-exchange potentials, as derived from the quark model. Possible $\Sigma_c^*\bar{\Sigma}$ bound states are discussed in Sec.~\ref{new result of OBE}.  Finally, we present a summary in Sec.~\ref{Sec:V}. Technical and pedagogical details are relegated to Appendices~\ref{Appendix:A}, \ref{Appendix:B}, \ref{Appendix:C} and \ref{Appendix:D}.

\section{Potential for the $\Sigma_c^*\bar{\Sigma}$ system}\label{sec:potential}

\begin{figure}[tbh]  
  \centering
   \begin{overpic}[width=0.9\linewidth]{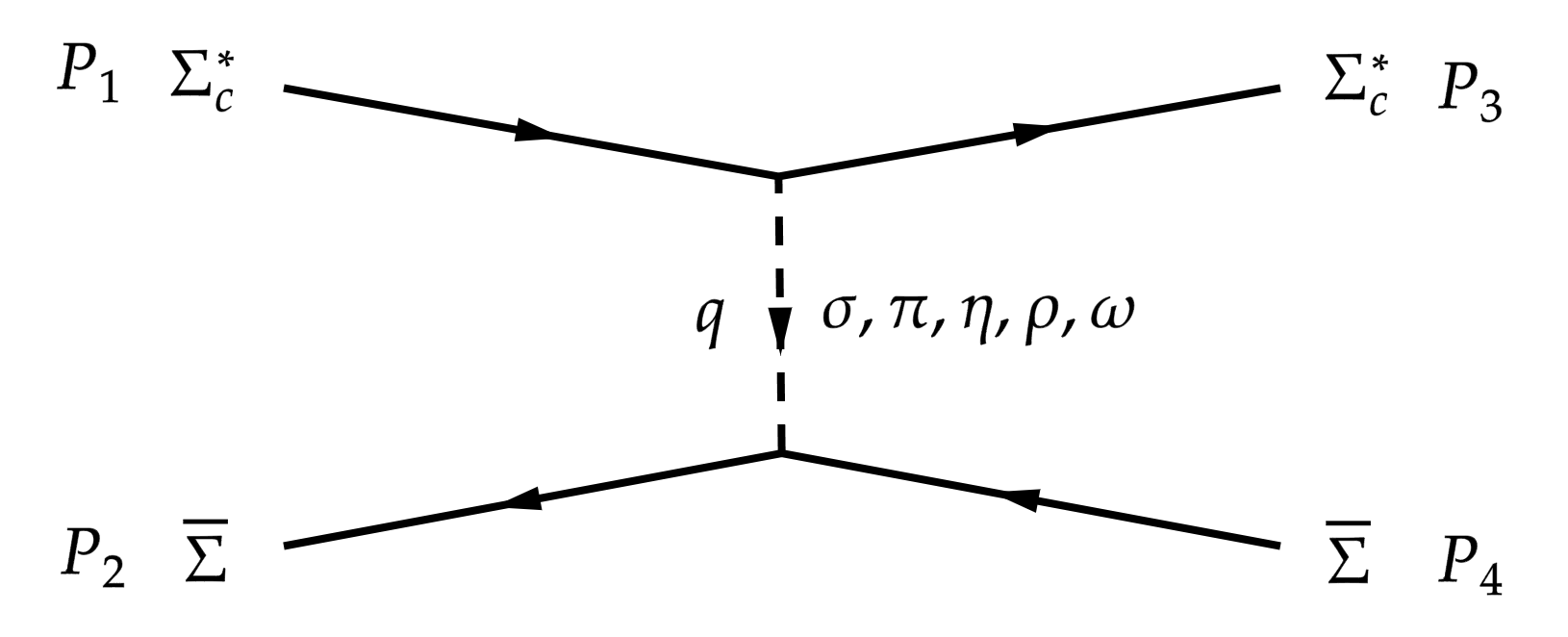}
    \end{overpic}
  \caption{Feynman diagram for the $\Sigma_c^*\bar{\Sigma} \to \Sigma_c^*\bar{\Sigma}$ process with $t$-channel meson exchanges, where $P_1=(M_{\Sigma_c^*}, \vec{p}\,)$, $P_2=(M_{\Sigma}, -\vec{p}\,)$, $P_3=(M_{\Sigma_c^*}, \vec{p\,}')$, $P_4=(M_{\Sigma}, -\vec{p\,}')$ and $q=P_1-P_3$ represent the four momenta of the corresponding particles.} 
  \label{img1} 
\end{figure}
In this section, we perform calculations to determine the OBE potential between $\Sigma_c^*\bar\Sigma$, as depicted in Fig.~\ref{img1}. The Lagrangians for the couplings of $\Sigma$ with the exchanged mesons ($\sigma, \pi,\eta,\rho$ and $\omega$) are adopted from Ref.~\cite{Ronchen:2012eg}, 
\begin{align}
\mathcal{L}_{\Sigma \Sigma \sigma}&=-g_{\Sigma \Sigma \sigma}\bar{\Sigma} \sigma \Sigma \label{lagangian for Sigma Sigma sigma},\\
\mathcal{L}_{\Sigma\Sigma\pi}&=-\frac{g_{\Sigma\Sigma\pi}}{m_\pi}\bar{\Sigma}\gamma^5\gamma^{\mu}\vec{\tau}\cdot\partial_{\mu}\vec{\pi}\Sigma \label{lagangian for Sigma Sigma pi},\\
\mathcal{L}_{\Sigma\Sigma\eta}&=-\frac{g_{\Sigma\Sigma\eta}}{m_\eta}\bar{\Sigma}\gamma^5\gamma^{\mu}\partial_{\mu}\eta\Sigma \label{lagangian for Sigma Sigma eta},\\
\mathcal{L}_{\Sigma \Sigma \rho}&=-g_{\Sigma \Sigma \rho}\bar{\Sigma}\left[ \gamma^{\mu}-\frac{k_{\Sigma \Sigma \rho}}{2M_\Sigma}\sigma^{\mu\nu}\partial_{\nu} \right]\vec{\tau}\cdot\vec{\rho}_\mu\Sigma \label{lagangian for Sigma Sigma rho},\\
\mathcal{L}_{\Sigma\Sigma\omega}&=-g_{\Sigma\Sigma\omega}\bar{\Sigma} \left[ \gamma^\mu -\frac{k_{\Sigma\Sigma\omega}}{2M_\Sigma}\sigma^{\mu\nu}\partial_\nu \right]\omega_{\mu}\Sigma,\label{lagangian for Sigma Sigma omega}
\end{align}
where the isospin multiplets are defined as 
\begin{align}
\Sigma&=\left( \Sigma^+, \Sigma^0, \Sigma^-\right)^T,\\
\vec{\pi}&=\left( \frac{\pi^+ + \pi^-}{\sqrt{2}},\frac{\pi^- -\pi^+}{i\sqrt{2}},\pi^0 \right),\\
\vec{\rho}_\mu&=\left( \frac{\rho^+_\mu + \rho^-_\mu}{\sqrt{2}}, \frac{\rho^-_\mu-\rho^+_\mu}{i\sqrt{2}} , \rho^0_\mu \right),
\end{align}
the tensor operator in spinor space is $\sigma^{\mu\nu}=i(\gamma^\mu\gamma^\nu-\gamma^\nu\gamma^\mu)/2$, the isospin operator $\vec{\tau}=\left( 
\tau_1,\tau_2,\tau_3 \right)$ with $\tau_i\ (i=1,2,3)$ the traceless isospin-1 matrices, and $m_\pi$, $m_\eta$, $M_\Sigma$ represent the respective masses of the corresponding particles.\footnote{Since we are not interested in isospin symmetry breaking effects, the isospin averaged masses are used for all particles within the same isospin multiplet. Regarding the $\sigma$, we select the mass value to be used in the OBE model, $m \simeq 519$~MeV, given in Ref.~\cite{Wu:2023uva} that corresponds to the coupling constant $g_{\Sigma\Sigma\sigma}$ listed in Table~\ref{couplings}.} In the heavy quark limit, $\Sigma_c^*$ belongs to the light flavor SU(3) sextet~\cite{Yan:1992gz},
\begin{align}
{{B}}_6^*=\left(
\begin{array}{ccc}
\Sigma_c^{*++}&\frac{\Sigma_c^{*+}}{\sqrt{2}}&\frac{{\Xi_c^{*+}}}{\sqrt{2}}\\
\frac{\Sigma_c^{*+}}{\sqrt{2}}&\Sigma_c^{*0}&\frac{{\Xi_c^{*0}}}{\sqrt{2}}\\
\frac{{\Xi_c^{*+}}}{\sqrt{2}}&\frac{{\Xi_c^{*0}}}{\sqrt{2}}&\Omega_c^{*0}
\end{array}
\right)
\end{align} 
and the related couplings satisfying heavy quark spin symmetry read~\cite{Liu:2011xc},\footnote{Indeed, Eqs.~(\ref{lagangian for Sigmac* Sigmac* sigma}-\ref{eq:BBV}) can be reformulated in a manner similar to Eqs.~(\ref{lagangian for Sigma Sigma sigma}-\ref{lagangian for Sigma Sigma omega}). Specifically, Eq.~(\ref{lagangian for Sigmac* Sigmac* sigma}) is of the form as Eq.~(\ref{lagangian for Sigma Sigma sigma}); Eq.~(\ref{lagangian for Sigmac* Sigmac* pi and eta}) aligns with Eqs.~(\ref{lagangian for Sigma Sigma pi},\ref{lagangian for Sigma Sigma eta}) in terms of axial vector coupling at the tree level~\cite{Meng:2017udf}; Eq.~(\ref{eq:BBV}) can be restructured into the form as Eqs.~(\ref{lagangian for Sigma Sigma rho},\ref{lagangian for Sigma Sigma omega}) using the Gordon identity, that is, the terms $i\bar{{B}}_{6\mu}^* (\partial^\mu {\mathcal{V}}^\nu-\partial^\nu {\mathcal{V}}^\mu) {B}_{6\nu}^*$ and $\bar{{B}}_{6\mu}^* (-2M_{6^*}\gamma^\mu {\mathcal{V}}^\nu+\sigma^{\mu\alpha}\partial_\alpha {\mathcal{V}}^\nu-2M_{6^*}\gamma^\nu {\mathcal{V}}^\mu+\sigma^{\nu\alpha}\partial_\alpha {\mathcal{V}}^\mu) {B}_{6\nu}^*$ are equivalent at the tree level.}
\begin{align}
\mathcal{L}_{B_6^*B_6^*\sigma }&=g_{ {B}_6^{*} {B}_6^{*}\sigma}{\rm{Tr}}\left[ \bar{{B}}_6^{*\mu}\sigma {B}_{6\mu}^* \right] \label{lagangian for Sigmac* Sigmac* sigma},\\
\mathcal{L}_{B_6^*B_6^* p}&=g_{{B}_6^*{B}_6^* p}{\rm{Tr}} \left[ \bar{B}_6^{*\mu}i\gamma_5 {\mathcal{P}}{B}^*_{6\mu}\right] \label{lagangian for Sigmac* Sigmac* pi and eta},\\
\mathcal{L}_{B_6^*B_6^*v}&=g_{{B}_6^*{B}_6^*v}{\rm{Tr}}\left[ \bar{{B}}_6^{*\mu}\gamma_\nu {\mathcal{V}}^\nu {B}_{6\mu}^* \right]\notag\\
&\quad+i\frac{f_{{B}_6^*{B}_6^*v}}{2M_{6^*}}{\rm{Tr}}\left[ \bar{{B}}_{6\mu}^* (\partial^\mu {\mathcal{V}}^\nu-\partial^\nu {\mathcal{V}}^\mu) {B}_{6\nu}^* \right],\label{eq:BBV}
\end{align}
where ${\rm{Tr}}\left[ \cdots \right]$ means the trace over flavor indices, and~\cite{Meissner:1987ge}
\begin{align}
{\mathcal{P}}&=\left(
\begin{array}{ccc}
\frac{\pi^0}{\sqrt{2}}+\frac{\eta}{\sqrt{6}}  &  \pi^+  &  K^+\\
\pi^-  &  -\frac{\pi^0}{\sqrt{2}}+\frac{\eta}{\sqrt{6}}  &  K^0\\
K^-  &  \bar{K}^0  &  -\frac{2}{\sqrt{6}}\eta
\end{array}
\right),\\
{\mathcal{V}^\mu}&=\left(
\begin{array}{ccc}
\frac{\rho^0}{\sqrt{2}}+\frac{\omega}{\sqrt{2}}  &  \rho^+  &  K^{*+}\\
\rho^-  &  -\frac{\rho^0}{\sqrt{2}}+\frac{\omega}{\sqrt{2}}  &  K^{*0}\\
K^{*-}  &  \bar{K}^{*0}  &  \phi
\end{array}
\right)^\mu.
\end{align} 
The pertinent coupling constants are listed in Table~\ref{couplings}.

\begin{table}[t]
\caption{Pertinent coupling constants for the $\Sigma_c^*\bar{\Sigma}\to\Sigma_c^*\bar{\Sigma}$ process~\cite{Ronchen:2012eg,Yang:2018amd,Wu:2023uva}. $g_{\Sigma\Sigma\sigma}$ is obtained by matching the amplitude of $\pi\pi$-exchange with that of the $\sigma$-exchange for the $t$-channel process of $\Sigma\bar{\Sigma}\to\bar{\Sigma}\Sigma$~\cite{Wu:2023uva}. For the vector-meson coupling constants, we use $g_{\Sigma\Sigma\rho}=g_{\Sigma\Sigma\omega}=g_{\Sigma\Sigma v}$ and $k_{\Sigma\Sigma\rho}=k_{\Sigma\Sigma\omega}=k_{\Sigma\Sigma v}$.}
\centering
\begin{tabular}{>{\centering\arraybackslash}p{5em}>{\centering\arraybackslash}p{3.5em} >{\centering\arraybackslash}p{3.5em} >{\centering\arraybackslash}p{3.5em} >{\centering\arraybackslash}p{3.5em} >{\centering\arraybackslash}p{3.5em} }
\toprule[2pt]
Couplings & $g_{\Sigma\Sigma\sigma}$ & $g_{\Sigma\Sigma\pi}$ & $g_{\Sigma\Sigma\eta}$ & $g_{\Sigma\Sigma v}$ & $k_{\Sigma\Sigma v}$  \\
\midrule[0.5pt]
Value & 3.50 & 0.79 & 0.69 & 7.48 & 1.33 \\
\toprule[1pt]
Couplings &  $g_{B_6^*B_6^*\sigma}$ & $g_{B_6^*B_6^* p}$ & $g_{B_6^*B_6^* v}$ & $f_{B_6^*B_6^* v}$& \\
\midrule[0.5pt]
Value &  5.64 & 59.50 & 9.19 & 95.80& \\
\bottomrule[2pt]
\end{tabular}
\label{couplings}
\end{table}

Utilizing the aforementioned Lagrangians, we can derive the $\Sigma_c^*\bar\Sigma$ scattering amplitude, and the details can be found in Appendix~\ref{Appendix:B}. The $\Sigma_c^*\bar\Sigma$  potential in the momentum space is linked to the scattering amplitude through 
\begin{align}
   \langle\vec{p}^{\,\prime}|{\hat V}|\vec{p}\,\rangle\approx-\frac{1}{(2\pi)^3}{\mathcal{M}\!\left(\Sigma_c^*\bar\Sigma\to \Sigma_c^*\bar\Sigma\right)},
\end{align}
with $\vec{p}$ and $\vec{p}^{\,\prime}$ the relative momenta of the incoming and outgoing particles; see Appendix~\ref{Appendix:C} for additional details. 
As usually done in the OBE model, we introduce a monopole form factor with a cutoff parameter $\Lambda$ at each vertex,
\begin{align}
F(q)=\frac{\Lambda^2-m_{\rm{ex}}^2}{\Lambda^2-q^2},
\label{eq:11}
\end{align}
which equals unity when the exchanged particle is on shell. Then one gets the effective potential in momentum space, which can be subsequently converted to the coordinate space potential utilizing the Fourier transformation; see Appendix~\ref{Appendix:A} for details. Consequently, we obtain the $S$-wave $\Sigma_c^*\bar{\Sigma}$ effective potential from exchanging the scalar meson ($\sigma$), pseudoscalar mesons ($p=\pi,\eta$) and vector mesons ($v=\rho,\omega$)  as $V = V_\sigma + \sum_{p=\pi,\eta} V_p + \sum_{v=\rho,\omega} V_v$, where
\begin{align}
{V}_\sigma&=-g_{ B_6^* B_6^*\sigma}g_{\Sigma\Sigma\sigma}F_\sigma(I)H_0(r,m_\sigma,\Lambda),\label{equ:12}\\
{V}_{p}&=-\frac{{g_{B_6^*B_6^*p}g_{\Sigma\Sigma p}}}{2M_{\Sigma_c^*} {m_p}}F_p(I)H_1(r,m_p,\Lambda)\Delta_{S_AS_B},\label{equ:13}\\
{V}_v&=F_v(I)\left({V}_v^{(1)}+{V}_v^{(2)}+{V}_v^{(3)}+{V}_v^{(4)}\right),\label{equ:14}
\end{align}
with
\begin{align}
{V}_v^{(1)}&=-g_{B_6^*B_6^*v}g_{\Sigma\Sigma v}H_0(r,m_v,\Lambda),\label{equ:15}\\
{V}_v^{(2)}&=\frac{g_{B_6^*B_6^*v}g_{\Sigma\Sigma v}k_{\Sigma\Sigma v}}{2M_{\Sigma_c^*}M_{\Sigma}}\left({\Delta_{S_AS_B}}-\frac{3M_{\Sigma_c^*}}{2M_{\Sigma}}\right)H_1(r,m_v,\Lambda),
\label{equ:16}\\
{V}_v^{(3)}&=\frac{f_{B_6^*B_6^*v}g_{\Sigma\Sigma v}}{2M_{\Sigma_c^*}M_{\Sigma}}\Delta_{S_AS_B}H_1(r,m_v,\Lambda),\label{equ:17}\\
{V}_v^{(4)}&=\frac{f_{B_6^*B_6^*v}g_{\Sigma\Sigma v}k_{\Sigma\Sigma v}}{2M_{\Sigma_c^*}M_{\Sigma}}\Delta_{S_AS_B}H_1(r,m_v,\Lambda),\label{equ:18}
\end{align}
and
\begin{align}
H_0(r,m,\Lambda)&=\frac{1}{4\pi}\left[ \frac{e^{-mr}-e^{-\Lambda r}}{r}-\frac{\Lambda^2-m^2}{2\Lambda}e^{-\Lambda r} \right],\\
H_1(r,m,\Lambda)&=\frac{2e^{-mr}m^2+e^{-\Lambda r}\left[ -r\Lambda^3+m^2(-2+\Lambda r) \right]}{24\pi r}.
\end{align}
For the $S$-wave $\Sigma_c^*\bar{\Sigma}$ systems, the spin factor $\Delta_{S_AS_B}$
outlined in Appendix \ref{Appendix:B} is defined as
\begin{align}
\Delta_{S_AS_B}=\frac{9-2S(S+1)}{3}=\left\{
\begin{array}{cc}
\frac{5}{3},&  S=1\\
-1  ,        &  S=2
\end{array} \right.
\end{align} 
with $S$ the total spin.
The pertinent isospin factors are listed in Table~\ref{isospin factor}.

\begin{table}[t]
\caption{The relevant isospin factors for exchanging $\sigma$, $\pi$, $\eta$, $\rho$, $\omega$ for the $\Sigma_c^*\bar{\Sigma}\to\Sigma_c^*\bar{\Sigma}$ process.}
\centering
\begin{tabular}{>{\centering\arraybackslash}p{7em}>{\centering\arraybackslash}p{3.5em} >{\centering\arraybackslash}p{3.5em} >{\centering\arraybackslash}p{3.5em} >{\centering\arraybackslash}p{3.5em} >{\centering\arraybackslash}p{3.5em}}
\toprule[2pt]
Isospin factors & $F_\sigma(I)$ & $F_\pi(I)$ &  $F_\eta(I)$ & $F_\rho(I)$ & $F_\omega(I)$ \\
\midrule[0.5pt]
$I=0$ & 1 & $\sqrt{2}$ & $1/\sqrt{6}$ & $\sqrt{2}$ & $1/\sqrt{2}$ \\
$I=1$ & 1 & $1/\sqrt{2}$ & $1/\sqrt{6}$ & $1/\sqrt{2}$ & $1/\sqrt{2}$ \\
$I=2$ & 1 & $-1/\sqrt{2}$ & $1/\sqrt{6}$ & $-1/\sqrt{2}$ & $1/\sqrt{2}$ \\
\bottomrule[2pt]
\end{tabular}
\label{isospin factor}
\end{table}

\section{OBE model}\label{Sec:III}

\subsection{Effects of $\vec{k}$ on the effective potential}\label{the effects of k}
The relation between the scattering amplitude and the effective potential in coordinate space, as demonstrated in Eq.~(\ref{equ:A16}), clearly indicates the necessity to perform the Fourier transformations of both $\vec{q}\equiv \vec p^{\,\prime} - \vec p$ and $\vec{k} \equiv \vec p^{\,\prime} + \vec p$, followed by integration with respect to $\vec{x}^{\,\prime}$ that is  defined in Eq.~(\ref{equ:A11}). However, altough this mathematical operation can be found in certain old references, e.g.,~\cite{Erkelenz:1971caz,Erkelenz:1974uj,Nagels:1975fb,Nagels:1977ze}, currently the majority of OBE models used for calculating the effective potential for hadronic molecules do not take into account the $\vec k$-dependent terms from the spinors of the initial and final states~\cite{Li:2012bt,Zhao:2013ffn,Liu:2017mrh,Liu:2018bkx,Liu:2019zvb,Yang:2018amd}. 
In the subsequent analysis, we specifically examine the influence of $\vec{k}$ on the final results, particularly on the binding energy of a specified bound state. From Eqs.~(\ref{equ:k^2}-\ref{equ:kxq}), one finds that $\vec{k}$ in the amplitude introduces the derivatives of the radial wavefunction and is thus a nonlocal contribution. Furthermore, considering Eq.~(\ref{equ:s-wave Schordinger equation}), for the $S$-wave, we need to solve the Schr\"odinger equation represented as
\begin{align}
\psi''(r)+2\mu E\psi(r)-2\mu r\hat{V}_{\mid^{2S+1}S_{J};I\rangle}^{\mathcal{M}(\vec{p},\vec{p}^{\,\prime})}(r)\frac{\psi(r)}{r}=0,
\end{align}
where $\hat{V}_{\mid^{2S+1}S_{J};I\rangle}^{\mathcal{M}(\vec{p},\vec{p}^{\,\prime})}(r)$ is the potential operator in the coordinate space, defined in Eq.~(\ref{equ:A21}). We can then proceed with the following substitution,
\begin{align}
r&\,\hat{V}_{\mid^{2S+1}S_{J};I\rangle}^{\mathcal{M}(\vec{p},\vec{p}^{\,\prime})}(r)\frac{\psi(r)}{r}=V_0^{\mathcal{M}(\vec{p},\vec{p}^{\,\prime})}(r)\psi(r)\notag\\
&\qquad+V_1^{\mathcal{M}(\vec{p},\vec{p}^{\,\prime})}(r)\psi'(r)+V_2^{\mathcal{M}(\vec{p},\vec{p}^{\,\prime})}(r)\psi''(r),\label{equ for dealing with k}
\end{align}
where the additional subscripts 0, 1 and 2 of $V^{\mathcal{M}(\vec{p},\vec{p}^{\,\prime})}(r)$ defined here represent the number of the derivatives of $\psi(r)$, specifically $\psi(r)$, $\psi'(r)$ and $\psi''(r)$, respectively. The momentum $\vec{k}$, from the spinor wavefunction of a spin-$1/2$ particle as given in Eq.~(\ref{equ:B2}), consistently appears as ${\vec{k}}/{(2M)}$ with $M$ the baryon mass, which would be small if the composite state was loosely bound. 
Via numerical calculations we find that the effects of $\psi'(r)$ and $\psi''(r)$ on the final binding energy are indeed negligible. However, the $\vec{k}$-dependent contribution in $V_0^{\mathcal{M}(\vec{p},\vec{p}^{\,\prime})}(r)$ could be sizable (see Appendix~\ref{Appendix:D}). In the following, we will keep the $\vec k$-dependent terms in our calculations,  i.e., we will compute the effective potential in the form of Eq.~(\ref{equ:B2}), rather than neglecting the ${\vec{\sigma}\cdot\vec{k}}/{(2M)}$ term, as was often done in literature.

\subsection{The $\delta^{3}(\vec{r}\,)$ potential}

As per Eq.~(\ref{equ:A24}), a Fourier transformation of the amplitude, denoted as $\mathcal{F}_{\vec{q}\to\vec{r}}^{-1}\!\left[\mathcal{M}(\vec{q}\,)\right]$, is required to derive the effective potential in the coordinate space. We now consider two distinct forms of amplitudes:
\begin{align}
\mathcal{M}_1(\vec{q}\,)&=\frac{1}{\vec{q}^{\,2}+m^2},\\
\mathcal{M}_2(\vec{q}\,)&=\frac{\vec{q}^{\,2}/M^2}{\vec{q}^{\,2}+m^2}=\frac{1}{M^2}\left(1-\frac{m^2}{\vec{q}^{\,2}+m^2}\right),
\end{align}
and the Fourier transformation yields
\begin{align}
&\mathcal{F}_{\vec{q}\to\vec{r}}^{-1}\!\left[\mathcal{M}_1(\vec{q}\,)\right]=\frac{1}{4\pi}\frac{e^{-mr}}{r},\\
&\mathcal{F}_{\vec{q}\to\vec{r}}^{-1}\!\left[\mathcal{M}_2(\vec{q}\,)\right]=\frac{1}{M^2}\left[\delta^{3}({\vec{r}})-\frac{m^2}{4\pi}\frac{e^{-mr}}{r}\right],\label{eq:22}
\end{align}
respectively. 
The zero-range $\delta^{3}(\vec{r}\,)$ potential in Eq.~(\ref{eq:22}) leads to a strong repulsion or attraction at $\vec{r}=0$ depending on the sign of the prefactor which has been neglected in the above. Being of short-distance in nature, the $\delta^{3}(\vec r)$ potential requires a regularization.
Considering the form factor in Eq.~(\ref{eq:11}), the potentials become
\begin{align}
\mathcal{F}_{\vec{q}\to\vec{r}}^{-1}\!\left[\mathcal{M}_1(\vec{q}\,)F^2\left(\vec{q}\right)\right]&=H_0(r,m,\Lambda),\\
\mathcal{F}_{\vec{q}\to\vec{r}}^{-1}\!\left[\mathcal{M}_2(\vec{q}\,)F^2\left(\vec{q}\right)\right]&=\frac{1}{M^2}\Bigg[\left(\frac{\Lambda^2-m^2}{4\pi}\right)^2\frac{2\pi}{\Lambda}e^{-\Lambda r}\notag\\
&\quad\quad\quad\quad-{m^2}H_0(r,m,\Lambda)\Bigg],\label{equ:29}
\end{align}
where
\[
\left(\frac{\Lambda^2-m^2}{4\pi}\right)^2\frac{2\pi}{\Lambda}e^{-\Lambda r}
\] 
 is the smeared form of $\delta^{3}\!\left(\vec{r}\right)$ in Eq.~(\ref{eq:22}). 
Not only does ${\vec{q}^{\,2}}/{(\vec{q}^{\,2}+m^2)}$ contribute to the $\delta^{3}\!\left(\vec{r}\right)$ potential for $S$-wave interactions, but also does ${\vec{A}\cdot\vec{q}\vec{B}\cdot\vec{q}}/{(\vec{q}^{\,2}+m^2)}$~\cite{Li:2012bt,Peng:2020xrf}. This observation aligns with Eq.~(\ref{equ:AqBq}), where for $S$-wave, we have
 \[\frac{\vec{A}\cdot\vec{q}\vec{B}\cdot\vec{q}}{\vec{q}^{\,2}+m^2}\sim\frac{\vec{A}\cdot\vec{B}}{3}\frac{\vec{q}^{\,2}}{\vec{q}^{\,2}+m^2}.
 \] 
 
In an effective field theory (EFT), one can introduce counterterms to absorb the cutoff dependence.\footnote{In Ref.~\cite{Reinert:2017usi}, the authors introduce a novel semi-local regularization approach for the chiral two-nucleon potentials. To minimize the short-range contributions in the regularized OPE potential, i.e., ensuring that the corresponding potential vanishes as $r\to 0$, they have incorporated a leading-order contact interaction within the momentum space representation. } 
However, due to the lack of data for most hadron-hadron scatterings, such counterterms can hardly be fixed. 
Thus, in the phenomenological OBE models, one normally does not bother introducing counterterms but rather plays with the $\delta^{3}(\vec r)$ term. The $\delta^{3}\!\left(\vec{r}\right)$ term is retained in its entirety in Refs.~\cite{Chen:2015loa,Liu:2018bkx,Yang:2018amd,Meng:2017udf,Liu:2007bf,Liu:2008fh,Liu:2019stu,Liu:2020nil}, while it is discarded in Ref.~\cite {Peng:2020xrf} and the authors simply make the following substitution\footnote{In fact, this substitution also triggers a substantial shift in the low-momentum part, even to the extent of changing its sign.}
\begin{align}
\frac{\vec{q}^{\,2}}{\vec{q}^{\,2}+m_\pi^2}\to-\frac{m_\pi^2}{\vec{q}^{\,2}+m_\pi^2}.\label{equ:30}
\end{align}
Moreover, in Ref.~\cite{Tornqvist:1993ng}, the $\delta^{3}(\vec{r}\,)$ term in the central potential is omitted. In Ref.~\cite{Li:2012bt}, the authors dismiss the $\delta^{3}(\vec{r}\,)$ term, arguing that in a loosely bound state, the zero-range components are not anticipated to be important.
Furthermore, in Ref.~\cite{Thomas:2008ja}, the authors explore the impacts of including or excluding the $\delta^{3}(\vec{r}\,)$ term in the OPE potential when solving the Schr\"odinger equation for the deuteron, and they find that the cutoff parameters need to be varied significantly to achieve the same binding energy. In Ref.~\cite{Liu:2019zvb}, the authors claim that the removal of the short-range $\delta^{3}(\vec{r}\,)$ contributions to the OBE potential is a necessary step for describing the pentaquark states consistently, and they argue that the behavior of the OBE potential at a distance shorter than the size of hadrons is not physical, so they remove these short-range $\delta$-potential contributions completely.
However, for a hadronic molecule close to threshold, its extended nature does not imply that the short-range potential is insignificant. In contrast, it indicates that the binding of molecular state can not probe details of the short-range binding force, which is distinct from being negligible.
In line with the EFT treatment, in Ref.~\cite{Yalikun:2021bfm} an additional parameter is introduced to adjust the strength of the $\delta^{3}\!\left(\vec{r}\right)$ term to reproduce the experimental masses of the $P_c$ states~\cite{LHCb:2019kea}. 

We can see from the above that the $\delta^{3}(\vec{r}\,)$ term is a contentious aspect within the OBE model for describing hadronic molecular states. 
It is an intrinsic defect of the OBE model and can be rectified as in EFT by introducing counterterms, which can be fixed only when sufficient data are available. Note that the coupling constants that will be used are taken from Refs.~\cite{Ronchen:2012eg,Yang:2018amd}, which fits to experimental data keeping full contributions from the $\delta^{3}\!\left(\vec{r}\right)$ potential. Hence, we will fully retain the $\delta^{3}(\vec{r})$ term in the subsequent calculations to maintain self-consistency.

\subsection{The tensor potential}

In this subsection, we concentrate on the contribution of the tensor term in the Lagrangian, i.e., the second term on the right-hand side of Eqs.~(\ref{lagangian for Sigma Sigma rho},\ref{lagangian for Sigma Sigma omega},\ref{eq:BBV}), to the effective potential. This term is to be distinguished from the vector term, which is the corresponding first term on the right-hand side of the same equations. Many papers have argued that the contribution of the tensor term to the effective potential is negligible~\cite{Manohar:1992nd,Dong:2019ofp,Dong:2021juy}, or it is ignored to simplify the calculation~\cite{Wu:2019adv,Kim:2019kef,Wang:2015jsa}. In general, the significance of the tensor term is case dependent and cutoff dependent.
As an illustration, here we consider the $\Sigma_c^* \bar \Sigma_c^*$ dibaryon systems composed of spin-${3}/{2}$ singly charmed baryons that have been studied in Ref.~\cite{Yang:2018amd}.

The Lagrangian utilized in Ref.~\cite{Yang:2018amd} for the vector meson exchange is given in Eq.~(\ref{eq:BBV}), with the associated coupling constants listed in Table~\ref{couplings}.\footnote{In Ref.~\cite{Yang:2018amd}, the following relations are used: $g_{vB_6^*B_6^*}=2\sqrt{2}g_{\rho NN}$ and $g_{vB_6^*B_6^*}+f_{vB_6^*B_6^*}={6\sqrt{2}}(g_{\rho NN}+f_{\rho NN}){\sqrt{M_iM_f}}/{(5M_N)}$ with $M_{i(f)}$ being the mass of the baryon in the initial (final) state. Using $g_{\rho N N}=3.25$ and $f_{\rho NN}=19.82$, they obtained $g_{vB_6^*B_6^*}=9.19$ and $f_{vB_6^*B_6^*}=95.80$ as listed in Table~\ref{couplings}. The large value of $f_{vB_6^*B_6^*}$ is attributed to the large mass of the charmed baryon.} 
The $S$-wave effective potentials for vector meson exchanges read
\begin{equation}
\begin{aligned}
V_C(r,v,g,f)&=C_v\Big[
g^2H_0(r,m_v,\Lambda)\notag\\
&\quad\quad+\frac{3}{8M_AM_B}(g^2+4gf)H_1(r,m_v,\Lambda)
\Big],\\
V_{SS}(r,v,g,f)&=C_v\frac{g^2+2gf+f^2
}{2M_AM_B} H_1(r,m_v,\Lambda)\Delta_{S_AS_B}^*,\label{eq:26}
\end{aligned} 
\end{equation}
where the subscripts $C$ and $SS$ denote the central and spin-spin potentials, respectively, $g$ and $f$ are the coupling constants of the vector and tensor coupling terms, respectively, $M_A$ and $M_B$ are the  baryon masses, $C_v$ is the isospin factor, $m_v$ ($v=\rho$, $\omega$, $\phi$) is the mass of the exchanged meson, and
\[
\Delta_{S_AS_B}^*=\frac{2S(S+1)-15}{9}.
\]
Taking the $\rho$-exchange potential as an example, we assess the contribution of the tensor term by comparing the following specific effective potentials:
\begin{equation}
\begin{aligned}
V_{\text{tot}}(r)&=V_C(r,\rho,g,f)+V_{SS}(r,\rho,g,f),\\
V_{\text{vector}}(r)&=V_C(r,\rho,g,0)+V_{SS}(r,\rho,g,0),\\
V_{\text{tensor}}(r)&=V_C(r,\rho,0,f)+V_{SS}(r,\rho,0,f),
\end{aligned} 
\end{equation}
where $V_{\text{vector}}(r)$ only contains the contribution of the vector coupling term in the Lagrangian, $V_{\text{tensor}}(r)$ only contains the contribution of the tensor term, and $V_{\text{tot}}(r)$ is the total effective potential. Note that $V_{\text{tot}}(r)\neq V_{\text{vector}}(r)+V_{\text{tensor}}(r)$ due to interference.

\begin{figure*}[tb]
 \centering
 \begin{overpic}[width=1\linewidth]{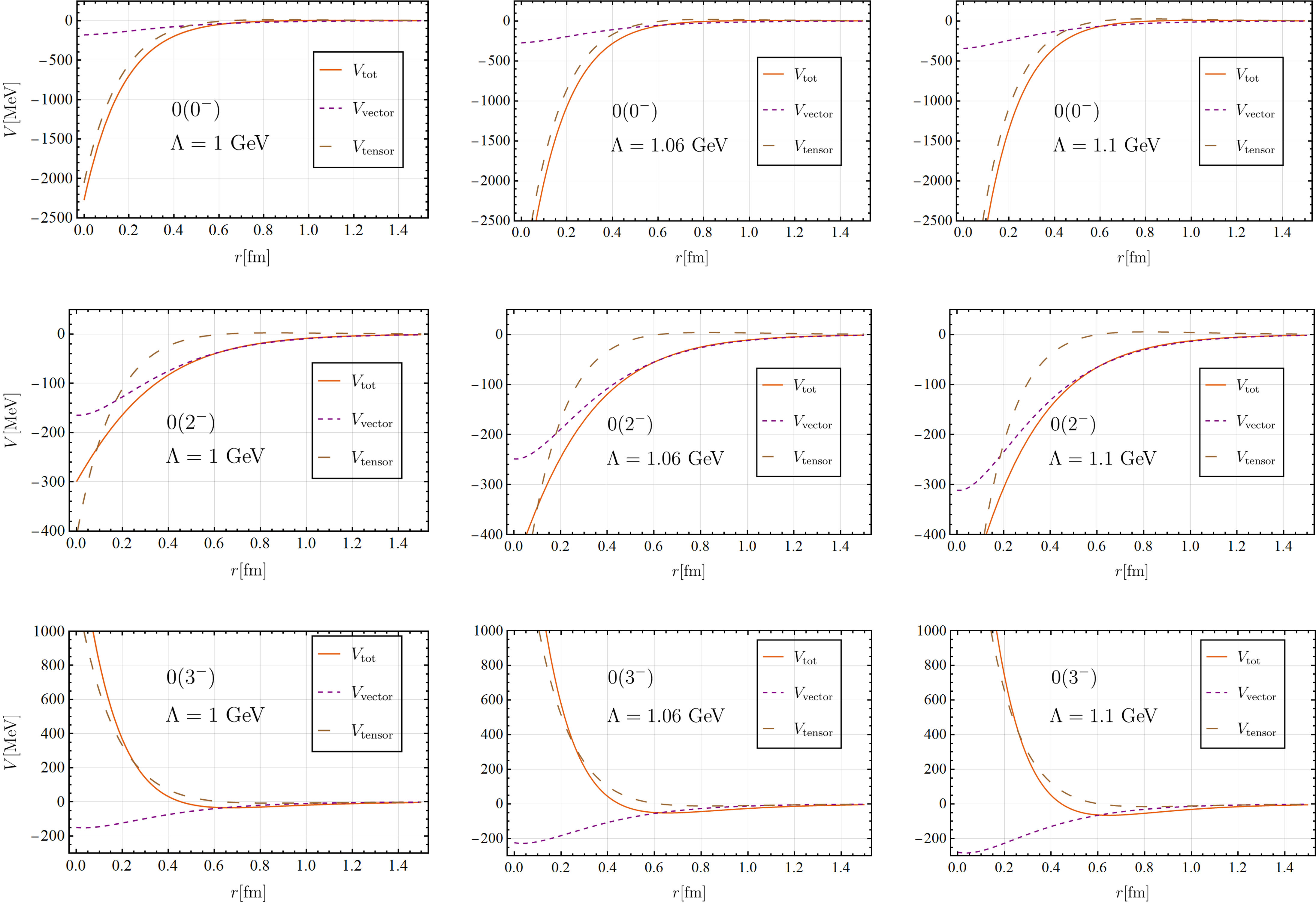}
    \end{overpic}
  \caption{Contributions of the vector and tensor coupling terms, $V_{\rm vector}$ and $V_{\rm tensor}$, respectively, in comparison to the total $\rho$-exchange potential for the $\Sigma_c^*\bar{\Sigma}_c^*$ systems with total spin $J=0$ (top row), $J=2$ (middle row), and $J=3$ (bottom row).
  Because of the $\delta$ potential in the tensor term, the relative importance is sensitive the cutoff. Here the $\Lambda$ values are those taken in Ref.~\cite{Yang:2018amd}.} 
  \label{FIG:2} 
\end{figure*}

The results for the isoscalar $J^P = 0^-$, $2^-$ and $3^-$ $\Sigma_c^* \bar \Sigma_c^*$ systems, using the chosen cutoffs as presented in Ref.~\cite{Yang:2018amd}, are depicted in Fig.~\ref{FIG:2}. It is observed that the tensor term, $V_{\text{tensor}}(r)$, plays a predominant role in the $J^P=0^-$ and $3^-$ cases. 
In particular, for the $I(J^P)=0(3^-)$ system, the total effective potential between the two particles becomes repulsive at short distances when the tensor term is included, despite the attractive nature of $V_{\text{vector}}(r)$. Note that the vector coupling does not lead to a $\delta^{3}(\vec r)$ term while the tensor coupling does. Thus, the relative importance of the tensor coupling contribution crucially depends on the form factor and cutoff. 

\section{Numerical results and discussion}
\label{Sec:IV}
\subsection{Results of the general OBE}\label{the general OBE}

\begin{figure*}[t]
 \centering
 \begin{overpic}[width=1\linewidth]{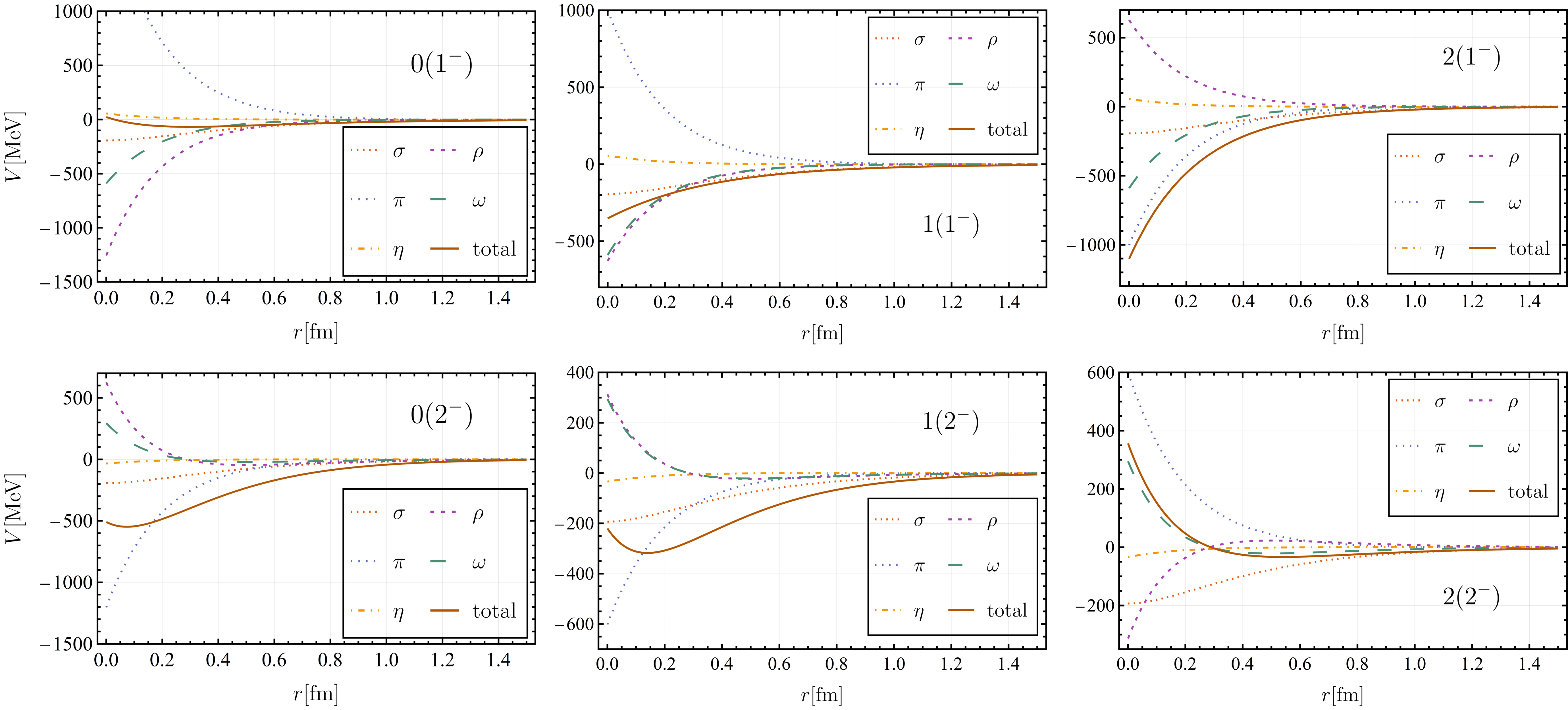}
    \end{overpic}
\caption{Effective potentials for the $S$-wave $\Sigma_c^*\bar{\Sigma}$ systems with $\Lambda=1$ GeV.}
\label{effective potential with tensor term}
\end{figure*}

\begin{figure}[tb]
 \centering
 \begin{overpic}[width=\linewidth]{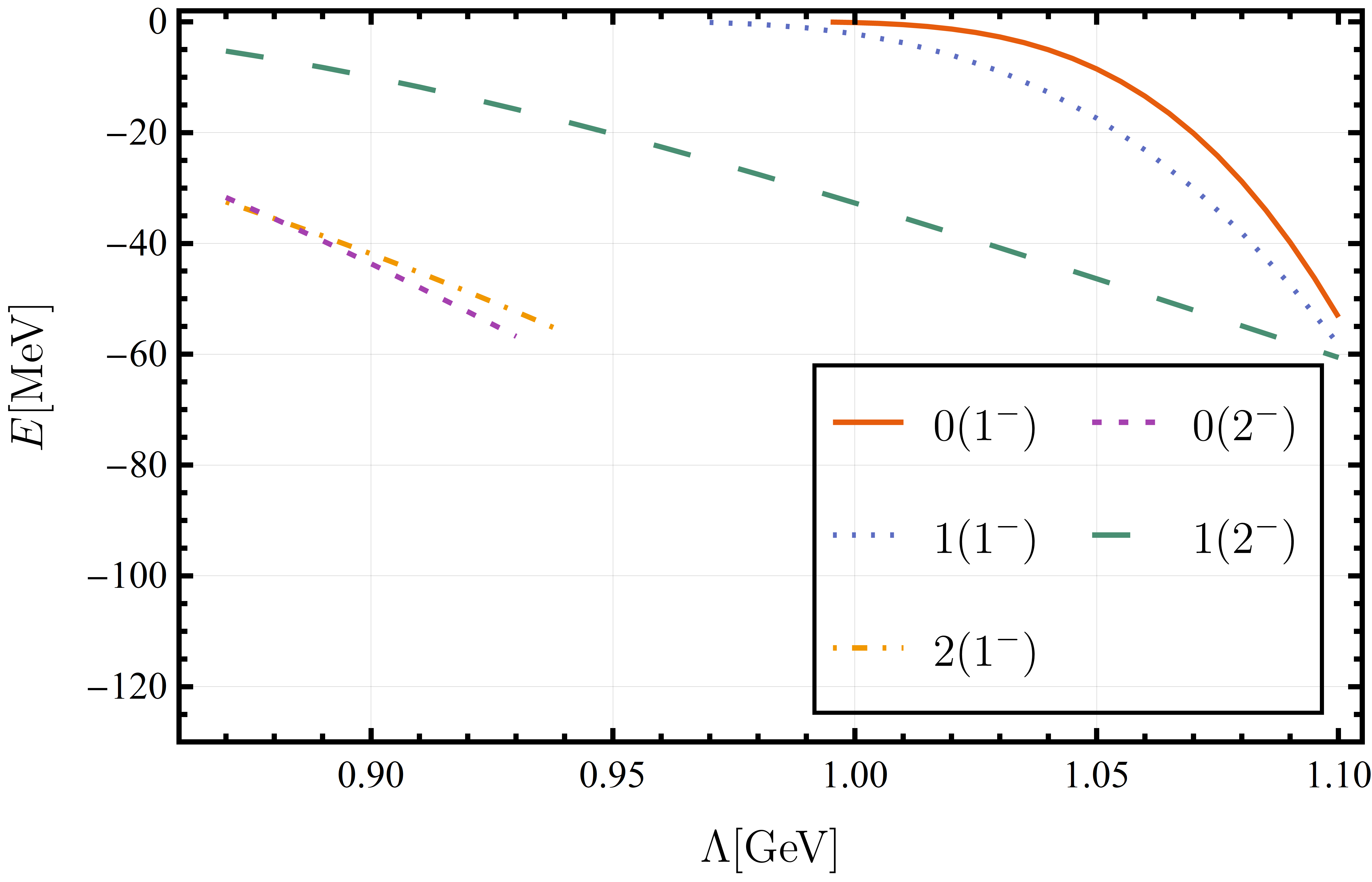}
    \end{overpic}
\caption{Dependence of the binding energy $E$ on the cutoff $\Lambda$ for the $S$-wave $\Sigma_c^*\bar{\Sigma}$ systems with the potential in Eq.~\eqref{eq:Vtotal}.}
\label{relation for E and cutoff for general OBE}
\end{figure}

The quantum numbers $I(J^P)$ of the $S$-wave $\Sigma_c^*\bar{\Sigma}$ systems encompass $0(1^-)$, $1(1^-)$, $2(1^-)$, $0(2^-)$, $1(2^-)$ and $2(2^-)$. Figure~\ref{effective potential with tensor term} showcases the effective potentials that include both the $\delta^{3}(\vec r)$ term and the vector-meson tensor coupling term. The total effective potential in our calculation comprises the exchanges of $\sigma$, $\pi$, $\eta$, $\rho$ and $\omega$, i.e.,
\begin{align}
V_{\text{total}}(r)=V_{\sigma}(r)+V_{\pi}(r)+V_{\eta}(r)+V_{\rho}(r)+V_{\omega}(r).
\label{eq:Vtotal}
\end{align} 
This effective potential is used to solve the Schr\"odinger equation~(\ref{equ:s-wave Schordinger equation}) to search for bound state solutions for the specific quantum numbers. 
The results obtained by varying $\Lambda$ from 0.8~GeV to 1.1~GeV are depicted in Fig.~\ref{relation for E and cutoff for general OBE}. It is evident that the employed potential supports $\Sigma_c^*\bar{\Sigma}$ bound state solutions when the cutoff is larger than certain values in the chosen range, except for the case of $2(2^-)$. 

\subsection{General relation between $\pi$- and $\rho$-exchange potentials in $S$-wave $\mathbb B\bar{\mathbb B}'$ systems}\label{general relation from quark model}

If we use the same form factor with the same cutoff for all the potentials of different mesons, as commonly done in literature, 
a distinct characteristic can be observed
from Fig.~\ref{effective potential with tensor term}: for the $S$-wave $\Sigma_c^*\bar{\Sigma}$ systems, the pion-exchange potential (including the $\delta^{3}(\vec{r}\,)$) and the $\rho$-exchange potential (including the tensor-term contribution) always have opposite signs, suggesting a mutual cancellation. A similar phenomenon is also noticeable in the $\Sigma_c^*\bar{\Sigma}_c^*$, $\Xi_c^*\bar{\Xi}_c^*$, $\Sigma_c\bar{\Sigma}_c$ and $\Xi'_c\bar{\Xi}'_c$ systems~\cite{Yang:2018amd,Lee:2011rka}. In the following, we will use the quark model to demonstrate that this pattern holds for any $S$-wave baryon-antibaryon ($\mathbb B\bar{\mathbb B}'$)  system: the total pion-exchange potential is comparable in magnitude to the tensor-term contribution in the $\rho$-exchange potential, but with opposite signs. This observation provides a theoretical substantiation for the model considering only the vector term for the vector-meson exchange potential~\cite{Dong:2021juy,Dong:2021bvy}. 

As per Refs.~\cite{Riska:2000gd,Meng:2017udf}, at the quark level, the Lagrangian for the coupling of  pseudoscalar ($\mathcal{P}$), vector ($\mathcal{V}$) and $\sigma$ mesons
and quarks reads 
\begin{align}
\mathcal{L}_q=g_{pqq}\bar{q}i\gamma_5 \mathcal{P}q+g_{vqq}\bar{q}\gamma_\mu {\mathcal{V}}^\mu q+g_{\sigma qq}\bar{q}\sigma q,\label{lagrangian at quark level}
\end{align}
where $q=(u,d,s)^T$ represents the light quark flavor triplet, and $g_{pqq}$, $g_{vqq}$, $g_{\sigma qq}$ are the couplings of the light quark to the light mesons. The Lagrangian in Eq.~\eqref{lagrangian at quark level}, assuming interaction vertices calculated at the quark and hadron levels to be identical, is frequently utilized to estimate certain coupling constants~\cite{Riska:2000gd,Meng:2017udf,Liu:2018bkx}. For instance, the relation between $g_{\pi \mathbb B \mathbb B}$ and $g_{\pi qq}$, the former of which represents the coupling constant between a baryon $\mathbb B$ and pion in $\mathcal{L}_{\pi \mathbb B \mathbb B}$, can be derived from 
\begin{align}
    \langle \mathbb B,\vec{s}\, \vert \mathcal{L}_{\pi \mathbb B \mathbb B}\vert \mathbb B,\vec{s}\,\rangle \equiv \langle \mathbb B,\vec{s}\, \vert \mathcal{L}_{\pi qq} \vert \mathbb B,\vec{s}\,\rangle,\label{quark level=hadron level}
\end{align}
where $\vec{s}$ represents the spin of $\mathbb B$. The calculation of the right-hand side of the above equation necessitates specific quark-model wavefunctions for the initial and final states. Following Ref.~\cite{Meng:2017udf}, we deduce 
\begin{align}
    g_{pqq}&=\frac{3\sqrt{2}}{5}\frac{m_q}{M_N}g_{\pi NN},\\
    g_{vqq}&=\sqrt{2}g_{\rho NN},\\
    g_{\sigma qq}&=\frac{1}{3}g_{\sigma NN},
\end{align}
where $g_{\pi NN}$, $g_{\rho NN}$ and $g_{\sigma NN}$ can be obtained by fitting to experimental data and $m_q\approx {M_N}/{3}\approx 313$ MeV~\cite{Riska:2000gd} is the constituent quark mass. Utilizing $g_{\pi NN}^2/4\pi=13.6$, $g_{\rho NN}^2/4\pi=0.84$~\cite{Machleidt:2000ge,Machleidt:1987hj}, and $g_{\sigma NN}=8.7$~\cite{Wu:2023uva}, we obtain $g_{pqq}\approx3.7$, $g_{vqq}\approx4.6$ and $g_{\sigma qq}\approx 2.9$.

\begin{figure}[t]
 \centering
 \begin{overpic}[width=\linewidth]{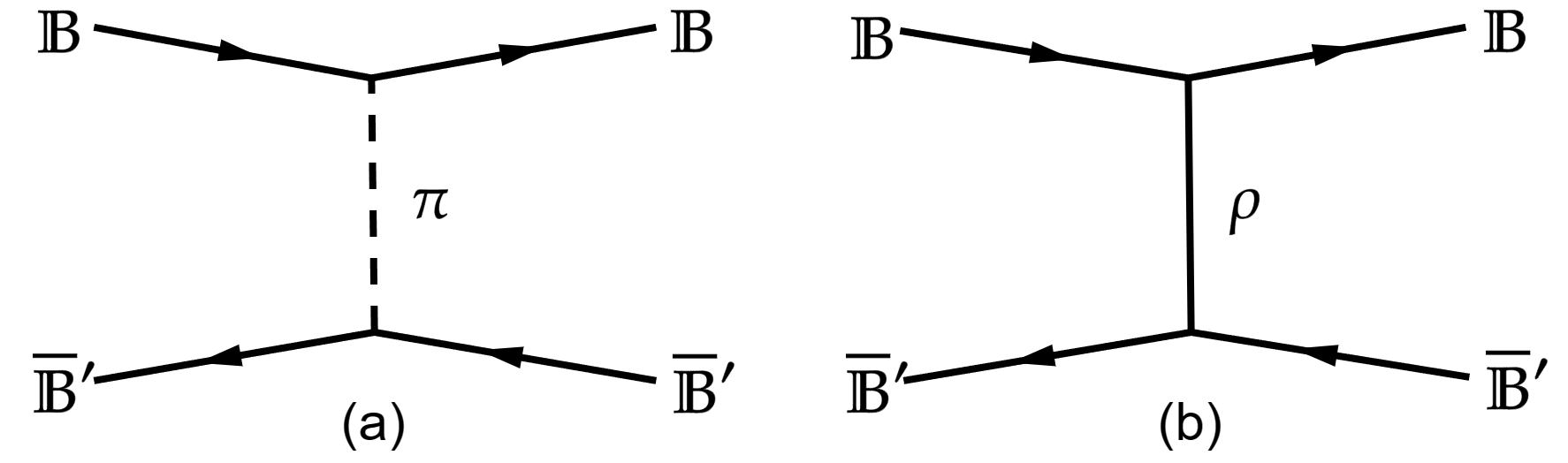}
    \end{overpic}
\caption{Diagrams for the $t$-channel pion and $\rho$-meson exchanges for $\mathbb B\bar{\mathbb B}'\to \mathbb B\bar{\mathbb B}'$.
}
\label{feynman diagram 2}
\end{figure}

In order to evaluate the contributions of the pion-exchange and the $\rho$-exchange in a generic $\mathbb B\bar{\mathbb B}'$ system, we will examine the amplitudes of the two processes depicted in Fig.~\ref{feynman diagram 2}(a) and (b). At the hadronic level, we have 
\begin{align}
    \mathcal{M}_\pi(\mathbb B\bar{\mathbb B}'\to \mathbb B\bar{\mathbb B}')
    =-\frac{\langle \mathbb B\vert \mathcal{L}_{\pi \mathbb B \mathbb B}\vert \mathbb B\rangle\langle \bar{\mathbb B}'\vert \mathcal{L}_{\pi \mathbb B'\mathbb B'} \vert \bar{\mathbb B}'\rangle}{Q^2-m_\pi^2} ,
\end{align}
where $Q$ denotes the four-momentum of the exchanged particle. Concurrently, with Eq.~(\ref{quark level=hadron level}), the above equation can be expressed at the quark level as
\begin{align}
    \mathcal{M}_\pi(\mathbb B\bar{\mathbb B}'\to \mathbb B\bar{\mathbb B}')=-\frac{\langle \mathbb B\vert \mathcal{L}_{\pi q_1q_1}\vert \mathbb B\rangle\langle \bar{\mathbb B}'\vert \mathcal{L}_{\pi q_2q_2} \vert \bar{\mathbb B}'\rangle}{Q^2-m_\pi^2} .
\end{align}
Utilizing Eq.~(\ref{lagrangian at quark level}), we obtain\footnote{Note that we omitted the flavor index in Eqs.~(\ref{quark-level amplitude of pi},\ref{quark-level amplitude of rho}) because it is evident from Eq.~(\ref{lagrangian at quark level}) that the pion and $\rho$ exchanges possess identical flavor structure, which does not influence the assessment of their relative strength.}
\begin{align}
    \mathcal{M}_\pi(\mathbb B\bar{\mathbb B}'\to \mathbb B\bar{\mathbb B}')=\langle \mathbb B\bar{\mathbb B}'\vert \frac{-g_{pqq}^2}{8m_q^2}\frac{\vec{\sigma}_1\cdot\vec{Q}\vec{\sigma}_2\cdot\vec{Q}}{Q^2-m_\pi^2}\vert \mathbb B\bar{\mathbb B}'\rangle.\label{quark-level amplitude of pi}
\end{align}
Similarly, we derive the amplitude of the $\rho$ exchange  as
\begin{align}
    \mathcal{M}_\rho(\mathbb B\bar{\mathbb B}'\to \mathbb B\bar{\mathbb B}')=\langle \mathbb B\bar{\mathbb B}'\vert \frac{g_{vqq}^2}{2}\frac{1}{Q^2-m_\rho^2} \vert \mathbb B\bar{\mathbb B}'\rangle \notag\\
    + \langle \mathbb B\bar{\mathbb B}'\vert\frac{g_{vqq}^2}{8m_q^2}\frac{(\vec{Q}\times\vec{\sigma}_1)\cdot(\vec{Q}\times\vec{\sigma}_2)}{Q^2-m_\rho^2} \vert \mathbb B\bar{\mathbb B}'\rangle,\label{quark-level amplitude of rho}
\end{align}
where the second term on the right-hand side corresponds to the contribution of the tensor term at the hadronic level. 

\begin{figure*}[tbhp]
 \centering
 \begin{overpic}[width=0.9\linewidth]{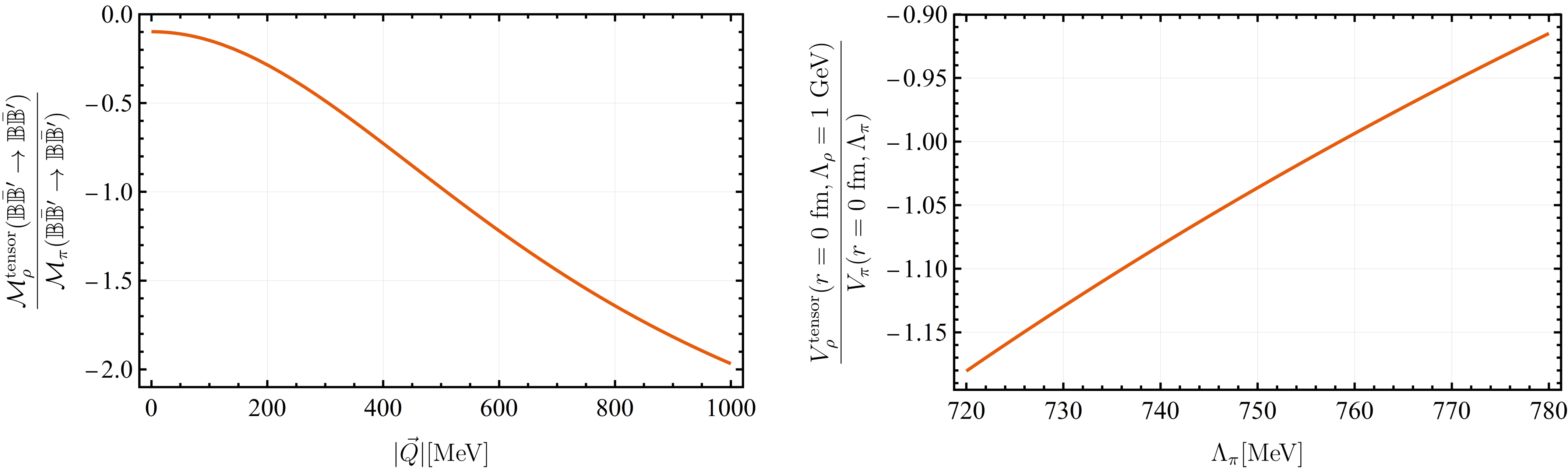}
 \put(43,26){\normalsize{(a)}}
 \put(63,26){\normalsize{(b)}}
    \end{overpic}
\caption{(a) Ratio of the tensor-term contribution in the $\rho$-exchange amplitude to the pion-exchange amplitude in the $S$-wave $\mathbb B\bar{\mathbb B}'\to \mathbb B\bar{\mathbb B}'$ process, and (b) ratio of the tensor-term contribution in the $\rho$-exchange potential to the pion-exchange potential at $r=0$~fm with $\Lambda_\rho=1$ GeV in the $S$-wave $\mathbb B\bar{\mathbb B}'$ system.}
\label{compare Vtensor in rho and Vpi}
\end{figure*}

Using $(\vec{a}_1\times\vec{a}_2)\cdot(\vec{a_1}\times\vec{a}_3)=\vec{a}_1^2(\vec{a}_2\cdot\vec{a}_3)-(\vec{a}_1\cdot\vec{a}_2)(\vec{a}_1\cdot\vec{a}_3)$ and Eq.~(\ref{equ:AqBq}),  for the $S$-wave $\mathbb B\bar{\mathbb B}'$ system we get,
\begin{align}
\mathcal{M}_\rho^{\rm{tensor}}(\mathbb B\bar{\mathbb B}'\to \mathbb B\bar{\mathbb B}')&=\langle \mathbb B\bar{\mathbb B}'\vert\frac{g_{vqq}^2}{8m_q^2}\frac{2}{3}\frac{\vec{\sigma}_1\cdot\vec{\sigma}_2\vec{Q}^2}{Q^2-m_\rho^2} \vert \mathbb B\bar{\mathbb B}'\rangle,\label{rho tensor term compare}\\
\mathcal{M}_\pi(\mathbb B\bar{\mathbb B}'\to \mathbb B\bar{\mathbb B}')&=\langle \mathbb B\bar{\mathbb B}'\vert \frac{-g_{pqq}^2}{8m_q^2}\frac{1}{3}\frac{\vec{\sigma}_1\cdot\vec{\sigma}_2\vec{Q}^2}{Q^2-m_\pi^2}\vert \mathbb B\bar{\mathbb B}'\rangle,\label{pi compare}
\end{align}
and their relative strength reads
\begin{align}
\frac{\mathcal{M}_\rho^{\rm{tensor}}(\mathbb B\bar{\mathbb B}'\to \mathbb B\bar{\mathbb B}')}{\mathcal{M}_\pi(\mathbb B\bar{\mathbb B}'\to \mathbb B\bar{\mathbb B}')}=-\frac{2g_{vqq}^2}{g_{pqq}^2}\frac{\vec{Q}^2+m_\pi^2}{\vec{Q}^2+m_\rho^2}.\label{the ratio of Vtensor in rho and Vpi}
\end{align}
As illustrated in Fig.~\ref{compare Vtensor in rho and Vpi}(a), the ratio lies between approximately $-0.1$ and $-2.0$ as $\vert\vec{Q}\vert$ varies from $0$ to $1$~GeV, indicating a certain degree of cancellation. To more accurately depict this mutual cancellation effect, we convert Eqs.~(\ref{rho tensor term compare},\ref{pi compare}) into the coordinate space using Eq.~(\ref{equ:q^2}). Consequently, the ratio of the contribution from the tensor term in the $\rho$-exchange potential to the pion-exchange potential in the $S$-wave $\mathbb B\bar{\mathbb B}'$ system reads
\begin{align}
    \frac{V_\rho^{\rm{tensor}}(r,\Lambda_\rho)}{V_\pi(r,\Lambda_\pi)}=\frac{-2g_{vqq}^2}{g_{pqq}^2}\frac{h(r,m_\rho,\Lambda_\rho)-m_\rho^2g(r,m_\rho,\Lambda_\rho)}{h(r,m_\pi,\Lambda_\pi)-m_\pi^2g(r,m_\pi,\Lambda_\pi)}.\label{equ for Vrho and Vpi at r=0}
\end{align}
At $r=0$ fm, $\Lambda_\rho=\Lambda_\pi=1$ GeV, we have
\begin{align}
    \frac{V_\rho^{\rm{tensor}}(r=0\ {\rm fm},\Lambda_\rho=1\ {\rm{GeV}})}{V_\pi(r=0\ {\rm fm},\Lambda_\pi=1\ {\rm{GeV}})}\approx-0.42,
\end{align}
in line with Fig.~\ref{effective potential with tensor term}. 
Varying the cutoff for the pion exchange to a smaller value, a larger cancellation may be achieved,
\begin{align}
    \frac{V_\rho^{\rm{tensor}}(r=0\ {\rm fm},\Lambda_\rho=1\ {\rm{GeV}})}{V_\pi(r=0\ {\rm fm},\Lambda_\pi=0.76\ {\rm{GeV}})}\approx-1.0,
\end{align}
as depicted in Fig.~\ref{compare Vtensor in rho and Vpi}(b). 

\begin{figure*}[tbhp]
 \centering
 \begin{overpic}[width=0.9\linewidth]{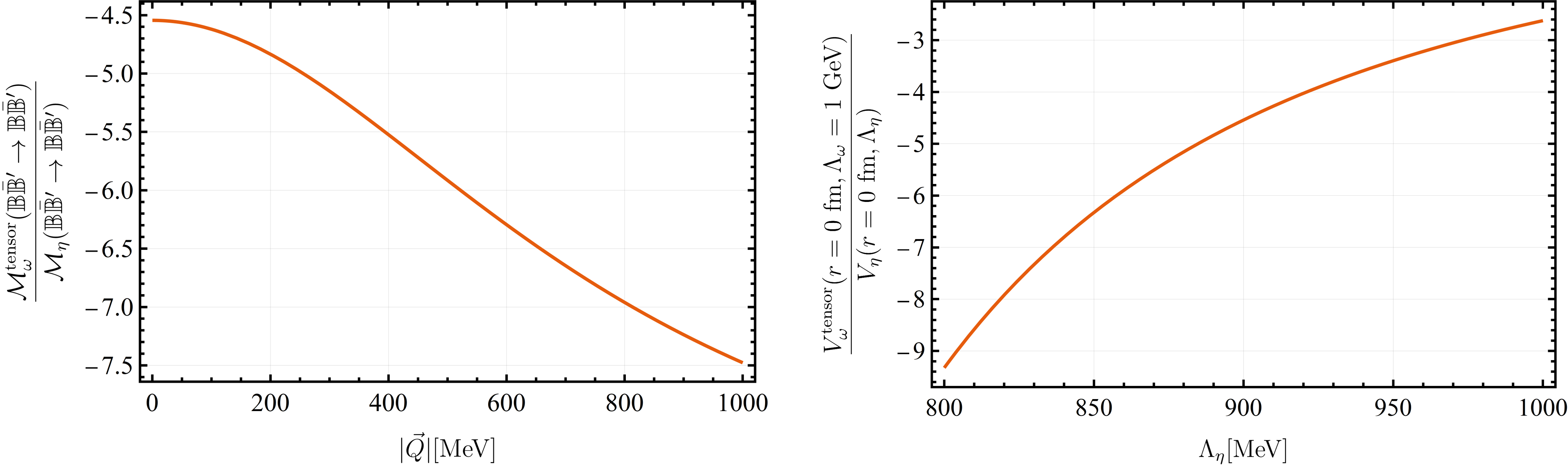}
 \put(43,26){\normalsize{(a)}}
 \put(61.5,26){\normalsize{(b)}}
    \end{overpic}
\caption{(a) Ratio of the tensor-term contribution in $\omega$-exchange amplitude to total $\eta$-exchange amplitude in the $S$-wave $\mathbb B\bar{\mathbb B}'\to \mathbb B\bar{\mathbb B}'$ process, and (b) ratio of the tensor-term contribution in $\omega$-exchange potential to total $\eta$-exchange potential at $r=0$~fm and $\Lambda_\omega=1$ GeV in the $S$-wave $\mathbb B\bar{\mathbb B}'$ system.}
\label{compare Vtensor in omega and Veta}
\end{figure*}

\begin{figure*}[tbhp]
 \centering
 \begin{overpic}[width=0.9\linewidth]{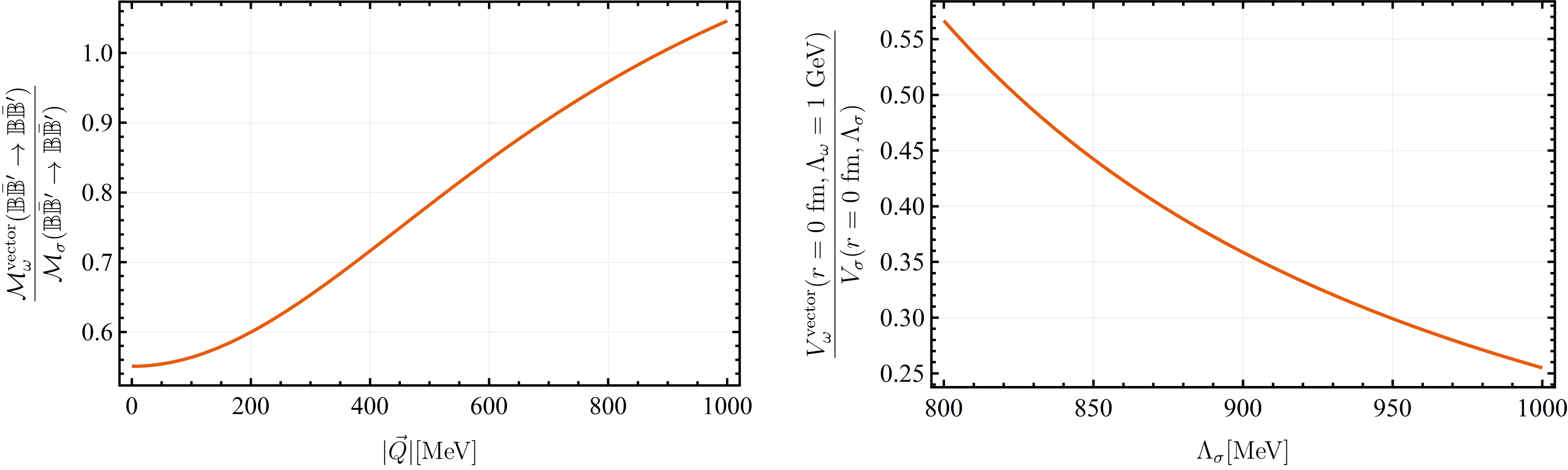}
 \put(10,26){\normalsize{(a)}}
 \put(94.5,26){\normalsize{(b)}}
    \end{overpic}
\caption{ (a) Ratio of the vector-term contribution in $\omega$-exchange amplitude to the total $\sigma$-exchange amplitude in the $S$-wave $\mathbb B\bar{\mathbb B}'\to \mathbb B\bar{\mathbb B}'$ process, and (b) ratio of vector-term contribution in $\omega$-exchange potential to total $\sigma$-exchange potential at $r=0$~fm and $\Lambda_\omega=1$~GeV in the $S$-wave $\mathbb B\bar{\mathbb B}'$ system.}
\label{compare VLO in omega and Vsigma}
\end{figure*}

The same analysis can be applied to other pseudoscalar mesons and  vector mesons, provided they share the same flavor structure. For instance, in the case of the $S$-wave $\mathbb B\bar{\mathbb B}'$ system where the light quark component includes only $u$, $d$, $\bar{u}$ and $\bar{d}$, we can conduct a similar analysis for $\eta$, $\omega$ and $\sigma$. The results are shown in Figs.~\ref{compare Vtensor in omega and Veta} and \ref{compare VLO in omega and Vsigma}. It is observed that the contribution of the tensor term in the $\omega$-exchange potential is opposite in sign to that of the $\eta$-exchange potential. Moreover, the former is significantly stronger than the latter, which further elucidates why the contribution of the $\eta$ is nearly negligible in the general OBE model. Concurrently, the vector coupling term in the $\omega$-exchange potential at short distances is comparable in magnitude to that of the $\sigma$-exchange potential and shares the same sign. 

In conclusion, we find that it is a plausible approximation to consider the contribution of the tensor term in the $\rho$-exchange potential and the pion-exchange potential as mutually cancelling, i.e., $V_\pi+V_\rho^{\rm{tensor}}\approx0$, in the OBE model for any $S$-wave $\mathbb B\bar{\mathbb B}'$ systems.  
In addition, if the light quark component comprises only $u$, $d$, $\bar{u}$ and $\bar{d}$, then the $\eta$-exchange potential becomes entirely negligible in comparison to the $\omega$-exchange potential. Given the spin-isospin independence of the $\sigma$ meson, which effectively leads to a single background term, this observation elucidates the rationality of the OBE model being dominated by the exchange of vector mesons. 

\subsection{Results after considering $V_\pi+V_\rho^{\rm{tensor}}\approx 0$}\label{new result of OBE}

\begin{figure*}[tbhp]
 \centering
 \begin{overpic}[width=1\linewidth]{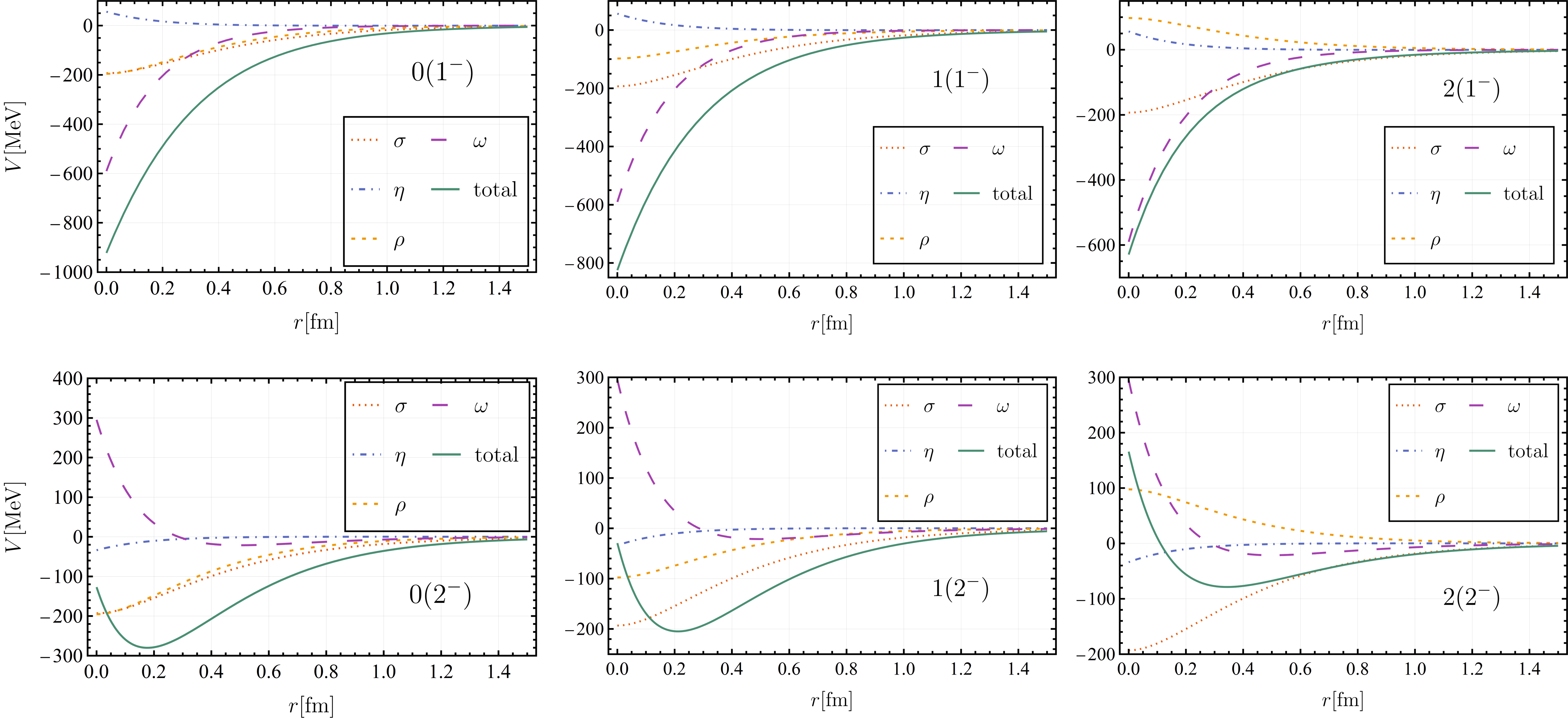}
    \end{overpic}
\caption{Effective potentials for the $S$-wave $\Sigma_c^*\bar{\Sigma}$ systems at $\Lambda=1$ GeV after dropping the pion-exchange potential and the contribution of the tensor term in the $\rho$-exchange potential.}
\label{effective potential without pi without tensor term in rho}
\end{figure*}

\begin{figure}[t]
 \centering
  \begin{overpic}[width=\linewidth]{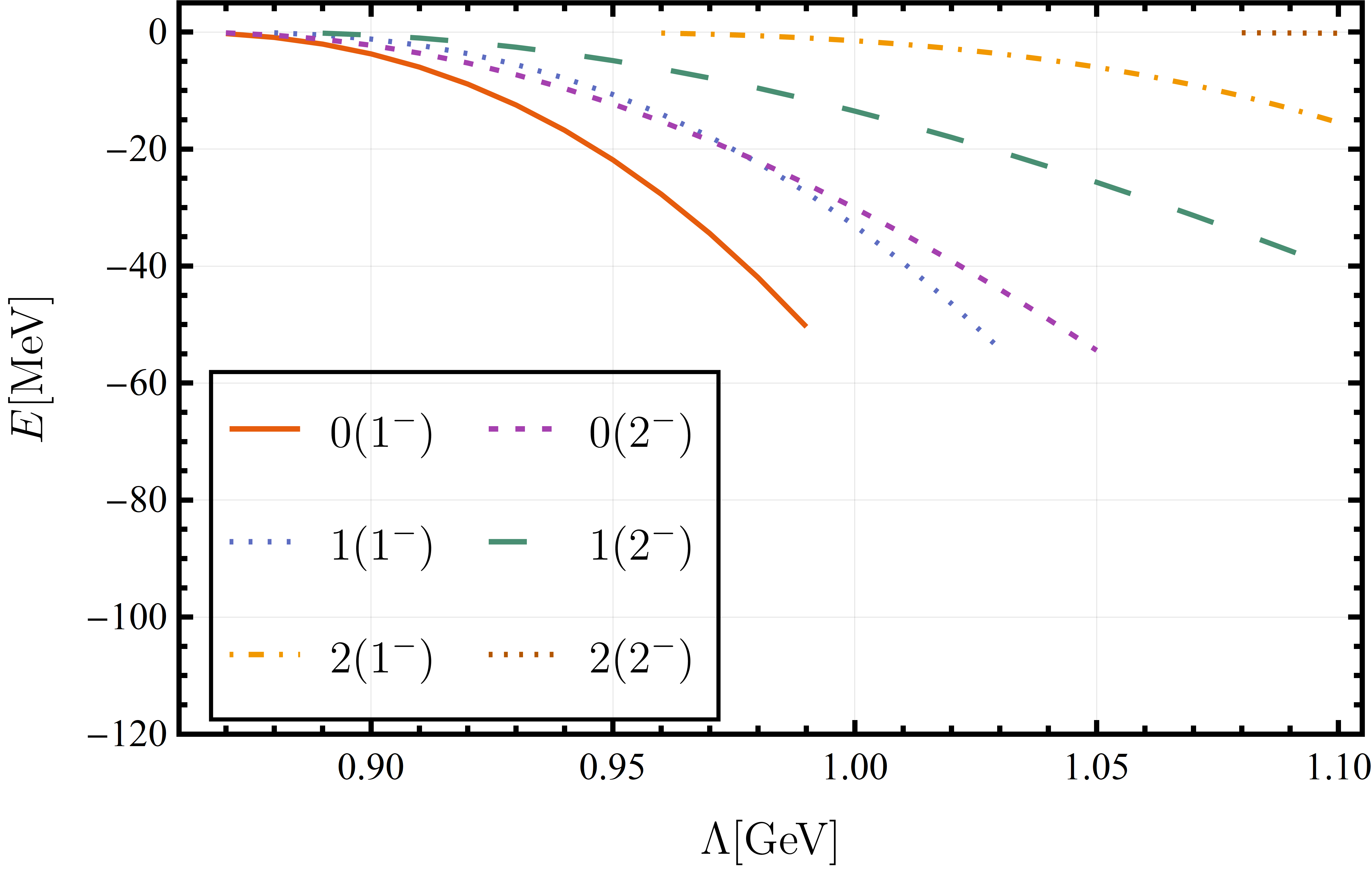}
    \end{overpic}
  \caption{Dependence of the binding energy $E$ on the cutoff $\Lambda$ for the $S$-wave $\Sigma_c^*\bar{\Sigma}$ systems with the potential in Eq.~\eqref{eq:Vreduced} which has dropped the pion-exchange potential and the contribution of the tensor term in the $\rho$-exchange potential.} 
  \label{relationship between cutoff and bounding energy} 
\end{figure}

From the above discussion, one may use the following approximation for the effective potential,
\begin{align}  V_{\rm{total}}(r)=V_{\sigma}(r)+V_\eta(r)+V_\rho^{{\rm{vector}}}(r)+V_\omega(r),
    \label{eq:Vreduced}
\end{align}
shown in Fig.~\ref{effective potential without pi without tensor term in rho}.
Results for the binding energies of the $S$-wave $\Sigma_c^*\bar{\Sigma}$ system with this potential are depicted in Fig.~\ref{relationship between cutoff and bounding energy}. 
The difference between the corresponding curves in Fig.~\ref{relation for E and cutoff for general OBE} and Fig.~\ref{relationship between cutoff and bounding energy} is an indication of the unavoidable model dependence of the OBE model.
Nevertheless, a $\Sigma_c^*\bar{\Sigma}$ bound state solution exists for  $0(2^-)$ and $1(2^-)$ for both potentials with the cutoff in the range between 0.9 to 1.1~GeV. 

\section{Summary}
\label{Sec:V}
In this work, we take the calculation of the $\Sigma_c^*\bar{\Sigma}$ bound states as an example and systematically clarify the complex issues encountered in the OBE model, including the effects of the sum of initial and final state momenta $\vec{k}$, the $\delta^{3}(\vec{r}\,)$ potential, and the contribution of the tensor term in the vector-meson exchange. 
The momentum $\vec{k}$ in the amplitude, which originates solely from the spinors and introduces derivatives of the radial wavefunction, is suppressed as $\mathcal{O}(\vec k^2/M^2)$ in the potential and thus negligible when the particle mass is significantly heavier than the binding momentum of the bound state. 
For the $\Sigma_c^*\bar{\Sigma}$ systems, we retain the $\vec{k}$ dependence as the $\Sigma$ is a light baryon. 

We find using quark model relations that for any $S$-wave baryon-antibaryon system the pion-exchange potential with the $\delta^{3}(\vec{r}\,)$ term and the tensor coupling contribution to the $\rho$-exchange potential have similar magnitudes but with different signs, indicating a tendency for mutual cancellation. 

Despite the model dependence of the results, we find that $I(J^P)=0(2^-)$ and $1(2^-)$ each emerge as the most probable quantum numbers to have a $\Sigma_c^*\bar{\Sigma}$ bound state, with mass around 3.7~GeV. They may be looked for in the final states of $\bar D_s\Sigma^*_c\bar\Sigma$, $\bar D_s\Sigma^*_c\bar\Lambda$, $\bar D_s\Lambda_c\bar\Sigma^{(*)}$, $\bar D_s\Lambda_c\bar\Lambda$, $\bar D_s D^*_s\pi$, $\bar D_s D^*_s\eta$, $\bar D_s D_s\rho$, $\bar D_s D_s\omega$, $\bar D_s D^* K$, etc. from the $e^+e^-$ annihilation process at Belle-II or experiments at other electron-positron colliders with higher luminosity in the future.

\bigskip

\begin{acknowledgements}
We extend our gratitude to Bing-Ran He, Hao-Jie Jing, Jia-Jun Wu, Shu-Ming Wu and Nijiati Yalikun for valuable discussions. We would like to thank Ulf-G. Mei{\ss}ner for a careful reading of the manuscript.
This work is supported in part by the National Key R\&D Program of China under Grant No. 2023YFA1606703; by the Chinese Academy of Sciences under Grants No.~XDB34030000 and No.~YSBR-101;
by the National Natural Science Foundation of China (NSFC) and the Deutsche Forschungsgemeinschaft (DFG) through the funds provided to the Sino-German Collaborative Research Center TRR110 ``Symmetries and the Emergence of Structure in QCD'' (NSFC Grant No. 12070131001, DFG Project-ID 196253076); and by the NSFC under Grants No. 12125507, No. 12361141819 and No. 12047503.
\end{acknowledgements}

\begin{widetext}
    
\begin{appendix}
\section{Basic formalism of the OBE Model}
\label{Appendix:A}
To find the bound state of two particles, we need to solve the relative-motion part of the Schr\"odinger equation for the two-body system in quantum mechanics (QM), given by
\begin{align}
\hat{H}|\varPsi \rangle=E|\varPsi\rangle.
\end{align}
Here, $\hat{H}$ represents the relative-motion part of the Hamiltonian of the system, and $\vert\varPsi\rangle$ is the wavefunction of the relative motion. 
Let us impose the constraint that the solution of this equation is given by
\begin{align}
\vert\varPsi\rangle=\vert f\rangle_r\vert ^{2S+1} L _J,J_z\rangle\vert I\  I_3\rangle.\label{equ:A2}
\end{align}
Here, the $\vert f\rangle_r$ represents the radial part of the relative-motion wavefunction $|\varPsi\rangle$, and the notation $\vert ^{2S+1} L _J,J_z\rangle\vert I\  I_3\rangle$ denotes that the quantum number of the total spin is $S$, the relative orbital angular momentum is $L$, the total angular momentum is $J$, the third component of total angular momentum is $J_z$, the total isospin is $I$ and the third component of the total isospin is $I_3$ of the system.
We can rewrite the Schr\"odinger equation as 
\begin{align}
\hat{H}\vert f\rangle_r\vert ^{2S+1} L _J,J_z\rangle\vert I\  I_3\rangle=E\vert f\rangle_r\vert ^{2S+1} L _J,J_z\rangle\vert I\  I_3\rangle.
\end{align}
Multiplying $ _r\langle r\vert \langle ^{2S+1}L _J,J_z\vert \langle I\  I_3\vert$ from the left to the above equation, we have
\begin{align}
_r\langle r\vert \langle ^{2S+1}L _J,J_z\vert \langle I\  I_3\vert\hat{H}\vert f \rangle _r\vert ^{2S+1} L _J,J_z\rangle\vert I\  I_3\rangle=E f(r).\label{equ:A4}
\end{align}
Taking into account
\begin{align}
\vert ^{2S+1} L _J,J_z\rangle&=\vert \left( \left( S_1S_2 \right) SL \right) JJ_z\rangle=\sum_{S_z~L_z}\sum_{S_{1z}~S_{2z}}C_{S_1S_{1z};S_2S_{2z}}^{SS_z}C_{SS_z;LL_z}^{JJ_z}\vert S_1S_{1z}\rangle \vert S_2S_{2z}\rangle\vert LL_z\rangle\notag\\
&=\sum_{L_z~S_{1z}}C_{S_1S_{1z};S_2(J_z-L_z-S_{1z})}^{S(J_z-L_z)}C_{S(J_z-L_z);LL_z}^{JJ_z}\vert S_1S_{1z}\rangle\vert S_2(J_z-L_z-S_{1z})\rangle \vert LL_z\rangle,
\end{align}
the complete bases
 \begin{align}
 \int {\rm{d}}^3\vec{x}|\vec{x}\rangle\langle\vec{x}|=\boldsymbol{1},\qquad \int {\rm{d}}^3\vec{p}\,|\vec{p}\,\rangle\langle\vec{p}\,|=\boldsymbol{1},\label{eq:completebasis}
 \end{align}
and $\hat{H}=\frac{\hat{\vec{p}}^{\,2}}{2\mu}+\hat{V}$,
where $C_{S_1S_{1z};S_2S_{2z}}^{SS_z}$ is the Clebsch-Gordan (CG) coefficient for the SU(2) group, and $\mu$ is the reduced mass of the two-body system, Eq.~(\ref{equ:A4}) can be rewritten as
\begin{equation}
\begin{aligned}
&-\frac{1}{2\mu r}\frac{{\rm{d}}^2}{{\rm{d}}r^2}\left( rf(r)\right)+\frac{L(L+1)}{2\mu r^2}f(r)+\sum_{L_z S_{1z}}\sum_{L'_z S_{3z}}C_{S_1 S_{1z};S_2 \left( J_z-L_z-S_{1z} \right)}^{S \left( J_z-L_z \right)}C_{S \left( J_z-L_z \right); L L_z}^{J J_z}C_{S_1 S_{3z};S_2 \left( J_z-L'_z-S_{3z} \right)}^{S \left( J_z-L'_z \right)}\\
&C_{S \left( J_z-L'_z \right); L L'_z}^{J J_z}\int {\rm{d}}x\delta (r-x)\Bigl( \int {\rm{d}}\Omega {\rm{d}}^3\vec{p}^{\,\prime}{\rm{d}}^3\vec{p}{\rm{d}}^3\vec{x}^{\,\prime}f(x')Y_L^{L_z}(\theta',\varphi')Y_L^{L'_z*}(\theta,\varphi)\\
&\langle\vec{x}|\vec{p}^{\,\prime}\rangle\langle\vec{p}\,|\vec{x}^{\,\prime}\rangle \langle\vec{p}^{\,\prime},S_1S_{3z},S_2 \left( J_z-L'_z-S_{3z} \right),II_3|\hat{V}|\vec{p},S_1S_{1z},S_2 \left( J_z-L_z-S_{1z} \right),II_3\rangle \Bigr)=Ef\!\left(r\right), \label{equ:A8}
\end{aligned}
\end{equation}
with the boundary conditions
\begin{align}
\lim_{r \to 0}rf(r)=0,\qquad\lim_{r \to \infty}rf(r)=0\label{equ:A9}.
\end{align}

We will solve Eq.~(\ref{equ:A8}) for the radial wavefunction $f(r)$, subject to the boundary conditions in Eq.~(\ref{equ:A9}), to find bound states. Furthermore, for simplicity, we define
\begin{align}
\hat{\sum}&\equiv \sum_{L_z S_{1z}}\sum_{L'_z S_{3z}}C_{S_1 S_{1z};S_2\left(J_z-L_z-S_{1z}\right)}^{S \left(J_z-L_z\right)}C_{S \left(J_z-L_z\right); L L_z}^{J J_z}C_{S_1 S_{3z};S_2\left(J_z-L'_z-S_{3z}\right)}^{S\left(J_z-L'_z\right)}C_{S \left(J_z-L'_z\right); L L'_z}^{J J_z}\label{operator},\\
\hat{\sum_{S-{\rm wave}}}&\equiv \sum_{S_{1z}}\sum_{S_{3z}}C_{S_1 S_{1z};S_2\left(J_z-S_{1z}\right)}^{S J_z}C_{S J_z; 0 0}^{J J_z}C_{S_1 S_{3z};S_2\left(J_z-S_{3z}\right)}^{S J_z}C_{S J_z; 0 0}^{J J_z}\label{operator for S-wave}.
\end{align}
Using the relation between the amplitude in Quantum Field Theory (QFT) and the potential in momentum space in QM, Eq.~(\ref{eq:VM}), the Schr\"odinger equation becomes
\begin{align}
-\frac{1}{2\mu r}\frac{{\rm{d}}^2}{{\rm{d}}r^2}\!&\left(rf(r)\right)+\frac{L(L+1)}{2\mu r^2}f(r)+\hat{\sum}\int {\rm{d}}x\delta (r-x)\biggl[\int {\rm{d}}\Omega {\rm{d}}^3\vec{p}^{\,\prime}{\rm{d}}^3\vec{p}{\rm{d}}^3\vec{x}^{\,\prime}f(x')Y_L^{L_z}\!\left(\theta',\varphi'\right)Y_L^{L'_z*}\!\left(\theta,\varphi\right)\langle\vec{x}|\vec{p}^{\,\prime}\rangle\langle\vec{p}\,|\vec{x}^{\,\prime}\rangle\notag\\
&\times\frac{-1}{(2\pi)^3}\mathcal{M}\!\left(\vec{p},S_1S_{1z},S_2 \left( J_z-L_z-S_{1z}\right),II_3\to\vec{p}^{\,\prime},S_1S_{3z},S_2 \left( J_z-L'_z-S_{3z} \right),II_3\right)\biggr]=Ef(r)\label{equ:A10}.
\end{align}
Considering $\langle\vec{r}\,|\vec{p}\,\rangle={(2\pi)^{-3/2}}e^{i\vec{p}\cdot\vec{r}}$, the variable transformations
\begin{equation}
\begin{aligned}
\left\{
\begin{array}{r}
\vec{q} =\vec{p}^{\,\prime}-\vec{p}\\
\vec{k} = \vec{p}^{\,\prime}+\vec{p}\\
\vec{x}_1 = \frac{\vec{x}-\vec{x}^{\,\prime}}{2}\\
\vec{x}_2 = \frac{\vec{x}+\vec{x}^{\,\prime}}{2}\\
\end{array}
\right. \qquad \Longleftrightarrow \qquad\left\{
\begin{array}{l}
\vec{p}=\frac{\vec{k}-\vec{q}}{2}\\
\vec{p}^{\,\prime}=\frac{\vec{k}+\vec{q}}{2}\\
\vec{x}=\vec{x}_1+\vec{x}_2\\
\vec{x}^{\,\prime}=\vec{x}_2-\vec{x}_1\\
\end{array}
\right.,\label{equ:A11}
\end{aligned}
\end{equation}
and the Fourier transformation
\begin{align}
\mathcal{F}_{\vec{x}\to\vec{q}}\!\left[f(\vec{x})\right]&=\int f(\vec{x})e^{-i\vec{q}\cdot\vec{x}}\mathrm{d}^3\vec{x},\\
\mathcal{F}_{\vec{q}\to\vec{x}}^{-1}\!\left[g(\vec{q}\,)\right]&=\frac{1}{(2\pi)^3}\int g(\vec{q}\,)e^{i\vec{q}\cdot\vec{x}}\mathrm{d}^3\vec{q},
\end{align}
the integrals of Eq.~(\ref{equ:A10}) in momentum space can be recast as 
\begin{align}
\int \mathrm{d}^3\vec{p}^{\,\prime}\mathrm{d}^3\vec{p}\langle\vec{x}|\vec{p}^{\,\prime}\rangle\langle\vec{p}\,|\vec{x}^{\,\prime}\rangle\frac{-1}{(2\pi)^3}\mathcal{M}\!\left( \vec{p},\vec{p}^{\,\prime}\right)=-\frac{1}{8}\mathcal{F}_{\vec{q}\to\vec{x}_2}^{-1}\!\left[\mathcal{F}_{\vec{k}\to\vec{x}_1}^{-1}\!\left[\mathcal{M}\!\left(\frac{\vec{k}-\vec{q}}{2},\frac{\vec{k}+\vec{q}}{2}\right)\right]\right],
\end{align}
where the $-1/8$ arises from the variable transformation. Furthermore, we introduce a new function $\psi(r)=rf(r)$ to simplify the calculation further. Finally, the Schr\"odinger equation can be rewritten in the following form
\begin{align}
\psi''(r)-\frac{L(L+1)}{r^2}\psi(r)+2\mu E\psi(r)-2\mu r\hat{V}_{\mid^{2S+1}L_{J},J_z;I,I_3\rangle}^{\mathcal{M}(\vec{p},\vec{p}^{\,\prime})}(r)\frac{\psi(r)}{r}=0\label{equ:A15},
\end{align}
where
\begin{align}
&\,\hat{V}_{\mid^{2S+1}L_{J},J_z;I,I_3\rangle}^{\mathcal{M}(\vec{p},\vec{p}^{\,\prime})}(r)f(r)=-\frac{1}{8}\hat{\sum}\int {\rm{d}}x\delta (r-x)\Biggl[ \int {\rm{d}}\Omega {\rm{d}}^3\vec{x}^{\,\prime}
f(x')Y_L^{L_z}\!\left(\theta',\varphi'\right)Y_L^{L'_z*}\!\left(\theta,\varphi\right)\notag\\
&\times\mathcal{F}_{\vec{q}\to\vec{x}_2}^{-1}\!\Biggl[\mathcal{F}_{\vec{k}\to\vec{x}_1}^{-1}\!\left[\mathcal{M}\!\left(\frac{\vec{k}-\vec{q}}{2},S_1S_{1z},S_2\left(J_z-L_z-S_{1z}\right),II_3\to\frac{\vec{k}+\vec{q}}{2},S_1S_{3z},S_2\left(J_z-L'_z-S_{3z}\right),II_3\right)\right]\Biggr]\Biggr]\label{equ:A16}.
\end{align}
The superscript $\mathcal{M}(\vec{p},\vec{p}^{\,\prime})$  denotes the amplitude corresponding to the effective potential, while the subscript $|^{2S+1}L_{J},J_z;I,I_3\rangle$ represents the state labeled by the corresponding quantum numbers of the two-body system. The quantum numbers of $J_z$ and $I_3$ are generally omitted since they do not affect final results. Similarly, the boundary conditions in Eq.~(\ref{equ:A9}) can be rewritten as
\begin{align}
\lim_{r \to 0}\psi(r)=0,\quad\lim_{r \to \infty}\psi(r)=0\label{equ:A17}.
\end{align}
For $S$-wave ($L$=0), the aforementioned formulas can be simplified as
\begin{align}
\psi''(r)+2\mu E\psi(r)-2\mu r\hat{V}_{\mid^{2S+1}S_{J};I\rangle}^{\mathcal{M}(\vec{p},\vec{p}^{\,\prime})}(r)\frac{\psi(r)}{r}=0,
\label{equ:s-wave Schordinger equation}
\end{align}
where
\begin{align}
&\,\hat{V}_{\mid^{2S+1}S_{J};I\rangle}^{\mathcal{M}(\vec{p},\vec{p}^{\,\prime})}(r)f(r)=-\frac{1}{32\pi}\hat{\sum_{S-{\rm wave}}}\int {\rm{d}}x\delta (r-x)\Biggl[ \int {\rm{d}}\Omega {\rm{d}}^3\vec{x}^{\,\prime}f(x')\notag\\
&\times\mathcal{F}_{\vec{q}\to\vec{x}_2}^{-1}\!\left[\mathcal{F}_{\vec{k}\to\vec{x}_1}^{-1}\!\left[\mathcal{M}\!\left(\frac{\vec{k}-\vec{q}}{2},S_1S_{1z},S_2\left(J_z-S_{1z}\right),II_3\to\frac{\vec{k}+\vec{q}}{2},S_1S_{3z},S_2\left(J_z-S_{3z}\right),II_3\right)\right]\right]\Biggr]\label{equ:A19}.
\end{align}
By further simplifying with the redefined amplitude
\begin{align}
\mathcal{M}_{\mid^{2S+1}S_{J};I\rangle}(\vec{p}\to\vec{p}^{\,\prime})\equiv \hat{\sum_{\rm S-wave}}\mathcal{M}\!\left(\vec{p},S_1S_{1z},S_2\left(J_z-S_{1z}\right),II_3\to\vec{p}^{\,\prime},S_1S_{3z},S_2\left(J_z-S_{3z}\right),II_3\right)\label{redefined amplitude},
\end{align}
Eq.~(\ref{equ:A19}) can be streamlined to
\begin{align}
\hat{V}_{\mid^{2S+1}S_{J};I\rangle}^{\mathcal{M}(\vec{p},\vec{p}^{\,\prime})}(r)f(r)=-\int {\rm{d}}x\delta (r-x)\Biggl[ \int {\rm{d}}\Omega {\rm{d}}^3\vec{x}^{\,\prime}
\frac{f(x')}{32\pi}\mathcal{F}_{\vec{q}\to\vec{x}_2}^{-1}\!\left[\mathcal{F}_{\vec{k}\to\vec{x}_1}^{-1}\!\left[\mathcal{M}_{\mid^{2S+1}S_{J};I\rangle}\left(\frac{\vec{k}-\vec{q}}{2}\to\frac{\vec{k}+\vec{q}}{2}\right)\right]\right]\Biggr]\label{equ:A21}.
\end{align}

It is worth noting that, in most papers concerning the OBE model, the amplitude generally does not include terms depending on the sum of the initial and final state c.m. momenta $\vec k$, i.e., setting $\vec k= \vec p + \vec p\,'=0$. 
As a result, only the momentum $\vec{q}$ of the exchanged meson from the propagator remains in the amplitude of Eq.~(\ref{equ:A16}). For this specific case, according to
\begin{align}
-\frac{1}{8}\mathcal{F}_{\vec{q}\to\vec{x}_2}^{-1}\!\left[ \mathcal{F}_{\vec{k}\to\vec{x}_1}^{-1}\!\left[\mathcal{M}(\vec{q}\,)\right]\right]=-\mathcal{F}_{\vec{q}\to\vec{x}_2}^{-1}\!\left[\mathcal{M}(\vec{q}\,)\right]\delta^{3}\!\left({\vec{x}-\vec{x}^{\,\prime}}\right),
\end{align}
Eq.~(\ref{equ:A16}) can be further simplified to
\begin{align}
\hat{V}_{\mid^{2S+1}L_{J};I\rangle}^{\mathcal{M}(\vec{q}\,)}(r)f(r)=-
\hat{\sum}\left[\int {\rm{d}}\Omega 
Y_L^{L_z}\!\left(\theta,\varphi\right)Y_L^{L'_z*}\!\left(\theta,\varphi\right)\mathcal{F}_{\vec{q}\to\vec{r}}^{-1}\!\left[\mathcal{M}\!\left(\vec{q}\right)\right]\right]f(r)\label{equ:A23}. 
\end{align}
Clearly, the impact of an effective potential operator on the radial wavefunction, i.e., $\hat{V}_{\mid^{2S+1}L_{J};I\rangle}^{\mathcal{M}(\vec{q}\,)}(r)f(r)$, can be simply regarded as an effective potential function
\begin{align} 
V_{\mid^{2S+1}L_{J};I\rangle}^{\mathcal{M}(\vec{q}\,)}(r)=\hat{\sum}\left[\int {\rm{d}}\Omega 
Y_L^{L_z}\!\left(\theta,\varphi\right)Y_L^{L'_z*}\!\left(\theta,\varphi\right)
(-1)\mathcal{F}_{\vec{q}\to\vec{r}}^{-1}\!\left[\mathcal{M}\!\left(\vec{q}\right)\right]\right]\label{equ:A24}.
\end{align}
Hence, in the subsequent discussion of the amplitude, which only contains the momenta $\vec{q}$, we may get rid of the hat on $\hat{V}$ to imply that its effect is equivalent to a function in Schr\"odinger equation.

In particular, with the redefined amplitude in Eq.~(\ref{redefined amplitude}), the corresponding case for $S$-wave is
\begin{align}
\hat{V}_{\mid^{2S+1}S_{J};I\rangle}^{\mathcal{M}(\vec{q}\,)}(r)f(r)=-\frac{1}{4\pi}\int {\rm{d}}\Omega \mathcal{F}_{\vec{q}\to\vec{r}}^{-1}\!\left[\mathcal{M}_{\mid^{2S+1}S_{J};I\rangle}(\vec{q}\,)\right]f(r).
\end{align}
In other words, when the amplitude contains only momentum $\vec{q}$, computing the $S$-wave effective potential boils down to taking the average of the redefined amplitude across the full solid angle space after applying a Fourier transformation, subject to a minus sign determined by the established convention within the relation between amplitude and potential.

We introduce an monopole form factor 
\begin{align}
F(q)=\frac{\Lambda^2-m_{\rm{ex}}^2}{\Lambda^2-q^2}\notag
\end{align}
at each vertex, where $\Lambda$ represents the cutoff parameter and $m_{\rm{ex}}$ denotes the mass of the exchanged meson. Since we are interested in near-threshold bound state, we disregard the term of $\mathcal{O}(\frac{1}{M^2})$. Actually, we only need to calculate the following cases of $\mathcal{M}$ in Eq.~(\ref{equ:A16}),
\begin{equation}
\begin{aligned}
&\frac{1}{\vec{q}^{\,2}+m^2}F^2(\vec{q}\,),
&1\cdot F^2(\vec{q}\,),\qquad\qquad&\frac{\vec{q}^{\,2}}{\vec{q}^{\,2}+m^2}F^2(\vec{q}\,),
&\frac{\vec{A}\cdot\vec{q}\vec{B}\cdot\vec{q}}{\vec{q}^{\,2}+m^2}F^2(\vec{q}\,),\\ 
&\frac{k^2}{\vec{q}^{\,2}+m^2}F^2(\vec{q}\,),
\qquad&\frac{\vec{A}\cdot\vec{k}\vec{B}\cdot\vec{k}}{\vec{q}^{\,2}+m^2}F^2(\vec{q}\,),\qquad
&\frac{\vec{A}\cdot\vec{q}\vec{B}\cdot\vec{k}}{\vec{q}^{\,2}+m^2}F^2(\vec{q}\,),
&\frac{\vec{k}\times\vec{q}}{\vec{q}^{\,2}+m^2}F^2(\vec{q}\,).\label{equ:A26}
\end{aligned}
\end{equation}
After lengthy derivations and using the following notations,
\begin{align}
    \psi(\vec{r},L,L_z)&=f(r)Y_L^{L_z}\!\left(\theta,\varphi\right),\\
    g(r,m,\Lambda)&=\mathcal{F}_{\vec{q}\to\vec{r}}^{-1}\left[\frac{F^2(\vec{q}\,)}{\vec{q}^{\,2}+m^2}\right]=\frac{1}{4\pi}\left(\frac{e^{-mr}-e^{-\Lambda r}}{r}-\frac{\Lambda^2-m^2}{2\Lambda}e^{-\Lambda r}\right),\\
    h(r,m,\Lambda)&=\mathcal{F}_{\vec{q}\to\vec{r}}^{-1}\left[1\cdot F^2(\vec{q}\,)\right]=\left(\frac{\Lambda^2-m^2}{4\pi}\right)^2\frac{2\pi}{\Lambda}e^{-\Lambda r},
\end{align}
we arrive at the following results:
\begin{align} 
\hat{V}_{\mid^{2S+1}L_{J};I\rangle}^{\frac{1}{\vec{q}^2+m^2}F^2(\vec{q}\,)}f(r)=&\,\hat{\sum}(-1)g(r,m,\Lambda)\delta_{L_zL_{z'}}f(r)\label{equ:A27},\\
\hat{V}_{\mid^{2S+1}L_{J};I\rangle}^{1\cdot F^2(\vec{q}\,)}f(r)=&\,\hat{\sum}(-1)h(r,m,\Lambda)\delta_{L_zL_{z'}}f(r)\label{equ:A28},\\
\hat{V}_{\mid^{2S+1}L_{J};I\rangle}^{\frac{\vec{q}^2}{\vec{q}^2+m^2} F^2(\vec{q}\,)}f(r)=&\,\hat{\sum}(-1)\left[h(r,m,\Lambda)-m^2g(r,m,\Lambda)\right]\delta_{L_zL_{z'}}f(r)\label{equ:A29},\\
\hat{V}_{\mid^{2S+1}L_{J};I\rangle}^{\frac{\vec{A}\cdot\vec{q}\vec{B}\cdot\vec{q}}{\vec{q}^2+m^2}F^2(\vec{q}\,)}f(r)=&\,\hat{\sum}\int\mathrm{d}\Omega Y_L^{L'_z*}\!\left(\theta,\varphi\right)\psi\!\left(\vec{r},L,L_z\right)\left(\vec{A}\cdot\nabla\vec{B}\cdot\nabla \right)g(r,m,\Lambda)\label{equ:A30},\\
\hat{V}_{\mid^{2S+1}L_{J};I\rangle}^{\frac{\vec{k}^2}{\vec{q}^2+m^2}F^2(\vec{q}\,)}f(r)=&\,\hat{\sum}\int\mathrm{d}\Omega Y_L^{L'_z*}\!\left(\theta,\varphi\right)4\Bigl[\nabla^2\left(\psi\!\left(\vec{r},L,L_z\right)g(r,m,\Lambda)\right)-\nabla\psi\left(\vec{r},L,L_z\right)\cdot\nabla g(r,m,\Lambda)\notag\\
&\hspace{4cm}-\frac{3}{4}\psi\left(\vec{r},L,L_z\right)\nabla^2g(r,m,\Lambda)\Bigr]\label{equ:A31},\\
\hat{V}_{\mid^{2S+1}L_{J};I\rangle}^{\frac{\vec{A}\cdot\vec{k}\vec{B}\cdot\vec{k}}{\vec{q}^2+m^2}F^2(\vec{q}\,)}f(r)=&\,\hat{\sum}\int\mathrm{d}\Omega Y_L^{L'_z*}\!\left(\theta,\varphi\right)\biggl[4\vec{A}\cdot\nabla\vec{B}\cdot\nabla\!\left(\psi\!\left(\vec{r},L,L_z\right)g(r,m,\Lambda)\right)-2\vec{A}\cdot\nabla\!\left(\psi\left(\vec{r},L,L_z\right)\vec{B}\cdot\nabla g(r,m,\Lambda)\right)\notag\\
&\qquad-2\vec{B}\cdot\nabla\!\left(\psi\!\left(\vec{r},L,L_z\right)\vec{A}\cdot\nabla g(r,m,\Lambda)\right)+\psi\!\left(\vec{r},L,L_z\right)\vec{B}\cdot\nabla\vec{A}\cdot\nabla g(r,m,\Lambda)\biggr]\label{equ:A32},\\
\hat{V}_{\mid^{2S+1}L_{J};I\rangle}^{\frac{\vec{A}\cdot\vec{q}\vec{B}\cdot\vec{k}}{\vec{q}^2+m^2}F^2(\vec{q}\,)}f(r)=&\,\hat{\sum}
\int\mathrm{d}\Omega Y_L^{L'_z*}\!\left(\theta,\varphi\right)\biggl[2\left(\vec{B}\cdot\nabla\psi\!\left(\vec{r},L,L_z\right)\right)\left(\vec{A}\cdot\nabla g(r,m,\Lambda)\right)\notag\\
&\qquad+\psi\!\left(\vec{r},L,L_z\right)\left(\vec{B}\cdot\nabla\vec{A}\cdot\nabla g(r,m,\Lambda)\right)\biggr]\label{equ:A33},\\
\hat{V}_{\mid^{2S+1}L_{J};I\rangle}^{\frac{\vec{k}\times\vec{q}}{\vec{q}^2+m^2}F^2(\vec{q}\,)}f(r)=&\,\hat{\sum}\int\mathrm{d}\Omega Y_L^{L'_z*}\!\left(\theta,\varphi\right)2\left(\nabla\psi\!\left(\vec{r},L,L_z\right)\right)\times\left(\nabla g(r,m,\Lambda)\right)\notag\\
=&\,\hat{\sum}(-2i)\frac{f(r)}{r}\frac{{\rm{d}}g(r,m,\Lambda)}{{\rm{d}}r}\int\mathrm{d}\Omega Y_L^{L'_z*}\!\left(\theta,\varphi\right)\hat{\vec{L}}Y_L^{L_z}\!\left(\theta,\varphi\right)\label{equ:A34},
\end{align}
where $\hat{\vec{L}}=\vec{r}\times\hat{\vec{p}}=\vec{r}\times(-i\nabla)$ is the orbital angular momentum operator.

It is worth noting that the momentum $\vec{k}$ introduces the derivative of the radial wavefunction, specifically $f'(r)$ and $f''(r)$. Furthermore, for $S$-wave, the aforementioned  Eqs.~(\ref{equ:A27}-\ref{equ:A34}) will be simplified as follows:
\begin{align}
\hat{V}_{\mid^{2S+1}S_{J};I\rangle}^{\frac{1}{\vec{q}^2+m^2}F^2(\vec{q}\,)}f(r)=&\,\hat{\sum_{S-{\rm wave}}}(-1)g(r,m,\Lambda)f(r)\label{equ:1/q^2},\\
\hat{V}_{\mid^{2S+1}S_{J};I\rangle}^{1\cdot F^2(\vec{q}\,)}f(r)=&\,\hat{\sum_{S-{\rm wave}}}(-1)h(r,m,\Lambda)f(r)\label{equ:1},\\
\hat{V}_{\mid^{2S+1}S_{J};I\rangle}^{\frac{\vec{q}^2}{\vec{q}^2+m^2} F^2(\vec{q}\,)}f(r)=&\,\hat{\sum_{S-{\rm wave}}}(-1)\left[h(r,m,\Lambda)-m^2g(r,m,\Lambda)\right]f(r)\label{equ:q^2},\\
\hat{V}_{\mid^{2S+1}S_{J};I\rangle}^{\frac{\vec{A}\cdot\vec{q}\vec{B}\cdot\vec{q}}{\vec{q}^2+m^2}F^2(\vec{q}\,)}f(r)=&\frac{\vec{A}\cdot\vec{B}}{3}\hat{V}_{\mid^{2S+1}0_J;I\rangle}^{\frac{\vec{q}^2}{\vec{q}^2+m^2} F^2(\vec{q}\,)}f(r)\label{equ:AqBq},\\
\hat{V}_{\mid^{2S+1}S_{J};I\rangle}^{\frac{\vec{k}^2}{\vec{q}^2+m^2}F^2(\vec{q}\,)}f(r)=&\,\hat{\sum_{S-{\rm wave}}}\frac{1}{8 \pi  \Lambda  r^2}\Biggl[e^{-r (\Lambda +m)}\Biggl(e^{m r}\biggl(4 f'(r)\left(r \left(m^2 \left(2-\Lambda  r\right)+\Lambda ^3 r\right)-2 \Lambda \right)\notag\\
&+4 r f''(r)\left(m^2 r-\Lambda \left(\Lambda  r+2\right)\right)+\Lambda  r f(r)\left(m^2 \left(\Lambda  r-2\right)-\Lambda ^3 r\right)\biggr)\notag\\
&+2 \Lambda  e^{\Lambda  r}\left(\left(4-4 m r\right) f'(r)+4 r f''(r)+m^2 r f(r)\right)\Biggr)\Biggr]\label{equ:k^2},\\
\hat{V}_{\mid^{2S+1}S_{J};I\rangle}^{\frac{\vec{A}\cdot\vec{k}\vec{B}\cdot\vec{k}}{\vec{q}^2+m^2}F^2(\vec{q}\,)}f(r)=&\frac{\vec{A}\cdot\vec{B}}{3}\hat{V}_{\mid^{2S+1}S_{J};I\rangle}^{\frac{\vec{k}^2}{\vec{q}^2+m^2}F^2\!\left(\vec{q}\right)}f(r)\label{equ:AkBk},\\
\hat{V}_{\mid^{2S+1}S_{J};I\rangle}^{\frac{\vec{q}\cdot\vec{k}}{\vec{q}^2+m^2}F^2(\vec{q}\,)}f(r)=&\,\hat{\sum_{S-{\rm wave}}}\biggl[\frac{2 e^{-m r} \left(m^2 r f(r)-2 \left(m r+1\right) f'(r)\right)}{8 \pi  r^2}\notag\\
&+\frac{e^{-\Lambda r} \left(2 f'(r) \left(-m^2 r^2+\Lambda  r \left(\Lambda  r+2\right)+2\right)+r f(r) \left(m^2 \left(\Lambda  r-2\right)-\Lambda ^3 r\right)\right)}{8 \pi  r^2}\biggr]\label{equ:q.k},\\
\hat{V}_{\mid^{2S+1}S_{J};I\rangle}^{\frac{\vec{A}\cdot\vec{q}\vec{B}\cdot\vec{k}}{\vec{q}^2+m^2}F^2(\vec{q}\,)}f(r)=&\frac{\vec{A}\cdot\vec{B}}{3}\hat{V}_{\mid^{2S+1}S_{J};I\rangle}^{\frac{\vec{q}\cdot\vec{k}}{\vec{q}^2+m^2}F^2(\vec{q}\,)}f(r)\label{equ:AqBk},\\
\hat{V}_{\mid^{2S+1}S_{J};I\rangle}^{\frac{\vec{k}\times\vec{q}}{\vec{q}^2+m^2}F^2(\vec{q}\,)}f(r)=&0\label{equ:kxq}.
\end{align}

In summary, the crucial computation in the OBE model can be broken down into three steps:
\begin{enumerate}
    \item Compute the amplitude $\mathcal{M}\!\left(\vec{p},S_1S_{1z},S_2S_{2z},II_3\to\vec{p}^{\,\prime},S_1S_{3z},S_2S_{4z},II_3\right)$ of the $t$-channel Feynman diagram. For $S$-wave, this step involves computing the redefined amplitude $\mathcal{M}_{\mid^{2S+1}S_{J};I\rangle}(\vec{p}\to\vec{p}^{\,\prime})$ as depicted in Eq.~(\ref{redefined amplitude}).
    \item Perform the integrals in Eq.~(\ref{equ:A16}) with the assistance of Eqs.~(\ref{equ:A27}-\ref{equ:A34}). For the $S$-wave scattering, this step is simplified to calculating Eq.~(\ref{equ:A21}) in light of Eqs.~(\ref{equ:1/q^2}-\ref{equ:kxq}).
    \item Solve the Schr\"odinger Eq.~(\ref{equ:A15}), or  Eq.~(\ref{equ:s-wave Schordinger equation}) for the $S$-wave case.
\end{enumerate}

\section{The amplitude for the $t$-channel process of $\Sigma_c^*\bar{\Sigma}\to\Sigma_c^*\bar{\Sigma}$}
\label{Appendix:B}
With the consideration that the spin of the particle $\Sigma_c^*$ is $\frac{3}{2}$, its vector-spinor wavefunction is formed through the coupling of the spin-$1/2$ spinor and spin-$1$ polarization vector~\cite{Ronchen:2012eg}, which can be expressed as
\begin{align}
u^\mu(\vec{p},\lambda)=\sum_{\lambda_1,\lambda_2}\left\langle 1\lambda_1,\frac{1}{2}\lambda_2\right|\left.\frac{3}{2}\lambda\right\rangle\epsilon^\mu(\vec{p},\lambda_1)u(\vec{p},\lambda_2)=\sum_{\lambda_1,\lambda_2}C_{1\lambda_1;\frac{1}{2}\lambda_2}^{\frac{3}{2}\lambda}\epsilon^\mu(\vec{p},\lambda_1)u(\vec{p},\lambda_2).
\end{align}
The wavefunctions for spin-$1/2$ and spin-$1$ particles are defined as follows: 
\begin{gather}
u\bigl(\vec{p},\alpha\bigr)=\left(\begin{array}{c}\varphi^\alpha\\
\frac{\vec{\sigma}\cdot\vec{p}}{2M}\varphi^\alpha
\end{array}\right),\qquad
v\bigl(\vec{p},\alpha\bigr)=\left(\begin{array}{c}\frac{\vec{\sigma}\cdot\vec{p}}{2M}\chi^\alpha\\
\chi^\alpha\end{array}\right)\label{equ:B2},\\
\epsilon\!\left(\pm1\right)=\mp\frac{1}{\sqrt{2}}\left(0,1,\pm i,0\right),\qquad
\epsilon\!\left(0\right)=\left(0,0,0,1\right),
\end{gather}
where $\varphi^\alpha$ and $\chi^\alpha$ are two-component spinors, $\vec{\sigma}$ represents the Pauli matrices, $M$ is the mass of the particle, and $\vec p$ is its momentum.

With these relations, one can derive the scattering amplitude for the process depicted in Fig.~\ref{img1}. We will neglect the $\mathcal O\left({\vec{p\,}^2}/{M^2}\right)$ terms, as their impact on the effective potential is minimal for non-relativistic systems. 
For a more detailed discussion, see Appendix~\ref{Appendix:D}. 
Below we derive the $S$-wave amplitudes.

For the $\sigma$ exchange, based on Eq.~(\ref{redefined amplitude}), it is straightforward to derive
\begin{align}
\mathcal{M}_{\mid^{2S+1}S_{J};I\rangle}^\sigma(\vec{p}\to\vec{p}^{\,\prime})\approx\frac{-C_{\sigma}(I)}{\vec{q}^{\,2}+m_\sigma^2},
\end{align}
where $C_\sigma(I)=-g_{\sigma B_6^*B_6^*}g_{\Sigma\Sigma\sigma}F_\sigma(I)$. For the pseudoscalar meson exchange, we have
\begin{align}
\mathcal{M}_{\mid^{2S+1}S_{J};I\rangle}^{p}(\vec{p}\to\vec{p}^{\,\prime})\approx\frac{C_p(I)}{2M_{\Sigma_c^*}}\frac{\Delta_{S_AS_B}}{3}\frac{\vec{q}^{\,2}}{\vec{q}^{\,2}+m_{p}^2},
\end{align}
where
\begin{align}
\Delta_{S_AS_B}=\frac{9-2S(S+1)}{3}=\left\{
\begin{array}{cc}
\frac{5}{3}&  (S=1)\\
-1          &  (S=2)
\end{array} \right.,
\end{align}
and $C_{p}(I)=-\frac{1}{m_p}g_{B_6^*B_6^*p}g_{\Sigma\Sigma p}F_p(I)$. Analogously, for the vector meson exchange, the corresponding redefined amplitude is
\begin{align}\mathcal{M}_{\mid^{2S+1}S_{J};I\rangle}^v(\vec{p}\to\vec{p}^{\,\prime})\approx F_v(I)g_{\Sigma\Sigma v}\left(g_{B_6^*B_6^*v}\mathcal{M}_1+g_{B_6^*B_6^*v}\frac{k_{\Sigma\Sigma v}}{2M_{\Sigma}}\mathcal{M}_2+\frac{f_{B_6^*B_6^*v}}{2M_{\Sigma^*_c}}\mathcal{M}_3+\frac{f_{B_6^*B_6^*v}}{2M_{\Sigma_c^*}}\frac{k_{\Sigma\Sigma v}}{2M_\Sigma}\mathcal{M}_4\right) ,
\end{align}
where
\begin{align}
&\mathcal{M}_1=\frac{1}{\vec{q}^{\,2}+m_v^2},\label{B8}\\
&\mathcal{M}_2=\frac{1}{\vec{q}^{\,2}+m_v^2}\left[-\frac{\vec{q}^{\,2}}{2M_{\Sigma}}+\frac{1}{8M_{\Sigma_c^*}}
\left(\frac{8\Delta_{S_AS_B}}{3}\vec{q}^{\,2}+\frac{\Delta_{\text{ten}}^{(1)}}{3}\vec{q}\cdot\vec{k}\right)\right],\label{B9}\\
&\mathcal{M}_3=\frac{1}{2M_{\Sigma}}\frac{1}{\vec{q}^{\,2}+m_v^2}\left( \Delta_{\text{ten}}^{(2)} \frac{\vec{q}\cdot\vec{k}}{3}+2\Delta_{S_AS_B}\frac{\vec{q}^{\,2}}{3} \right),\label{B10}\\
&\mathcal{M}_4=\frac{1}{\vec{q}^{\,2}+m_v^2}\left(-\frac{1}{2}\right)\frac{-4\Delta_{S_AS_B}}{3}\vec{q}^{\,2} \label{B11},
\end{align}
and for the $S$-wave, one has
\begin{align}
\Delta_{\text{ten}}^{(1)}=\Delta_{\text{ten}}^{(2)}=0.
\end{align}

Let us comment on one subtle detail in the derivation. The $\bar{\Sigma}$ belongs to the $\bar{2}$ representation of the spin SU(2) group, whereas the $\Sigma$ belongs to the $2$ representation. The $2$ and $\bar{2}$ representations of the SU(2) group are equivalent. However, to uphold consistency in the application of CG coefficients, a similarity transformation on the $\bar{2}$ representation is required. We adopt the following convention for the two-component spinors in Eq.~(\ref{equ:B2}):
\begin{align}
\varphi^{\frac{1}{2}}=\left(
\begin{array}{c}
1\\0
\end{array}
\right), 
\quad
\varphi^{-\frac{1}{2}}=\left(
\begin{array}{c}
0\\1
\end{array}
\right), 
\quad
\chi^{\frac{1}{2}}=\left(
\begin{array}{c}
0\\1
\end{array}
\right),
\quad
\chi^{-\frac{1}{2}}=\left(
\begin{array}{c}
-1\\0
\end{array}
\right).
\end{align} 

\section{Relation between the momentum-space potential and the QFT amplitude}
\label{Appendix:C}

The relation between the amplitude and the potential can be established by comparing the $S$-matrix elements in QFT and in QM. Specifically, for a two-to-two elastic scattering process, the QFT representation is given by
\begin{align}
\langle\vec{p}_1\vec{p}_2|\hat{{S}}|\vec{p}_A\vec{p}_B\rangle\ {=} \ \langle\vec{p}_1\vec{p}_2|\vec{p}_A\vec{p}_B\rangle
+(2\pi)^4\delta^{4}\!\left( p_A+p_B-p_1-p_2 \right) \cdot i\mathcal{M}\!\left(p_A+p_B\to p_1+p_2\right)\label{equ:C1}.
\end{align}
In the QM context, it is represented as
\begin{align}
\langle\vec{p}^{\,\prime}|{\hat{{S}}}|\vec{p}\,\rangle&=\delta^{3} \!( \vec{p}^{\,\prime}-\vec{p} \,)-2\pi i\delta \!\left(E_{\vec{p}^{\,\prime}}-E_{\vec{p}}\right)\langle\vec{p}^{\,\prime}|{\hat{T}}\!\left(E_{\vec{p}+i0^+}\right)|\vec{p}\,\rangle{\approx}\ \delta^{3}(\vec{p}^{\,\prime}-\vec{p}\,)-2\pi i\delta\!\left({E_{\vec{p}^{\,\prime}}-E_{\vec{p}}}\right)\langle\vec{p}^{\,\prime}|\hat{V}|\vec{p}\,\rangle\label{equ:C2}.
\end{align}
Here $\vec{p}=({M_2\vec{p}_A-M_1\vec{p}_B})/{M}$ and $\vec{p}^{\,\prime}=({M_2\vec{p}_1-M_1\vec{p}_2})/{M}$ denote the relative momentum of the initial and final two-body systems, respectively. 
Furthermore, $p_A^2=p_1^2=M_1^2$, $p_B^2=p_2^2=M_2^2$ and $M=M_1+M_2$. 
The $S$-matrix elements in QFT and QM should be the same up to the normalization, i.e.,
\begin{align}
\frac{\langle\vec{p}_1\vec{p}_2|\hat{{S}}|\vec{p}_A\vec{p}_B\rangle_{\rm{QFT}}}
{\langle\vec{p}_1\vec{p}_2|\vec{p}_A\vec{p}_B\rangle_{\rm{QFT}}}=\frac{\langle\vec{p}^{\,\prime}|\hat{{S}}|\vec{p}\,\rangle_{\rm{QM}}}{\langle\vec{p}^{\,\prime}|\vec{p}\,\rangle_{\rm{QM}}},
\end{align}
which implies that
\begin{align}
\langle\vec{p}_1\vec{p}_2|\hat{{S}}|\vec{p}_A\vec{p}_B\rangle_{\rm{QFT}}
=\frac{\langle\vec{p}_1\vec{p}_2|\vec{p}_A\vec{p}_B\rangle_{\rm{QFT}}}{\langle\vec{p}^{\,\prime}|\vec{p}\,\rangle_{\rm{QM}}}\langle\vec{p}^{\,\prime}|\hat{{S}}|\vec{p}\,\rangle_{\rm{QM}}\label{equ:C4}.
\end{align}
In QFT, we adopt the normalization $|\vec{p},r\rangle=\sqrt{2E_{\vec{p}}}\,a_{\vec{p}}^{r\dag}|0\rangle$ and 
\begin{align}
    \{ a_{\vec{p}}^r,a_{\vec{q}}^{s\dag}\}=\{ b_{\vec{p}}^r,b_{\vec{q}}^{s\dag}\}=\left(2\pi\right)^3\delta^{3}\!\left(\vec{p}-\vec{q}\,\right)\delta^{rs}
\end{align} for the Dirac field. In QM, we have $\langle\vec{x}|\vec{p}\,\rangle={(2\pi)^{-3/2}}e^{i\vec{p}\cdot\vec{x}}$ and Eq.~(\ref{eq:completebasis}). So we obtain 
\begin{align}
{\rm{QFT}} \qquad\qquad  \langle \vec{p},r|\vec{q},s\rangle &=\sqrt{2E_{\vec{p}}}\sqrt{2E_{\vec{q}}}\left(2\pi\right)^3\delta^{3}\!\left(\vec{p}-\vec{q}\,\right)\delta^{rs}\label{equ:C5},\\
{\rm{QM}}  \qquad\qquad   \langle \vec{p},r|\vec{q},s\rangle&=\delta^{3}\!\left(\vec{p}-\vec{q}\,\right)\delta^{rs}\label{equ:C6}.
\end{align}
Clearly, the spin of a particle only generates the identical term $\delta^{rs}$ in both QM and QFT. For simplicity, we will disregard the particle's spin in the following. Thus, we have
\begin{align}
\frac{\langle\vec{p}_1\vec{p}_2|\vec{p}_A\vec{p}_B\rangle_{\rm{QFT}}}{\langle\vec{p}^{\,\prime}|\vec{p}\,\rangle_{\rm{QM}}}=\frac{\sqrt{2E_{\vec{p}_1}2E_{\vec{p}_2}2E_{\vec{p}_A}2E_{\vec{p}_B}}\left(2\pi \right)^6\delta^{3}\!\left(\vec{p}_1-\vec{p}_A\right)\delta^{3}\!\left(\vec{p}_2-\vec{p}_B\right)}{\delta^{3}\!\left( \vec{p}^{\,\prime}-\vec{p}\, \right) }.\label{equ:C7}
\end{align}
We define $\vec P=\vec{p}_A+\vec{p}_B$ and $\vec P'=\vec{p}_1+\vec{p}_2$ as the total three-momenta of the initial and final two-body systems, respectively. Utilizing the property of the Dirac-$\delta$ function, i.e., $f\!\left(x\right)\delta\!\left(x-x_0\right)=f(x_0)\delta\!\left(x-x_0\right)$, we get
\begin{align}
\delta^{3}( \vec{p}^{\,\prime}-\vec{p} \,)\delta^{3}( \vec{P}'-\vec{P}\,)&=\delta^{3}\!\left( \frac{M_2}{M}\left(\vec{p}_1-\vec{p}_A \right)-\frac{M_1}{M}\left(\vec{p}_2-\vec{p}_B\right)\right)\delta^{3}\!\left(\left(\vec{p}_1-\vec{p}_A\right)-\left(\vec{p}_B-\vec{p}_2\right)\right)\notag\\
&=\delta^{3}\!\left(\frac{M_2}{M}\left(\vec{p}_1-\vec{p}_A\right)+\frac{M_1}{M}\left(\vec{p}_1-\vec{p}_A\right)\right)\delta^{3}\!\left(\left(\vec{p}_1-\vec{p}_A\right)-\left(\vec{p}_B-\vec{p}_2\right)\right)\notag\\
&=\delta^{3}\!\left(\vec{p}_1-\vec{p}_A\right)\delta^{3}\!\left(\vec{p}_2-\vec{p}_B\right)\label{equ:C8}.
\end{align}
Then, we have
\begin{align}
\frac{\langle\vec{p}_1\vec{p}_2|\vec{p}_A\vec{p}_B\rangle_{\rm{QFT}}}{\langle\vec{p}^{\,\prime}|\vec{p}\,\rangle_{\rm{QM}}}{=}\sqrt{2E_{\vec{p}_1}2E_{\vec{p}_2}2E_{\vec{p}_A}2E_{\vec{p}_B}}\left(2\pi\right)^6\delta^{3}(\vec{P}'-\vec{P}),\label{equ:C9}
\end{align}
which just means that the total momentum is conserved and in the usual QM treatment the c.m. motion has been factored out.
Substituting Eq.~(\ref{equ:C9}) into Eq.~(\ref{equ:C4}), we obtain
\begin{align}
\langle\vec{p}_1\vec{p}_2|\hat{{S}}|\vec{p}_A\vec{p}_B\rangle_{\rm{QFT}}=\sqrt{2E_{\vec{p}_1}2E_{\vec{p}_2}2E_{\vec{p}_A}2E_{\vec{p}_B}}\left(2\pi\right)^6\delta^{3}(\vec{P}'-\vec{P})\langle\vec{p}^{\,\prime}|\hat{{S}}|\vec{p}\,\rangle_{\rm{QM}}\label{equ:C10}.
\end{align}
Considering Eqs.~(\ref{equ:C1},\ref{equ:C5}), we have
\begin{align}
\langle\vec{p}_1\vec{p}_2|\hat{{S}}|\vec{p}_A\vec{p}_B\rangle_{\rm{QFT}}= &\, \sqrt{2E_{\vec{p}_1}2E_{\vec{p}_2}2E_{\vec{p}_A}2E_{\vec{p}_B}}\left(2\pi\right)^6\delta^{3}\!\left(\vec{p}_1-\vec{p}_A\right)\delta^{3}\!\left(\vec{p}_2-\vec{p}_B\right)\notag\\
&+\left(2\pi\right)^4\delta^{4}\!\left(p_A+p_B-p_1-p_2\right)\cdot i\mathcal{M}\!\left(p_A+p_B\to p_1+p_2\right).
\end{align}
Substituting Eq.~(\ref{equ:C2}) into the right half part of Eq.~(\ref{equ:C10}), we obtain
\begin{align}
\delta^{3}(\vec{P}'-\vec{P})\langle\vec{p}^{\,\prime}|\hat{{S}}|\vec{p}\,\rangle_{\rm{QM}}&=
\delta^{3}(\vec{P}^{\prime}-\vec{P})\delta^{3}(\vec{p}^{\,\prime}-\vec{p}\,)-\left(2\pi\right)i\delta\!\left(E_{\vec{p}^{\,\prime}}-E_{\vec{p}}\right)\delta^{3}(\vec{P}'-\vec{P})\langle\vec{p}^{\,\prime}|\hat{V}|\vec{p}\,\rangle
\notag\\
&=
\delta^{3}\!\left(\vec{p}_1-\vec{p}_A\right)\delta^{3}\!\left(\vec{p}_2-\vec{p}_B\right)-\left(2\pi\right)i\delta^{4}\!\left(p_A+p_B-p_1-p_2\right)\langle\vec{p}^{\,\prime}|\hat{V}|\vec{p}\,\rangle
\label{equ:C12}.
\end{align}
From Eqs.~(\ref{equ:C10}-\ref{equ:C12}), it is easy to get
\begin{align}
-\sqrt{2E_{\vec{p}_1}2E_{\vec{p}_2}2E_{\vec{p}_A}2E_{\vec{p}_B}}\left(2\pi\right)^7i\langle\vec{p}^{\,\prime}|\hat{V}|\vec{p}\,\rangle=\left(2\pi\right)^4i\mathcal{M}\!\left(p_A+p_B\to p_1+p_2\right).
\end{align}
Finally, we obtain
\begin{align}
\langle\vec{p}^{\,\prime}|{V}|\vec{p}\,\rangle=-\frac{\mathcal{M}\!\left( p_A+p_B\to p_1+p_2 \right)}{\left( 2\pi \right)^3\sqrt{2E_{\vec{p}_1}2E_{\vec{p}_2}2E_{\vec{p}_A}2E_{\vec{p}_B}}}\approx-\frac{\mathcal{M}\!\left(p_A+p_B\to p_1+p_2\right)}{\left(2\pi\right)^3\sqrt{2M_12M_22M_A2M_B}}\label{equ:C14}.
\end{align}
This relation differs from those in Refs.~\cite{Zhao:2013ffn,Liu:2007bf,Liu:2017mrh,Sun:2011uh}, due to the distinct normalization relation of $\langle \vec x|\vec p\, \rangle$ in QM.  
However, when using Eq.~(\ref{equ:B2}) as the spin-$1/2$ particle wavefunction, where $|\vec{p},r\rangle=a_{\vec{p}}^{r\dag}|0\rangle$, Eq.~(\ref{equ:C14}) will not contain $\sqrt{2M_i}$. Therefore, in this paper, the relation between amplitude and potential in momentum space reads
\begin{align}
\langle\vec{p}^{\,\prime}|{V}|\vec{p}\,\rangle\approx-\frac{1}{(2\pi)^3}{\mathcal{M}\!\left(p_A+p_B\to p_1+p_2\right)}.\label{eq:VM}
\end{align}

\section{Relative importance of each term in the scattering amplitude}
\label{Appendix:D}

In this appendix, we scrutinize the contribution from each term in the amplitude to the effective potential with the intention of simplifying the computation by eliminating insignificant quantities. 

First, we compare the following four distinct terms: 
\begin{align}
\mathcal M_0=\frac{1}{\vec{q}^{\,2}+m^2}F^2(\vec{q}\,),\quad\mathcal M_1=\frac{1}{M_1^2}F^2(\vec{q}\,),\quad\mathcal M_2=\frac{1}{M_2^2}\frac{\vec{q}^{\,2}}{\vec{q}^{\,2}+m^2}F^2(\vec{q}\,),\quad\mathcal M_3=\frac{1}{M_3^2}\frac{\vec{A}\cdot\vec{q}\vec{B}\cdot\vec{q}}{\vec{q}^{\,2}+m^2}F^2(\vec{q}\,),\label{equ:D1}
\end{align}
where we introduce additional masses $M_i(i=1,2,3)$ to match the dimensions of all terms. To compare the relative importance of the effective potentials from the terms in 
Eq.~(\ref{equ:D1}), we normalize them at $r=0$ fm:
\begin{align}
\lim_{r\to0}V ^{\mathcal M_0}(r)=
\lim_{r\to0}V ^{\mathcal M_1}(r)=
\lim_{r\to0}V ^{\mathcal M_2}(r)=
\lim_{r\to0}V ^{\mathcal M_3}(r)\label{equ:D2},
\end{align}
which results in
\begin{align}
M_1^2=\left(\Lambda+m\right)^2,\quad M_2^2=\Lambda\left(\Lambda+2m\right),\quad M_3^2=\frac{\vec{A}\cdot\vec{B}\Lambda\left(\Lambda+2m\right)}{3}.
\end{align}
The resulting $S$-wave effective potentials obtained by 
assigning $\Lambda=1$ GeV, $m=0.5$ GeV and $\vec{A}\cdot\vec{B}=1$ are depicted in Fig.~\ref{img11}(a).
To a certain extent, from the results we can estimate that 
\begin{align}
\vec{q}^{\,2}\to\Lambda\left(\Lambda+2m\right).
\end{align} 
We note that the average effect of the exchanged-meson momentum is too large, even though we have incorporated a form factor to suppress the contribution from high-momentum transition.
\begin{figure}[t]
 \centering
 \begin{overpic}[width=1\linewidth]{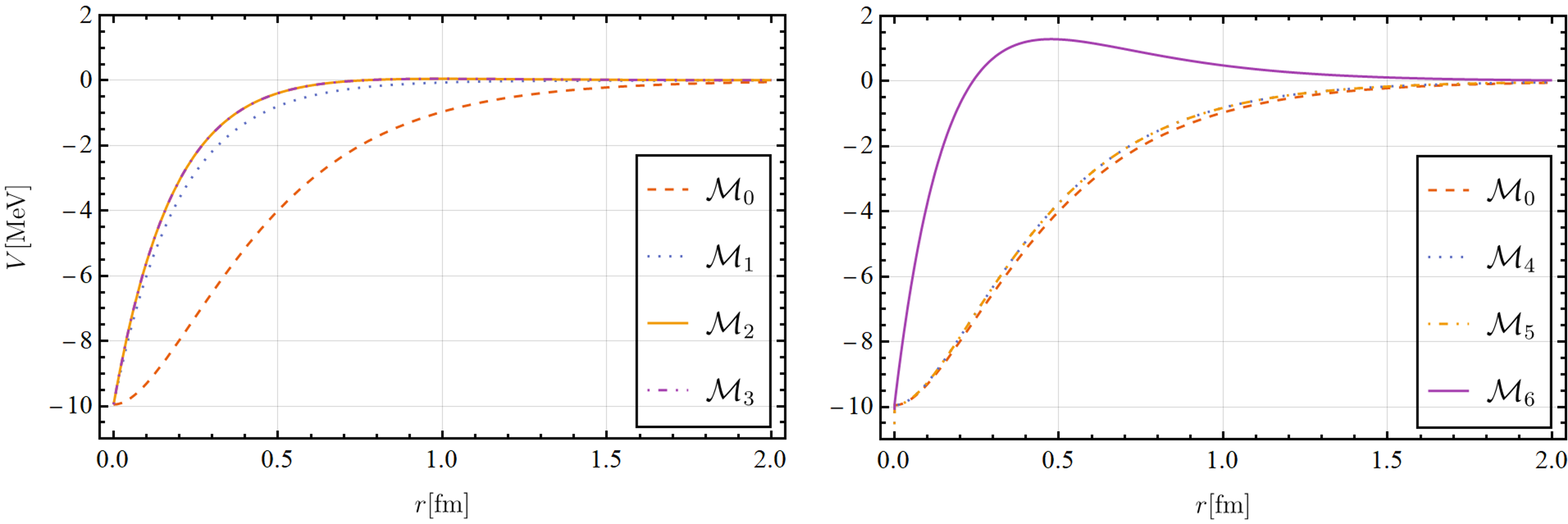}
    \put(45,29.5){\normalsize{(a)}}
    \put(95,29.5){\normalsize{(b)}}
    \end{overpic}
  \caption{The different $S$-wave effective potential $V(r)$ from different amplitude $\mathcal{M}(\vec{q}\,)$ (a) or the different $S$-wave effective potential $V_0(r)$ from different amplitude $\mathcal{M}(\vec{q},\vec{k})$ (b), where $\Lambda=1$ GeV, $m=0.5$ GeV and $\vec{A}\cdot\vec{B}=1$. The curves of $\mathcal{M}_2$ and $\mathcal{M}_3$ align perfectly, as they should according to Eq.~(\ref{equ:AqBq}); similarly the $\mathcal{M}_4$ and $\mathcal{M}_5$ curves are identical due to Eq.~(\ref{equ:AkBk}).} 
  \label{img11} 
\end{figure}
\par
Next, we consider the other terms in amplitude, which contain momentum $\vec{k}$,
\begin{align}
\mathcal M_4=\frac{1}{M_4^2}\frac{\vec{k}^2}{\vec{q}^{\,2}+m^2}F^2(\vec{q}\,),\quad \mathcal M_5=\frac{1}{M_5^2}\frac{\vec{A}\cdot\vec{k}\vec{B}\cdot\vec{k}}{\vec{q}^{\,2}+m^2}F^2(\vec{q}\,),\quad \mathcal M_6=\frac{1}{M_6^2}\frac{\vec{A}\cdot\vec{q}\vec{B}\cdot\vec{k}}{\vec{q}^{\,2}+m^2}F^2(\vec{q}\,).
\end{align}
Given that $\vec{k}$ introduces the derivatives of the radial wavefunction, we adopt Eq.~(\ref{equ for dealing with k}) in the form of
\begin{align}
r\hat{V} ^{\mathcal{M}(\vec{p},\vec{p}^{\,\prime})}(r)\frac{\psi(r)}{r}=V_{0}^{\mathcal{M}(\vec{p},\vec{p}^{\,\prime})}(r)\psi(r)+V_{1}^{\mathcal{M}(\vec{p},\vec{p}^{\,\prime})}(r)\psi'(r)+V_{2}^{\mathcal{M}(\vec{p},\vec{p}^{\,\prime})}(r)\psi''(r),
\end{align}
where the additional subscripts 0, 1 and 2 correspond to the number of derivatives on $\psi(r)$. For terms containing only $\vec{q}$, they can also be expanded in this manner, where only $V_0^{\mathcal{M}(\vec{q}\,)}(r)\psi(r)$ appears, i.e., $V_1^{\mathcal{M}(\vec{q}\,)}(r)=V_2^{\mathcal{M}(\vec{q}\,)}(r)=0$.
It can be verified that the effects from $\psi'(r)$ and $\psi''(r)$ in the Schr\"odinger equation are marginal and do not alter the existence of bound states as discussed in Section~\ref{the effects of k}. Therefore, we primarily focus on the magnitude of $V_{0}^{\mathcal{M}(\vec{p},\vec{p}^{\,\prime})}(r)$ as the main contribution of the corresponding term to the effective potential.

For a comparison of the behavior of each term, we take the following normalization at $r=0$ fm
\begin{align}
\lim_{r\to0}V^{\mathcal M_0}(r)=
\lim_{r\to0}V_{0}^{\mathcal M_4}(r)=
\lim_{r\to0}V_{0}^{\mathcal M_5}(r)=
\lim_{r\to0}V_{0}^{\mathcal M_6}(r),
\end{align}
which results in
\begin{align}
M_4^2=-\frac{\Lambda(\Lambda+2m)}{3},\qquad
M_5^2=-\frac{\vec{A}\cdot\vec{B}\Lambda(\Lambda+2m)}{9},\qquad
M_6^2=\frac{\vec{A}\cdot\vec{B}\Lambda(\Lambda+2m)}{9}.
\end{align}
The results obtained by setting $\Lambda=1$ GeV, $m=0.5$ GeV and $\vec{A}\cdot\vec{B}=1$ are shown in Fig.~\ref{img11}(b). 

Let us take the terms in $\bar{u}(\vec{p}^{\,\prime},S_{3z})u(\vec{p},S_{1z})$ as an example. With Eq.~(\ref{equ:B2}) and the $\gamma$-matrices in Bjorken-Drell representation, 
\begin{align}
\gamma^0=\left(
\begin{array}{cc}
\mathbf{1}&0\\
0&-\mathbf{1}
\end{array}
\right),\quad
\gamma^i=\left(
\begin{array}{cc}
0&\sigma^i\\
-\sigma^i&0
\end{array}
\right),\quad
\gamma_5=\left(
\begin{array}{cc}
0&\mathbf{1}\\
\mathbf{1}&0
\end{array}
\right),
\end{align}
we obtain
\begin{align}
\bar{u}(\vec{p}^{\,\prime},S_{3z})u(\vec{p},S_{1z})=\phi^{S_{3z}\dagger} \left(
1-\frac{\vec{\sigma}\cdot\vec{p}^{\,\prime}\vec{\sigma}\cdot\vec{p}}{4M^2}
\right) \phi^{S_{1z}} \label{equ:D12}.
\end{align}
With $(\vec{\sigma}\cdot\vec{a}_1)(\vec{\sigma}\cdot\vec{a}_2)=\vec{a}_1\cdot\vec{a}_2+i\vec{\sigma}\cdot(\vec{a}_1\times\vec{a}_2)$
for $[\vec{\sigma},\vec{a}_1]=[\vec{\sigma},\vec{a}_2]=0$, Eqs.~(\ref{equ:A11},~\ref{equ:kxq}) 
and the above normalizations at $r=0$ fm,
for the $S$-wave we have
\begin{align}
V_0^{\frac{\bar{u}(\vec{p}^{\,\prime},S_{3z})u(\vec{p},S_{1z})}{\vec{q}^{\,2}+m^2}F^2(\vec{q}\,)}(r)&= V_0^{\frac{\phi^{S_{3z}\dagger}\left(1-\frac{\vec{k}^2-\vec{q}^2}{16M^2}\right)\phi^{S_{1z}}}{\vec{q}^{\,2}+m^2}F^2(\vec{q}\,)}(r)\xrightarrow{r\to0}
V_0^{\frac{\phi^{S_{3z}\dagger}\left( 1+\frac{\Lambda(\Lambda+2m)}{12M^2} \right)\phi^{S_{1z}}}{\vec{q}^{\,2}+m^2}F^2(\vec{q}\,)}(r)\notag\\
&=\phi^{S_{3z}\dagger}\left( 1+\frac{\Lambda(\Lambda+2m)}{12M^2} \right)\phi^{S_{1z}}V_0^{\frac{1}{\vec{q}^{\,2}+m^2}F^2(\vec{q}\,)}(r).
\end{align}
In the computation of the $\Sigma_c^*\bar{\Sigma}$ bound state with the $\sigma$, $\pi$, $\eta$, $\rho$ and $\omega$ exchanges, we take the cutoff range to be $\Lambda\in[0.8,1.5]$ GeV. Consequently, the value of $\frac{\Lambda(\Lambda+2m)}{12M^2}$ reaches a maximum of about 0.27 for $\Sigma$ ($M\approx$~1190 MeV) and 0.06 for $\Sigma_c^*$ ($M\approx$~2520 MeV). Therefore, it is reasonable to neglect the term $\frac{\vec{\sigma}\cdot\vec{p}^{\,\prime}\vec{\sigma}\cdot\vec{p}}{4M^2}$ for the $\Sigma_c^*$ vertex in Eq.~(\ref{equ:D12}). As for the $\bar{\Sigma}$ vertex, the approximation may not be precise enough. However, considering that the primary purpose of OBE is to explore the potential existence of a molecule state, this approximation is also acceptable. The main reason for this difference is the significantly larger mass of $\Sigma_c^*$, compared to that of $\Sigma$. As a consequence, this reminds us that the non-relativistic limit ${|\vec{p}\,|}/{M}\ll1$ only holds well if the mass is considerably larger than the typical energy scale of the interaction.
\end{appendix}

\end{widetext}

\bibliography{ref}

\begin{thebibliography}{90}%
\makeatletter
\providecommand \@ifxundefined [1]{%
 \@ifx{#1\undefined}
}%
\providecommand \@ifnum [1]{%
 \ifnum #1\expandafter \@firstoftwo
 \else \expandafter \@secondoftwo
 \fi
}%
\providecommand \@ifx [1]{%
 \ifx #1\expandafter \@firstoftwo
 \else \expandafter \@secondoftwo
 \fi
}%
\providecommand \natexlab [1]{#1}%
\providecommand \enquote  [1]{``#1''}%
\providecommand \bibnamefont  [1]{#1}%
\providecommand \bibfnamefont [1]{#1}%
\providecommand \citenamefont [1]{#1}%
\providecommand \href@noop [0]{\@secondoftwo}%
\providecommand \href [0]{\begingroup \@sanitize@url \@href}%
\providecommand \@href[1]{\@@startlink{#1}\@@href}%
\providecommand \@@href[1]{\endgroup#1\@@endlink}%
\providecommand \@sanitize@url [0]{\catcode `\\12\catcode `\$12\catcode
  `\&12\catcode `\#12\catcode `\^12\catcode `\_12\catcode `\%12\relax}%
\providecommand \@@startlink[1]{}%
\providecommand \@@endlink[0]{}%
\providecommand \url  [0]{\begingroup\@sanitize@url \@url }%
\providecommand \@url [1]{\endgroup\@href {#1}{\urlprefix }}%
\providecommand \urlprefix  [0]{URL }%
\providecommand \Eprint [0]{\href }%
\providecommand \doibase [0]{https://doi.org/}%
\providecommand \selectlanguage [0]{\@gobble}%
\providecommand \bibinfo  [0]{\@secondoftwo}%
\providecommand \bibfield  [0]{\@secondoftwo}%
\providecommand \translation [1]{[#1]}%
\providecommand \BibitemOpen [0]{}%
\providecommand \bibitemStop [0]{}%
\providecommand \bibitemNoStop [0]{.\EOS\space}%
\providecommand \EOS [0]{\spacefactor3000\relax}%
\providecommand \BibitemShut  [1]{\csname bibitem#1\endcsname}%
\let\auto@bib@innerbib\@empty
\bibitem [{\citenamefont {Choi}\ \emph {et~al.}(2003)\citenamefont {Choi} \emph
  {et~al.}}]{Belle:2003nnu}%
  \BibitemOpen
  \bibfield  {author} {\bibinfo {author} {\bibfnamefont {S.~K.}\ \bibnamefont
  {Choi}} \emph {et~al.} (\bibinfo {collaboration} {Belle}),\ }\bibfield
  {title} {\bibinfo {title} {{Observation of a narrow charmonium-like state in
  exclusive $B^\pm \to K^\pm \pi^+ \pi^- J/\psi$ decays}},\ }\href
  {https://doi.org/10.1103/PhysRevLett.91.262001} {\bibfield  {journal}
  {\bibinfo  {journal} {Phys. Rev. Lett.}\ }\textbf {\bibinfo {volume} {91}},\
  \bibinfo {pages} {262001} (\bibinfo {year} {2003})},\ \Eprint
  {https://arxiv.org/abs/hep-ex/0309032} {arXiv:hep-ex/0309032} \BibitemShut
  {NoStop}%
\bibitem [{\citenamefont {Chen}\ \emph {et~al.}(2016)\citenamefont {Chen},
  \citenamefont {Chen}, \citenamefont {Liu},\ and\ \citenamefont
  {Zhu}}]{Chen:2016qju}%
  \BibitemOpen
  \bibfield  {author} {\bibinfo {author} {\bibfnamefont {H.-X.}\ \bibnamefont
  {Chen}}, \bibinfo {author} {\bibfnamefont {W.}~\bibnamefont {Chen}}, \bibinfo
  {author} {\bibfnamefont {X.}~\bibnamefont {Liu}},\ and\ \bibinfo {author}
  {\bibfnamefont {S.-L.}\ \bibnamefont {Zhu}},\ }\bibfield  {title} {\bibinfo
  {title} {{The hidden-charm pentaquark and tetraquark states}},\ }\href
  {https://doi.org/10.1016/j.physrep.2016.05.004} {\bibfield  {journal}
  {\bibinfo  {journal} {Phys. Rept.}\ }\textbf {\bibinfo {volume} {639}},\
  \bibinfo {pages} {1} (\bibinfo {year} {2016})},\ \Eprint
  {https://arxiv.org/abs/1601.02092} {arXiv:1601.02092 [hep-ph]} \BibitemShut
  {NoStop}%
\bibitem [{\citenamefont {Hosaka}\ \emph {et~al.}(2016)\citenamefont {Hosaka},
  \citenamefont {Iijima}, \citenamefont {Miyabayashi}, \citenamefont {Sakai},\
  and\ \citenamefont {Yasui}}]{Hosaka:2016pey}%
  \BibitemOpen
  \bibfield  {author} {\bibinfo {author} {\bibfnamefont {A.}~\bibnamefont
  {Hosaka}}, \bibinfo {author} {\bibfnamefont {T.}~\bibnamefont {Iijima}},
  \bibinfo {author} {\bibfnamefont {K.}~\bibnamefont {Miyabayashi}}, \bibinfo
  {author} {\bibfnamefont {Y.}~\bibnamefont {Sakai}},\ and\ \bibinfo {author}
  {\bibfnamefont {S.}~\bibnamefont {Yasui}},\ }\bibfield  {title} {\bibinfo
  {title} {{Exotic hadrons with heavy flavors: $X, Y, Z,$ and related
  states}},\ }\href {https://doi.org/10.1093/ptep/ptw045} {\bibfield  {journal}
  {\bibinfo  {journal} {PTEP}\ }\textbf {\bibinfo {volume} {2016}},\ \bibinfo
  {pages} {062C01} (\bibinfo {year} {2016})},\ \Eprint
  {https://arxiv.org/abs/1603.09229} {arXiv:1603.09229 [hep-ph]} \BibitemShut
  {NoStop}%
\bibitem [{\citenamefont {Richard}(2016)}]{Richard:2016eis}%
  \BibitemOpen
  \bibfield  {author} {\bibinfo {author} {\bibfnamefont {J.-M.}\ \bibnamefont
  {Richard}},\ }\bibfield  {title} {\bibinfo {title} {{Exotic hadrons: review
  and perspectives}},\ }\href {https://doi.org/10.1007/s00601-016-1159-0}
  {\bibfield  {journal} {\bibinfo  {journal} {Few Body Syst.}\ }\textbf
  {\bibinfo {volume} {57}},\ \bibinfo {pages} {1185} (\bibinfo {year}
  {2016})},\ \Eprint {https://arxiv.org/abs/1606.08593} {arXiv:1606.08593
  [hep-ph]} \BibitemShut {NoStop}%
\bibitem [{\citenamefont {Lebed}\ \emph {et~al.}(2017)\citenamefont {Lebed},
  \citenamefont {Mitchell},\ and\ \citenamefont {Swanson}}]{Lebed:2016hpi}%
  \BibitemOpen
  \bibfield  {author} {\bibinfo {author} {\bibfnamefont {R.~F.}\ \bibnamefont
  {Lebed}}, \bibinfo {author} {\bibfnamefont {R.~E.}\ \bibnamefont
  {Mitchell}},\ and\ \bibinfo {author} {\bibfnamefont {E.~S.}\ \bibnamefont
  {Swanson}},\ }\bibfield  {title} {\bibinfo {title} {{Heavy-Quark QCD
  Exotica}},\ }\href {https://doi.org/10.1016/j.ppnp.2016.11.003} {\bibfield
  {journal} {\bibinfo  {journal} {Prog. Part. Nucl. Phys.}\ }\textbf {\bibinfo
  {volume} {93}},\ \bibinfo {pages} {143} (\bibinfo {year} {2017})},\ \Eprint
  {https://arxiv.org/abs/1610.04528} {arXiv:1610.04528 [hep-ph]} \BibitemShut
  {NoStop}%
\bibitem [{\citenamefont {Esposito}\ \emph {et~al.}(2017)\citenamefont
  {Esposito}, \citenamefont {Pilloni},\ and\ \citenamefont
  {Polosa}}]{Esposito:2016noz}%
  \BibitemOpen
  \bibfield  {author} {\bibinfo {author} {\bibfnamefont {A.}~\bibnamefont
  {Esposito}}, \bibinfo {author} {\bibfnamefont {A.}~\bibnamefont {Pilloni}},\
  and\ \bibinfo {author} {\bibfnamefont {A.~D.}\ \bibnamefont {Polosa}},\
  }\bibfield  {title} {\bibinfo {title} {{Multiquark Resonances}},\ }\href
  {https://doi.org/10.1016/j.physrep.2016.11.002} {\bibfield  {journal}
  {\bibinfo  {journal} {Phys. Rept.}\ }\textbf {\bibinfo {volume} {668}},\
  \bibinfo {pages} {1} (\bibinfo {year} {2017})},\ \Eprint
  {https://arxiv.org/abs/1611.07920} {arXiv:1611.07920 [hep-ph]} \BibitemShut
  {NoStop}%
\bibitem [{\citenamefont {Guo}\ \emph {et~al.}(2018)\citenamefont {Guo},
  \citenamefont {Hanhart}, \citenamefont {Mei\ss{}ner}, \citenamefont {Wang},
  \citenamefont {Zhao},\ and\ \citenamefont {Zou}}]{Guo:2017jvc}%
  \BibitemOpen
  \bibfield  {author} {\bibinfo {author} {\bibfnamefont {F.-K.}\ \bibnamefont
  {Guo}}, \bibinfo {author} {\bibfnamefont {C.}~\bibnamefont {Hanhart}},
  \bibinfo {author} {\bibfnamefont {U.-G.}\ \bibnamefont {Mei\ss{}ner}},
  \bibinfo {author} {\bibfnamefont {Q.}~\bibnamefont {Wang}}, \bibinfo {author}
  {\bibfnamefont {Q.}~\bibnamefont {Zhao}},\ and\ \bibinfo {author}
  {\bibfnamefont {B.-S.}\ \bibnamefont {Zou}},\ }\bibfield  {title} {\bibinfo
  {title} {{Hadronic molecules}},\ }\href
  {https://doi.org/10.1103/RevModPhys.90.015004} {\bibfield  {journal}
  {\bibinfo  {journal} {Rev. Mod. Phys.}\ }\textbf {\bibinfo {volume} {90}},\
  \bibinfo {pages} {015004} (\bibinfo {year} {2018})},\ \Eprint
  {https://arxiv.org/abs/1705.00141} {arXiv:1705.00141 [hep-ph]} \BibitemShut
  {NoStop}%
\bibitem [{\citenamefont {Ali}\ \emph {et~al.}(2017)\citenamefont {Ali},
  \citenamefont {Lange},\ and\ \citenamefont {Stone}}]{Ali:2017jda}%
  \BibitemOpen
  \bibfield  {author} {\bibinfo {author} {\bibfnamefont {A.}~\bibnamefont
  {Ali}}, \bibinfo {author} {\bibfnamefont {J.~S.}\ \bibnamefont {Lange}},\
  and\ \bibinfo {author} {\bibfnamefont {S.}~\bibnamefont {Stone}},\ }\bibfield
   {title} {\bibinfo {title} {{Exotics: Heavy Pentaquarks and Tetraquarks}},\
  }\href {https://doi.org/10.1016/j.ppnp.2017.08.003} {\bibfield  {journal}
  {\bibinfo  {journal} {Prog. Part. Nucl. Phys.}\ }\textbf {\bibinfo {volume}
  {97}},\ \bibinfo {pages} {123} (\bibinfo {year} {2017})},\ \Eprint
  {https://arxiv.org/abs/1706.00610} {arXiv:1706.00610 [hep-ph]} \BibitemShut
  {NoStop}%
\bibitem [{\citenamefont {Olsen}\ \emph {et~al.}(2018)\citenamefont {Olsen},
  \citenamefont {Skwarnicki},\ and\ \citenamefont {Zieminska}}]{Olsen:2017bmm}%
  \BibitemOpen
  \bibfield  {author} {\bibinfo {author} {\bibfnamefont {S.~L.}\ \bibnamefont
  {Olsen}}, \bibinfo {author} {\bibfnamefont {T.}~\bibnamefont {Skwarnicki}},\
  and\ \bibinfo {author} {\bibfnamefont {D.}~\bibnamefont {Zieminska}},\
  }\bibfield  {title} {\bibinfo {title} {{Non-standard heavy mesons and
  baryons, an Experimental review}},\ }\href
  {https://doi.org/10.1103/RevModPhys.90.015003} {\bibfield  {journal}
  {\bibinfo  {journal} {Rev. Mod. Phys.}\ }\textbf {\bibinfo {volume} {90}},\
  \bibinfo {pages} {015003} (\bibinfo {year} {2018})},\ \Eprint
  {https://arxiv.org/abs/1708.04012} {arXiv:1708.04012 [hep-ph]} \BibitemShut
  {NoStop}%
\bibitem [{\citenamefont {Altmannshofer}\ \emph {et~al.}(2019)\citenamefont
  {Altmannshofer} \emph {et~al.}}]{Kou:2018nap}%
  \BibitemOpen
  \bibfield  {author} {\bibinfo {author} {\bibfnamefont {W.}~\bibnamefont
  {Altmannshofer}} \emph {et~al.} (\bibinfo {collaboration} {Belle-II}),\
  }\bibfield  {title} {\bibinfo {title} {{The Belle II Physics Book}},\ }\href
  {https://doi.org/10.1093/ptep/ptz106} {\bibfield  {journal} {\bibinfo
  {journal} {PTEP}\ }\textbf {\bibinfo {volume} {2019}},\ \bibinfo {pages}
  {123C01} (\bibinfo {year} {2019})},\ \bibinfo {note} {[Erratum: PTEP 2020,
  029201 (2020)]},\ \Eprint {https://arxiv.org/abs/1808.10567}
  {arXiv:1808.10567 [hep-ex]} \BibitemShut {NoStop}%
\bibitem [{\citenamefont {Kalashnikova}\ and\ \citenamefont
  {Nefediev}(2019)}]{Kalashnikova:2018vkv}%
  \BibitemOpen
  \bibfield  {author} {\bibinfo {author} {\bibfnamefont {Y.~S.}\ \bibnamefont
  {Kalashnikova}}\ and\ \bibinfo {author} {\bibfnamefont {A.~V.}\ \bibnamefont
  {Nefediev}},\ }\bibfield  {title} {\bibinfo {title} {{$X(3872)$ in the
  molecular model}},\ }\href {https://doi.org/10.3367/UFNe.2018.08.038411}
  {\bibfield  {journal} {\bibinfo  {journal} {Phys. Usp.}\ }\textbf {\bibinfo
  {volume} {62}},\ \bibinfo {pages} {568} (\bibinfo {year} {2019})},\ \Eprint
  {https://arxiv.org/abs/1811.01324} {arXiv:1811.01324 [hep-ph]} \BibitemShut
  {NoStop}%
\bibitem [{\citenamefont {Cerri}\ \emph {et~al.}(2019)\citenamefont {Cerri}
  \emph {et~al.}}]{Cerri:2018ypt}%
  \BibitemOpen
  \bibfield  {author} {\bibinfo {author} {\bibfnamefont {A.}~\bibnamefont
  {Cerri}} \emph {et~al.},\ }\bibfield  {title} {\bibinfo {title}
  {{Opportunities in Flavour Physics at the HL-LHC and HE-LHC}},\ }\href
  {https://doi.org/10.23731/CYRM-2019-007.867} {\bibfield  {journal} {\bibinfo
  {journal} {CERN Yellow Rep. Monogr.}\ }\textbf {\bibinfo {volume} {7}},\
  \bibinfo {pages} {867} (\bibinfo {year} {2019})},\ \Eprint
  {https://arxiv.org/abs/1812.07638} {arXiv:1812.07638 [hep-ph]} \BibitemShut
  {NoStop}%
\bibitem [{\citenamefont {Liu}\ \emph {et~al.}(2019{\natexlab{a}})\citenamefont
  {Liu}, \citenamefont {Chen}, \citenamefont {Chen}, \citenamefont {Liu},\ and\
  \citenamefont {Zhu}}]{Liu:2019zoy}%
  \BibitemOpen
  \bibfield  {author} {\bibinfo {author} {\bibfnamefont {Y.-R.}\ \bibnamefont
  {Liu}}, \bibinfo {author} {\bibfnamefont {H.-X.}\ \bibnamefont {Chen}},
  \bibinfo {author} {\bibfnamefont {W.}~\bibnamefont {Chen}}, \bibinfo {author}
  {\bibfnamefont {X.}~\bibnamefont {Liu}},\ and\ \bibinfo {author}
  {\bibfnamefont {S.-L.}\ \bibnamefont {Zhu}},\ }\bibfield  {title} {\bibinfo
  {title} {{Pentaquark and Tetraquark states}},\ }\href
  {https://doi.org/10.1016/j.ppnp.2019.04.003} {\bibfield  {journal} {\bibinfo
  {journal} {Prog. Part. Nucl. Phys.}\ }\textbf {\bibinfo {volume} {107}},\
  \bibinfo {pages} {237} (\bibinfo {year} {2019}{\natexlab{a}})},\ \Eprint
  {https://arxiv.org/abs/1903.11976} {arXiv:1903.11976 [hep-ph]} \BibitemShut
  {NoStop}%
\bibitem [{\citenamefont {Brambilla}\ \emph {et~al.}(2020)\citenamefont
  {Brambilla}, \citenamefont {Eidelman}, \citenamefont {Hanhart}, \citenamefont
  {Nefediev}, \citenamefont {Shen}, \citenamefont {Thomas}, \citenamefont
  {Vairo},\ and\ \citenamefont {Yuan}}]{Brambilla:2019esw}%
  \BibitemOpen
  \bibfield  {author} {\bibinfo {author} {\bibfnamefont {N.}~\bibnamefont
  {Brambilla}}, \bibinfo {author} {\bibfnamefont {S.}~\bibnamefont {Eidelman}},
  \bibinfo {author} {\bibfnamefont {C.}~\bibnamefont {Hanhart}}, \bibinfo
  {author} {\bibfnamefont {A.}~\bibnamefont {Nefediev}}, \bibinfo {author}
  {\bibfnamefont {C.-P.}\ \bibnamefont {Shen}}, \bibinfo {author}
  {\bibfnamefont {C.~E.}\ \bibnamefont {Thomas}}, \bibinfo {author}
  {\bibfnamefont {A.}~\bibnamefont {Vairo}},\ and\ \bibinfo {author}
  {\bibfnamefont {C.-Z.}\ \bibnamefont {Yuan}},\ }\bibfield  {title} {\bibinfo
  {title} {{The $XYZ$ states: experimental and theoretical status and
  perspectives}},\ }\href {https://doi.org/10.1016/j.physrep.2020.05.001}
  {\bibfield  {journal} {\bibinfo  {journal} {Phys. Rept.}\ }\textbf {\bibinfo
  {volume} {873}},\ \bibinfo {pages} {1} (\bibinfo {year} {2020})},\ \Eprint
  {https://arxiv.org/abs/1907.07583} {arXiv:1907.07583 [hep-ex]} \BibitemShut
  {NoStop}%
\bibitem [{\citenamefont {Guo}\ \emph {et~al.}(2020)\citenamefont {Guo},
  \citenamefont {Liu},\ and\ \citenamefont {Sakai}}]{Guo:2019twa}%
  \BibitemOpen
  \bibfield  {author} {\bibinfo {author} {\bibfnamefont {F.-K.}\ \bibnamefont
  {Guo}}, \bibinfo {author} {\bibfnamefont {X.-H.}\ \bibnamefont {Liu}},\ and\
  \bibinfo {author} {\bibfnamefont {S.}~\bibnamefont {Sakai}},\ }\bibfield
  {title} {\bibinfo {title} {{Threshold cusps and triangle singularities in
  hadronic reactions}},\ }\href {https://doi.org/10.1016/j.ppnp.2020.103757}
  {\bibfield  {journal} {\bibinfo  {journal} {Prog. Part. Nucl. Phys.}\
  }\textbf {\bibinfo {volume} {112}},\ \bibinfo {pages} {103757} (\bibinfo
  {year} {2020})},\ \Eprint {https://arxiv.org/abs/1912.07030}
  {arXiv:1912.07030 [hep-ph]} \BibitemShut {NoStop}%
\bibitem [{\citenamefont {Yang}\ \emph {et~al.}(2020)\citenamefont {Yang},
  \citenamefont {Ping},\ and\ \citenamefont {Segovia}}]{Yang:2020atz}%
  \BibitemOpen
  \bibfield  {author} {\bibinfo {author} {\bibfnamefont {G.}~\bibnamefont
  {Yang}}, \bibinfo {author} {\bibfnamefont {J.}~\bibnamefont {Ping}},\ and\
  \bibinfo {author} {\bibfnamefont {J.}~\bibnamefont {Segovia}},\ }\bibfield
  {title} {\bibinfo {title} {{Tetra- and penta-quark structures in the
  constituent quark model}},\ }\href {https://doi.org/10.3390/sym12111869}
  {\bibfield  {journal} {\bibinfo  {journal} {Symmetry}\ }\textbf {\bibinfo
  {volume} {12}},\ \bibinfo {pages} {1869} (\bibinfo {year} {2020})},\ \Eprint
  {https://arxiv.org/abs/2009.00238} {arXiv:2009.00238 [hep-ph]} \BibitemShut
  {NoStop}%
\bibitem [{\citenamefont {Ortega}\ and\ \citenamefont
  {Entem}(2021)}]{Ortega:2020tng}%
  \BibitemOpen
  \bibfield  {author} {\bibinfo {author} {\bibfnamefont {P.~G.}\ \bibnamefont
  {Ortega}}\ and\ \bibinfo {author} {\bibfnamefont {D.~R.}\ \bibnamefont
  {Entem}},\ }\bibfield  {title} {\bibinfo {title} {{Coupling hadron-hadron
  thresholds within a chiral quark model approach}},\ }\href
  {https://doi.org/10.3390/sym13020279} {\bibfield  {journal} {\bibinfo
  {journal} {Symmetry}\ }\textbf {\bibinfo {volume} {13}},\ \bibinfo {pages}
  {279} (\bibinfo {year} {2021})},\ \Eprint {https://arxiv.org/abs/2012.10105}
  {arXiv:2012.10105 [hep-ph]} \BibitemShut {NoStop}%
\bibitem [{\citenamefont {Dong}\ \emph
  {et~al.}(2021{\natexlab{a}})\citenamefont {Dong}, \citenamefont {Guo},\ and\
  \citenamefont {Zou}}]{Dong:2021juy}%
  \BibitemOpen
  \bibfield  {author} {\bibinfo {author} {\bibfnamefont {X.-K.}\ \bibnamefont
  {Dong}}, \bibinfo {author} {\bibfnamefont {F.-K.}\ \bibnamefont {Guo}},\ and\
  \bibinfo {author} {\bibfnamefont {B.-S.}\ \bibnamefont {Zou}},\ }\bibfield
  {title} {\bibinfo {title} {{A survey of heavy-antiheavy hadronic
  molecules}},\ }\href {https://doi.org/10.13725/j.cnki.pip.2021.02.001}
  {\bibfield  {journal} {\bibinfo  {journal} {Progr. Phys.}\ }\textbf {\bibinfo
  {volume} {41}},\ \bibinfo {pages} {65} (\bibinfo {year}
  {2021}{\natexlab{a}})},\ \Eprint {https://arxiv.org/abs/2101.01021}
  {arXiv:2101.01021 [hep-ph]} \BibitemShut {NoStop}%
\bibitem [{\citenamefont {Dong}\ \emph
  {et~al.}(2021{\natexlab{b}})\citenamefont {Dong}, \citenamefont {Guo},\ and\
  \citenamefont {Zou}}]{Dong:2021bvy}%
  \BibitemOpen
  \bibfield  {author} {\bibinfo {author} {\bibfnamefont {X.-K.}\ \bibnamefont
  {Dong}}, \bibinfo {author} {\bibfnamefont {F.-K.}\ \bibnamefont {Guo}},\ and\
  \bibinfo {author} {\bibfnamefont {B.-S.}\ \bibnamefont {Zou}},\ }\bibfield
  {title} {\bibinfo {title} {{A survey of heavy\textendash{}heavy hadronic
  molecules}},\ }\href {https://doi.org/10.1088/1572-9494/ac27a2} {\bibfield
  {journal} {\bibinfo  {journal} {Commun. Theor. Phys.}\ }\textbf {\bibinfo
  {volume} {73}},\ \bibinfo {pages} {125201} (\bibinfo {year}
  {2021}{\natexlab{b}})},\ \Eprint {https://arxiv.org/abs/2108.02673}
  {arXiv:2108.02673 [hep-ph]} \BibitemShut {NoStop}%
\bibitem [{\citenamefont {Ablikim}\ \emph
  {et~al.}(2013{\natexlab{a}})\citenamefont {Ablikim} \emph
  {et~al.}}]{BESIII:2013ris}%
  \BibitemOpen
  \bibfield  {author} {\bibinfo {author} {\bibfnamefont {M.}~\bibnamefont
  {Ablikim}} \emph {et~al.} (\bibinfo {collaboration} {BESIII}),\ }\bibfield
  {title} {\bibinfo {title} {{Observation of a Charged Charmoniumlike Structure
  in $e^+e^-$ \textrightarrow{} $\pi^+\pi^-J/\psi$ at $\sqrt{s}$=4.26 GeV}},\
  }\href {https://doi.org/10.1103/PhysRevLett.110.252001} {\bibfield  {journal}
  {\bibinfo  {journal} {Phys. Rev. Lett.}\ }\textbf {\bibinfo {volume} {110}},\
  \bibinfo {pages} {252001} (\bibinfo {year} {2013}{\natexlab{a}})},\ \Eprint
  {https://arxiv.org/abs/1303.5949} {arXiv:1303.5949 [hep-ex]} \BibitemShut
  {NoStop}%
\bibitem [{\citenamefont {Liu}\ \emph {et~al.}(2013)\citenamefont {Liu} \emph
  {et~al.}}]{Belle:2013yex}%
  \BibitemOpen
  \bibfield  {author} {\bibinfo {author} {\bibfnamefont {Z.~Q.}\ \bibnamefont
  {Liu}} \emph {et~al.} (\bibinfo {collaboration} {Belle}),\ }\bibfield
  {title} {\bibinfo {title} {{Study of $e^+e^- \to \pi^+ \pi^- J/\psi$ and
  Observation of a Charged Charmonium-like State at Belle}},\ }\href
  {https://doi.org/10.1103/PhysRevLett.110.252002} {\bibfield  {journal}
  {\bibinfo  {journal} {Phys. Rev. Lett.}\ }\textbf {\bibinfo {volume} {110}},\
  \bibinfo {pages} {252002} (\bibinfo {year} {2013})},\ \bibinfo {note}
  {[Erratum: Phys.Rev.Lett. 111, 019901 (2013)]},\ \Eprint
  {https://arxiv.org/abs/1304.0121} {arXiv:1304.0121 [hep-ex]} \BibitemShut
  {NoStop}%
\bibitem [{\citenamefont {Ablikim}\ \emph
  {et~al.}(2014{\natexlab{a}})\citenamefont {Ablikim} \emph
  {et~al.}}]{BESIII:2013qmu}%
  \BibitemOpen
  \bibfield  {author} {\bibinfo {author} {\bibfnamefont {M.}~\bibnamefont
  {Ablikim}} \emph {et~al.} (\bibinfo {collaboration} {BESIII}),\ }\bibfield
  {title} {\bibinfo {title} {{Observation of a charged $(D\bar{D}^{*})^\pm$
  mass peak in $e^{+}e^{-} \to \pi D\bar{D}^{*}$ at $\sqrt{s} =$ 4.26 GeV}},\
  }\href {https://doi.org/10.1103/PhysRevLett.112.022001} {\bibfield  {journal}
  {\bibinfo  {journal} {Phys. Rev. Lett.}\ }\textbf {\bibinfo {volume} {112}},\
  \bibinfo {pages} {022001} (\bibinfo {year} {2014}{\natexlab{a}})},\ \Eprint
  {https://arxiv.org/abs/1310.1163} {arXiv:1310.1163 [hep-ex]} \BibitemShut
  {NoStop}%
\bibitem [{\citenamefont {Aaij}\ \emph
  {et~al.}(2022{\natexlab{a}})\citenamefont {Aaij} \emph
  {et~al.}}]{LHCb:2021vvq}%
  \BibitemOpen
  \bibfield  {author} {\bibinfo {author} {\bibfnamefont {R.}~\bibnamefont
  {Aaij}} \emph {et~al.} (\bibinfo {collaboration} {LHCb}),\ }\bibfield
  {title} {\bibinfo {title} {{Observation of an exotic narrow doubly charmed
  tetraquark}},\ }\href {https://doi.org/10.1038/s41567-022-01614-y} {\bibfield
   {journal} {\bibinfo  {journal} {Nature Phys.}\ }\textbf {\bibinfo {volume}
  {18}},\ \bibinfo {pages} {751} (\bibinfo {year} {2022}{\natexlab{a}})},\
  \Eprint {https://arxiv.org/abs/2109.01038} {arXiv:2109.01038 [hep-ex]}
  \BibitemShut {NoStop}%
\bibitem [{\citenamefont {Aaij}\ \emph
  {et~al.}(2022{\natexlab{b}})\citenamefont {Aaij} \emph
  {et~al.}}]{LHCb:2021auc}%
  \BibitemOpen
  \bibfield  {author} {\bibinfo {author} {\bibfnamefont {R.}~\bibnamefont
  {Aaij}} \emph {et~al.} (\bibinfo {collaboration} {LHCb}),\ }\bibfield
  {title} {\bibinfo {title} {{Study of the doubly charmed tetraquark
  $T_{cc}^{+}$}},\ }\href {https://doi.org/10.1038/s41467-022-30206-w}
  {\bibfield  {journal} {\bibinfo  {journal} {Nature Commun.}\ }\textbf
  {\bibinfo {volume} {13}},\ \bibinfo {pages} {3351} (\bibinfo {year}
  {2022}{\natexlab{b}})},\ \Eprint {https://arxiv.org/abs/2109.01056}
  {arXiv:2109.01056 [hep-ex]} \BibitemShut {NoStop}%
\bibitem [{\citenamefont {Ablikim}\ \emph
  {et~al.}(2013{\natexlab{b}})\citenamefont {Ablikim} \emph
  {et~al.}}]{BESIII:2013ouc}%
  \BibitemOpen
  \bibfield  {author} {\bibinfo {author} {\bibfnamefont {M.}~\bibnamefont
  {Ablikim}} \emph {et~al.} (\bibinfo {collaboration} {BESIII}),\ }\bibfield
  {title} {\bibinfo {title} {{Observation of a Charged Charmoniumlike Structure
  $Z_c$(4020) and Search for the $Z_c$(3900) in $e^+e^- \to \pi^+\pi^-h_c$}},\
  }\href {https://doi.org/10.1103/PhysRevLett.111.242001} {\bibfield  {journal}
  {\bibinfo  {journal} {Phys. Rev. Lett.}\ }\textbf {\bibinfo {volume} {111}},\
  \bibinfo {pages} {242001} (\bibinfo {year} {2013}{\natexlab{b}})},\ \Eprint
  {https://arxiv.org/abs/1309.1896} {arXiv:1309.1896 [hep-ex]} \BibitemShut
  {NoStop}%
\bibitem [{\citenamefont {Ablikim}\ \emph
  {et~al.}(2014{\natexlab{b}})\citenamefont {Ablikim} \emph
  {et~al.}}]{BESIII:2013mhi}%
  \BibitemOpen
  \bibfield  {author} {\bibinfo {author} {\bibfnamefont {M.}~\bibnamefont
  {Ablikim}} \emph {et~al.} (\bibinfo {collaboration} {BESIII}),\ }\bibfield
  {title} {\bibinfo {title} {{Observation of a charged charmoniumlike structure
  in $e^+e^- \to (D^{*} \bar{D}^{*})^{\pm} \pi^\mp$ at $\sqrt{s}=4.26$ GeV}},\
  }\href {https://doi.org/10.1103/PhysRevLett.112.132001} {\bibfield  {journal}
  {\bibinfo  {journal} {Phys. Rev. Lett.}\ }\textbf {\bibinfo {volume} {112}},\
  \bibinfo {pages} {132001} (\bibinfo {year} {2014}{\natexlab{b}})},\ \Eprint
  {https://arxiv.org/abs/1308.2760} {arXiv:1308.2760 [hep-ex]} \BibitemShut
  {NoStop}%
\bibitem [{\citenamefont {Bondar}\ \emph {et~al.}(2012)\citenamefont {Bondar}
  \emph {et~al.}}]{Belle:2011aa}%
  \BibitemOpen
  \bibfield  {author} {\bibinfo {author} {\bibfnamefont {A.}~\bibnamefont
  {Bondar}} \emph {et~al.} (\bibinfo {collaboration} {Belle}),\ }\bibfield
  {title} {\bibinfo {title} {{Observation of two charged bottomonium-like
  resonances in $\Upsilon(5S)$ decays}},\ }\href
  {https://doi.org/10.1103/PhysRevLett.108.122001} {\bibfield  {journal}
  {\bibinfo  {journal} {Phys. Rev. Lett.}\ }\textbf {\bibinfo {volume} {108}},\
  \bibinfo {pages} {122001} (\bibinfo {year} {2012})},\ \Eprint
  {https://arxiv.org/abs/1110.2251} {arXiv:1110.2251 [hep-ex]} \BibitemShut
  {NoStop}%
\bibitem [{\citenamefont {Garmash}\ \emph {et~al.}(2016)\citenamefont {Garmash}
  \emph {et~al.}}]{Belle:2015upu}%
  \BibitemOpen
  \bibfield  {author} {\bibinfo {author} {\bibfnamefont {A.}~\bibnamefont
  {Garmash}} \emph {et~al.} (\bibinfo {collaboration} {Belle}),\ }\bibfield
  {title} {\bibinfo {title} {{Observation of $Z_b(10610)$ and $Z_b(10650)$
  Decaying to $B$ Mesons}},\ }\href
  {https://doi.org/10.1103/PhysRevLett.116.212001} {\bibfield  {journal}
  {\bibinfo  {journal} {Phys. Rev. Lett.}\ }\textbf {\bibinfo {volume} {116}},\
  \bibinfo {pages} {212001} (\bibinfo {year} {2016})},\ \Eprint
  {https://arxiv.org/abs/1512.07419} {arXiv:1512.07419 [hep-ex]} \BibitemShut
  {NoStop}%
\bibitem [{\citenamefont {Ablikim}\ \emph {et~al.}(2021)\citenamefont {Ablikim}
  \emph {et~al.}}]{BESIII:2020qkh}%
  \BibitemOpen
  \bibfield  {author} {\bibinfo {author} {\bibfnamefont {M.}~\bibnamefont
  {Ablikim}} \emph {et~al.} (\bibinfo {collaboration} {BESIII}),\ }\bibfield
  {title} {\bibinfo {title} {{Observation of a Near-Threshold Structure in the
  $K^+$ Recoil-Mass Spectra in $e^+e^- \to K^+(D_s^-D^{*0}+D_s^{*-}D^0$)}},\
  }\href {https://doi.org/10.1103/PhysRevLett.126.102001} {\bibfield  {journal}
  {\bibinfo  {journal} {Phys. Rev. Lett.}\ }\textbf {\bibinfo {volume} {126}},\
  \bibinfo {pages} {102001} (\bibinfo {year} {2021})},\ \Eprint
  {https://arxiv.org/abs/2011.07855} {arXiv:2011.07855 [hep-ex]} \BibitemShut
  {NoStop}%
\bibitem [{\citenamefont {Ablikim}\ \emph {et~al.}(2022)\citenamefont {Ablikim}
  \emph {et~al.}}]{BESIII:2022qzr}%
  \BibitemOpen
  \bibfield  {author} {\bibinfo {author} {\bibfnamefont {M.}~\bibnamefont
  {Ablikim}} \emph {et~al.} (\bibinfo {collaboration} {BESIII}),\ }\bibfield
  {title} {\bibinfo {title} {{Evidence for a Neutral Near-Threshold Structure
  in the $K^{0}_{S}$ recoil-mass spectra in $e^+e^- \to K^{0}_{S}D_s^+D^{*-}$
  and $e^+e^- \to K^{0}_{S}D_s^{*+}D^{-}$}},\ }\href
  {https://doi.org/10.1103/PhysRevLett.129.112003} {\bibfield  {journal}
  {\bibinfo  {journal} {Phys. Rev. Lett.}\ }\textbf {\bibinfo {volume} {129}},\
  \bibinfo {pages} {112003} (\bibinfo {year} {2022})},\ \Eprint
  {https://arxiv.org/abs/2204.13703} {arXiv:2204.13703 [hep-ex]} \BibitemShut
  {NoStop}%
\bibitem [{\citenamefont {Aaij}\ \emph
  {et~al.}(2021{\natexlab{a}})\citenamefont {Aaij} \emph
  {et~al.}}]{LHCb:2021uow}%
  \BibitemOpen
  \bibfield  {author} {\bibinfo {author} {\bibfnamefont {R.}~\bibnamefont
  {Aaij}} \emph {et~al.} (\bibinfo {collaboration} {LHCb}),\ }\bibfield
  {title} {\bibinfo {title} {{Observation of New Resonances Decaying to $J/\psi
  K^+$ and $J/\psi \phi$}},\ }\href
  {https://doi.org/10.1103/PhysRevLett.127.082001} {\bibfield  {journal}
  {\bibinfo  {journal} {Phys. Rev. Lett.}\ }\textbf {\bibinfo {volume} {127}},\
  \bibinfo {pages} {082001} (\bibinfo {year} {2021}{\natexlab{a}})},\ \Eprint
  {https://arxiv.org/abs/2103.01803} {arXiv:2103.01803 [hep-ex]} \BibitemShut
  {NoStop}%
\bibitem [{\citenamefont {Aaij}\ \emph
  {et~al.}(2023{\natexlab{a}})\citenamefont {Aaij} \emph
  {et~al.}}]{LHCb:2023hxg}%
  \BibitemOpen
  \bibfield  {author} {\bibinfo {author} {\bibfnamefont {R.}~\bibnamefont
  {Aaij}} \emph {et~al.} (\bibinfo {collaboration} {LHCb}),\ }\bibfield
  {title} {\bibinfo {title} {{Evidence of a $J/\psi K_s^0$ Structure in $
  B^0\to J/\psi \phi K_s^0 $ Decays}},\ }\href
  {https://doi.org/10.1103/PhysRevLett.131.131901} {\bibfield  {journal}
  {\bibinfo  {journal} {Phys. Rev. Lett.}\ }\textbf {\bibinfo {volume} {131}},\
  \bibinfo {pages} {131901} (\bibinfo {year} {2023}{\natexlab{a}})},\ \Eprint
  {https://arxiv.org/abs/2301.04899} {arXiv:2301.04899 [hep-ex]} \BibitemShut
  {NoStop}%
\bibitem [{\citenamefont {Aaij}\ \emph {et~al.}(2019)\citenamefont {Aaij} \emph
  {et~al.}}]{LHCb:2019kea}%
  \BibitemOpen
  \bibfield  {author} {\bibinfo {author} {\bibfnamefont {R.}~\bibnamefont
  {Aaij}} \emph {et~al.} (\bibinfo {collaboration} {LHCb}),\ }\bibfield
  {title} {\bibinfo {title} {{Observation of a narrow pentaquark state,
  $P_c(4312)^+$, and of two-peak structure of the $P_c(4450)^+$}},\ }\href
  {https://doi.org/10.1103/PhysRevLett.122.222001} {\bibfield  {journal}
  {\bibinfo  {journal} {Phys. Rev. Lett.}\ }\textbf {\bibinfo {volume} {122}},\
  \bibinfo {pages} {222001} (\bibinfo {year} {2019})},\ \Eprint
  {https://arxiv.org/abs/1904.03947} {arXiv:1904.03947 [hep-ex]} \BibitemShut
  {NoStop}%
\bibitem [{\citenamefont {Aaij}\ \emph
  {et~al.}(2021{\natexlab{b}})\citenamefont {Aaij} \emph
  {et~al.}}]{LHCb:2020jpq}%
  \BibitemOpen
  \bibfield  {author} {\bibinfo {author} {\bibfnamefont {R.}~\bibnamefont
  {Aaij}} \emph {et~al.} (\bibinfo {collaboration} {LHCb}),\ }\bibfield
  {title} {\bibinfo {title} {{Evidence of a $J/\psi\Lambda$ structure and
  observation of excited $\Xi^-$ states in the $\Xi^-_b \to J/\psi\Lambda K^-$
  decay}},\ }\href {https://doi.org/10.1016/j.scib.2021.02.030} {\bibfield
  {journal} {\bibinfo  {journal} {Sci. Bull.}\ }\textbf {\bibinfo {volume}
  {66}},\ \bibinfo {pages} {1278} (\bibinfo {year} {2021}{\natexlab{b}})},\
  \Eprint {https://arxiv.org/abs/2012.10380} {arXiv:2012.10380 [hep-ex]}
  \BibitemShut {NoStop}%
\bibitem [{\citenamefont {Aaij}\ \emph
  {et~al.}(2023{\natexlab{b}})\citenamefont {Aaij} \emph
  {et~al.}}]{LHCb:2022ogu}%
  \BibitemOpen
  \bibfield  {author} {\bibinfo {author} {\bibfnamefont {R.}~\bibnamefont
  {Aaij}} \emph {et~al.} (\bibinfo {collaboration} {LHCb}),\ }\bibfield
  {title} {\bibinfo {title} {{Observation of a $J/\psi\Lambda$ Resonance
  Consistent with a Strange Pentaquark Candidate in $B^-\to J/\psi\Lambda \bar
  p$ Decays}},\ }\href {https://doi.org/10.1103/PhysRevLett.131.031901}
  {\bibfield  {journal} {\bibinfo  {journal} {Phys. Rev. Lett.}\ }\textbf
  {\bibinfo {volume} {131}},\ \bibinfo {pages} {031901} (\bibinfo {year}
  {2023}{\natexlab{b}})},\ \Eprint {https://arxiv.org/abs/2210.10346}
  {arXiv:2210.10346 [hep-ex]} \BibitemShut {NoStop}%
\bibitem [{\citenamefont {Dong}\ \emph
  {et~al.}(2021{\natexlab{c}})\citenamefont {Dong}, \citenamefont {Guo},\ and\
  \citenamefont {Zou}}]{Dong:2020hxe}%
  \BibitemOpen
  \bibfield  {author} {\bibinfo {author} {\bibfnamefont {X.-K.}\ \bibnamefont
  {Dong}}, \bibinfo {author} {\bibfnamefont {F.-K.}\ \bibnamefont {Guo}},\ and\
  \bibinfo {author} {\bibfnamefont {B.-S.}\ \bibnamefont {Zou}},\ }\bibfield
  {title} {\bibinfo {title} {{Explaining the Many Threshold Structures in the
  Heavy-Quark Hadron Spectrum}},\ }\href
  {https://doi.org/10.1103/PhysRevLett.126.152001} {\bibfield  {journal}
  {\bibinfo  {journal} {Phys. Rev. Lett.}\ }\textbf {\bibinfo {volume} {126}},\
  \bibinfo {pages} {152001} (\bibinfo {year} {2021}{\natexlab{c}})},\ \Eprint
  {https://arxiv.org/abs/2011.14517} {arXiv:2011.14517 [hep-ph]} \BibitemShut
  {NoStop}%
\bibitem [{\citenamefont {Voloshin}\ and\ \citenamefont
  {Okun}(1976)}]{Voloshin:1976ap}%
  \BibitemOpen
  \bibfield  {author} {\bibinfo {author} {\bibfnamefont {M.~B.}\ \bibnamefont
  {Voloshin}}\ and\ \bibinfo {author} {\bibfnamefont {L.~B.}\ \bibnamefont
  {Okun}},\ }\bibfield  {title} {\bibinfo {title} {{Hadron Molecules and
  Charmonium Atom}},\ }\href@noop {} {\bibfield  {journal} {\bibinfo  {journal}
  {JETP Lett.}\ }\textbf {\bibinfo {volume} {23}},\ \bibinfo {pages} {333}
  (\bibinfo {year} {1976})}\BibitemShut {NoStop}%
\bibitem [{\citenamefont {De~R\'ujula}\ \emph {et~al.}(1977)\citenamefont
  {De~R\'ujula}, \citenamefont {Georgi},\ and\ \citenamefont
  {Glashow}}]{DeRujula:1976zlg}%
  \BibitemOpen
  \bibfield  {author} {\bibinfo {author} {\bibfnamefont {A.}~\bibnamefont
  {De~R\'ujula}}, \bibinfo {author} {\bibfnamefont {H.}~\bibnamefont
  {Georgi}},\ and\ \bibinfo {author} {\bibfnamefont {S.~L.}\ \bibnamefont
  {Glashow}},\ }\bibfield  {title} {\bibinfo {title} {{Molecular Charmonium: A
  New Spectroscopy?}},\ }\href {https://doi.org/10.1103/PhysRevLett.38.317}
  {\bibfield  {journal} {\bibinfo  {journal} {Phys. Rev. Lett.}\ }\textbf
  {\bibinfo {volume} {38}},\ \bibinfo {pages} {317} (\bibinfo {year}
  {1977})}\BibitemShut {NoStop}%
\bibitem [{\citenamefont {Nagels}\ \emph {et~al.}(1975)\citenamefont {Nagels},
  \citenamefont {Rijken},\ and\ \citenamefont {de~Swart}}]{Nagels:1975fb}%
  \BibitemOpen
  \bibfield  {author} {\bibinfo {author} {\bibfnamefont {M.~M.}\ \bibnamefont
  {Nagels}}, \bibinfo {author} {\bibfnamefont {T.~A.}\ \bibnamefont {Rijken}},\
  and\ \bibinfo {author} {\bibfnamefont {J.~J.}\ \bibnamefont {de~Swart}},\
  }\bibfield  {title} {\bibinfo {title} {{Baryon-baryon scattering in an
  one-boson-exchange-potential approach. I. Nucleon-nucleon scattering}},\
  }\href {https://doi.org/10.1103/PhysRevD.12.744} {\bibfield  {journal}
  {\bibinfo  {journal} {Phys. Rev. D}\ }\textbf {\bibinfo {volume} {12}},\
  \bibinfo {pages} {744} (\bibinfo {year} {1975})}\BibitemShut {NoStop}%
\bibitem [{\citenamefont {Machleidt}\ \emph {et~al.}(1987)\citenamefont
  {Machleidt}, \citenamefont {Holinde},\ and\ \citenamefont
  {Elster}}]{Machleidt:1987hj}%
  \BibitemOpen
  \bibfield  {author} {\bibinfo {author} {\bibfnamefont {R.}~\bibnamefont
  {Machleidt}}, \bibinfo {author} {\bibfnamefont {K.}~\bibnamefont {Holinde}},\
  and\ \bibinfo {author} {\bibfnamefont {C.}~\bibnamefont {Elster}},\
  }\bibfield  {title} {\bibinfo {title} {{The Bonn Meson Exchange Model for the
  Nucleon Nucleon Interaction}},\ }\href
  {https://doi.org/10.1016/S0370-1573(87)80002-9} {\bibfield  {journal}
  {\bibinfo  {journal} {Phys. Rept.}\ }\textbf {\bibinfo {volume} {149}},\
  \bibinfo {pages} {1} (\bibinfo {year} {1987})}\BibitemShut {NoStop}%
\bibitem [{\citenamefont {Machleidt}(1989)}]{Machleidt:1989tm}%
  \BibitemOpen
  \bibfield  {author} {\bibinfo {author} {\bibfnamefont {R.}~\bibnamefont
  {Machleidt}},\ }\bibfield  {title} {\bibinfo {title} {{The Meson theory of
  nuclear forces and nuclear structure}},\ }\href@noop {} {\bibfield  {journal}
  {\bibinfo  {journal} {Adv. Nucl. Phys.}\ }\textbf {\bibinfo {volume} {19}},\
  \bibinfo {pages} {189} (\bibinfo {year} {1989})}\BibitemShut {NoStop}%
\bibitem [{\citenamefont {Stoks}\ \emph {et~al.}(1994)\citenamefont {Stoks},
  \citenamefont {Klomp}, \citenamefont {Terheggen},\ and\ \citenamefont
  {de~Swart}}]{Stoks:1994wp}%
  \BibitemOpen
  \bibfield  {author} {\bibinfo {author} {\bibfnamefont {V.~G.~J.}\
  \bibnamefont {Stoks}}, \bibinfo {author} {\bibfnamefont {R.~A.~M.}\
  \bibnamefont {Klomp}}, \bibinfo {author} {\bibfnamefont {C.~P.~F.}\
  \bibnamefont {Terheggen}},\ and\ \bibinfo {author} {\bibfnamefont {J.~J.}\
  \bibnamefont {de~Swart}},\ }\bibfield  {title} {\bibinfo {title}
  {{Construction of high quality $N N$ potential models}},\ }\href
  {https://doi.org/10.1103/PhysRevC.49.2950} {\bibfield  {journal} {\bibinfo
  {journal} {Phys. Rev. C}\ }\textbf {\bibinfo {volume} {49}},\ \bibinfo
  {pages} {2950} (\bibinfo {year} {1994})},\ \Eprint
  {https://arxiv.org/abs/nucl-th/9406039} {arXiv:nucl-th/9406039} \BibitemShut
  {NoStop}%
\bibitem [{\citenamefont {T{\"o}rnqvist}(1991)}]{Tornqvist:1991ks}%
  \BibitemOpen
  \bibfield  {author} {\bibinfo {author} {\bibfnamefont {N.~A.}\ \bibnamefont
  {T{\"o}rnqvist}},\ }\bibfield  {title} {\bibinfo {title} {{Possible large
  deuteronlike meson-meson states bound by pions}},\ }\href
  {https://doi.org/10.1103/PhysRevLett.67.556} {\bibfield  {journal} {\bibinfo
  {journal} {Phys. Rev. Lett.}\ }\textbf {\bibinfo {volume} {67}},\ \bibinfo
  {pages} {556} (\bibinfo {year} {1991})}\BibitemShut {NoStop}%
\bibitem [{\citenamefont {T{\"o}rnqvist}(1994)}]{Tornqvist:1993ng}%
  \BibitemOpen
  \bibfield  {author} {\bibinfo {author} {\bibfnamefont {N.~A.}\ \bibnamefont
  {T{\"o}rnqvist}},\ }\bibfield  {title} {\bibinfo {title} {{From the deuteron
  to deusons, an analysis of deuteronlike meson-meson bound states}},\ }\href
  {https://doi.org/10.1007/BF01413192} {\bibfield  {journal} {\bibinfo
  {journal} {Z. Phys. C}\ }\textbf {\bibinfo {volume} {61}},\ \bibinfo {pages}
  {525} (\bibinfo {year} {1994})},\ \Eprint
  {https://arxiv.org/abs/hep-ph/9310247} {arXiv:hep-ph/9310247} \BibitemShut
  {NoStop}%
\bibitem [{\citenamefont {Eichten}\ \emph {et~al.}(1980)\citenamefont
  {Eichten}, \citenamefont {Gottfried}, \citenamefont {Kinoshita},
  \citenamefont {Lane},\ and\ \citenamefont {Yan}}]{Eichten:1979ms}%
  \BibitemOpen
  \bibfield  {author} {\bibinfo {author} {\bibfnamefont {E.}~\bibnamefont
  {Eichten}}, \bibinfo {author} {\bibfnamefont {K.}~\bibnamefont {Gottfried}},
  \bibinfo {author} {\bibfnamefont {T.}~\bibnamefont {Kinoshita}}, \bibinfo
  {author} {\bibfnamefont {K.~D.}\ \bibnamefont {Lane}},\ and\ \bibinfo
  {author} {\bibfnamefont {T.-M.}\ \bibnamefont {Yan}},\ }\bibfield  {title}
  {\bibinfo {title} {{Charmonium: Comparison with Experiment}},\ }\href
  {https://doi.org/10.1103/PhysRevD.21.203} {\bibfield  {journal} {\bibinfo
  {journal} {Phys. Rev. D}\ }\textbf {\bibinfo {volume} {21}},\ \bibinfo
  {pages} {203} (\bibinfo {year} {1980})}\BibitemShut {NoStop}%
\bibitem [{\citenamefont {Godfrey}\ and\ \citenamefont
  {Isgur}(1985)}]{Godfrey:1985xj}%
  \BibitemOpen
  \bibfield  {author} {\bibinfo {author} {\bibfnamefont {S.}~\bibnamefont
  {Godfrey}}\ and\ \bibinfo {author} {\bibfnamefont {N.}~\bibnamefont
  {Isgur}},\ }\bibfield  {title} {\bibinfo {title} {{Mesons in a Relativized
  Quark Model with Chromodynamics}},\ }\href
  {https://doi.org/10.1103/PhysRevD.32.189} {\bibfield  {journal} {\bibinfo
  {journal} {Phys. Rev. D}\ }\textbf {\bibinfo {volume} {32}},\ \bibinfo
  {pages} {189} (\bibinfo {year} {1985})}\BibitemShut {NoStop}%
\bibitem [{\citenamefont {T{\"o}rnqvist}(2003)}]{Tornqvist:2003na}%
  \BibitemOpen
  \bibfield  {author} {\bibinfo {author} {\bibfnamefont {N.~A.}\ \bibnamefont
  {T{\"o}rnqvist}},\ }\bibfield  {title} {\bibinfo {title} {{Comment on the
  narrow charmonium state of Belle at 3871.8 MeV as a deuson}},\ }\href@noop {}
  {\  (\bibinfo {year} {2003})},\ \Eprint
  {https://arxiv.org/abs/hep-ph/0308277} {arXiv:hep-ph/0308277} \BibitemShut
  {NoStop}%
\bibitem [{\citenamefont {Close}\ and\ \citenamefont
  {Page}(2004)}]{Close:2003sg}%
  \BibitemOpen
  \bibfield  {author} {\bibinfo {author} {\bibfnamefont {F.~E.}\ \bibnamefont
  {Close}}\ and\ \bibinfo {author} {\bibfnamefont {P.~R.}\ \bibnamefont
  {Page}},\ }\bibfield  {title} {\bibinfo {title} {{The $D^{*0}\bar{D}^0$
  threshold resonance}},\ }\href
  {https://doi.org/10.1016/j.physletb.2003.10.032} {\bibfield  {journal}
  {\bibinfo  {journal} {Phys. Lett. B}\ }\textbf {\bibinfo {volume} {578}},\
  \bibinfo {pages} {119} (\bibinfo {year} {2004})},\ \Eprint
  {https://arxiv.org/abs/hep-ph/0309253} {arXiv:hep-ph/0309253} \BibitemShut
  {NoStop}%
\bibitem [{\citenamefont {Pakvasa}\ and\ \citenamefont
  {Suzuki}(2004)}]{Pakvasa:2003ea}%
  \BibitemOpen
  \bibfield  {author} {\bibinfo {author} {\bibfnamefont {S.}~\bibnamefont
  {Pakvasa}}\ and\ \bibinfo {author} {\bibfnamefont {M.}~\bibnamefont
  {Suzuki}},\ }\bibfield  {title} {\bibinfo {title} {{On the hidden charm state
  at 3872 MeV}},\ }\href {https://doi.org/10.1016/j.physletb.2003.11.005}
  {\bibfield  {journal} {\bibinfo  {journal} {Phys. Lett. B}\ }\textbf
  {\bibinfo {volume} {579}},\ \bibinfo {pages} {67} (\bibinfo {year} {2004})},\
  \Eprint {https://arxiv.org/abs/hep-ph/0309294} {arXiv:hep-ph/0309294}
  \BibitemShut {NoStop}%
\bibitem [{\citenamefont {Braaten}\ and\ \citenamefont
  {Kusunoki}(2004)}]{Braaten:2003he}%
  \BibitemOpen
  \bibfield  {author} {\bibinfo {author} {\bibfnamefont {E.}~\bibnamefont
  {Braaten}}\ and\ \bibinfo {author} {\bibfnamefont {M.}~\bibnamefont
  {Kusunoki}},\ }\bibfield  {title} {\bibinfo {title} {{Low-energy universality
  and the new charmonium resonance at 3870 MeV}},\ }\href
  {https://doi.org/10.1103/PhysRevD.69.074005} {\bibfield  {journal} {\bibinfo
  {journal} {Phys. Rev. D}\ }\textbf {\bibinfo {volume} {69}},\ \bibinfo
  {pages} {074005} (\bibinfo {year} {2004})},\ \Eprint
  {https://arxiv.org/abs/hep-ph/0311147} {arXiv:hep-ph/0311147} \BibitemShut
  {NoStop}%
\bibitem [{\citenamefont {Wong}(2004)}]{Wong:2003xk}%
  \BibitemOpen
  \bibfield  {author} {\bibinfo {author} {\bibfnamefont {C.-Y.}\ \bibnamefont
  {Wong}},\ }\bibfield  {title} {\bibinfo {title} {{Molecular states of heavy
  quark mesons}},\ }\href {https://doi.org/10.1103/PhysRevC.69.055202}
  {\bibfield  {journal} {\bibinfo  {journal} {Phys. Rev. C}\ }\textbf {\bibinfo
  {volume} {69}},\ \bibinfo {pages} {055202} (\bibinfo {year} {2004})},\
  \Eprint {https://arxiv.org/abs/hep-ph/0311088} {arXiv:hep-ph/0311088}
  \BibitemShut {NoStop}%
\bibitem [{\citenamefont {T{\"o}rnqvist}(2004)}]{Tornqvist:2004qy}%
  \BibitemOpen
  \bibfield  {author} {\bibinfo {author} {\bibfnamefont {N.~A.}\ \bibnamefont
  {T{\"o}rnqvist}},\ }\bibfield  {title} {\bibinfo {title} {{Isospin breaking
  of the narrow charmonium state of Belle at 3872 MeV as a deuson}},\ }\href
  {https://doi.org/10.1016/j.physletb.2004.03.077} {\bibfield  {journal}
  {\bibinfo  {journal} {Phys. Lett. B}\ }\textbf {\bibinfo {volume} {590}},\
  \bibinfo {pages} {209} (\bibinfo {year} {2004})},\ \Eprint
  {https://arxiv.org/abs/hep-ph/0402237} {arXiv:hep-ph/0402237} \BibitemShut
  {NoStop}%
\bibitem [{\citenamefont {Gamermann}\ and\ \citenamefont
  {Oset}(2009)}]{Gamermann:2009fv}%
  \BibitemOpen
  \bibfield  {author} {\bibinfo {author} {\bibfnamefont {D.}~\bibnamefont
  {Gamermann}}\ and\ \bibinfo {author} {\bibfnamefont {E.}~\bibnamefont
  {Oset}},\ }\bibfield  {title} {\bibinfo {title} {{Isospin breaking effects in
  the $X(3872)$ resonance}},\ }\href
  {https://doi.org/10.1103/PhysRevD.80.014003} {\bibfield  {journal} {\bibinfo
  {journal} {Phys. Rev. D}\ }\textbf {\bibinfo {volume} {80}},\ \bibinfo
  {pages} {014003} (\bibinfo {year} {2009})},\ \Eprint
  {https://arxiv.org/abs/0905.0402} {arXiv:0905.0402 [hep-ph]} \BibitemShut
  {NoStop}%
\bibitem [{\citenamefont {Gamermann}\ \emph {et~al.}(2010)\citenamefont
  {Gamermann}, \citenamefont {Nieves}, \citenamefont {Oset},\ and\
  \citenamefont {Ruiz~Arriola}}]{Gamermann:2009uq}%
  \BibitemOpen
  \bibfield  {author} {\bibinfo {author} {\bibfnamefont {D.}~\bibnamefont
  {Gamermann}}, \bibinfo {author} {\bibfnamefont {J.}~\bibnamefont {Nieves}},
  \bibinfo {author} {\bibfnamefont {E.}~\bibnamefont {Oset}},\ and\ \bibinfo
  {author} {\bibfnamefont {E.}~\bibnamefont {Ruiz~Arriola}},\ }\bibfield
  {title} {\bibinfo {title} {{Couplings in coupled channels versus wave
  functions: Application to the $X(3872)$ resonance}},\ }\href
  {https://doi.org/10.1103/PhysRevD.81.014029} {\bibfield  {journal} {\bibinfo
  {journal} {Phys. Rev. D}\ }\textbf {\bibinfo {volume} {81}},\ \bibinfo
  {pages} {014029} (\bibinfo {year} {2010})},\ \Eprint
  {https://arxiv.org/abs/0911.4407} {arXiv:0911.4407 [hep-ph]} \BibitemShut
  {NoStop}%
\bibitem [{\citenamefont {Thomas}\ and\ \citenamefont
  {Close}(2008)}]{Thomas:2008ja}%
  \BibitemOpen
  \bibfield  {author} {\bibinfo {author} {\bibfnamefont {C.~E.}\ \bibnamefont
  {Thomas}}\ and\ \bibinfo {author} {\bibfnamefont {F.~E.}\ \bibnamefont
  {Close}},\ }\bibfield  {title} {\bibinfo {title} {{Is $X(3872)$ a
  molecule?}},\ }\href {https://doi.org/10.1103/PhysRevD.78.034007} {\bibfield
  {journal} {\bibinfo  {journal} {Phys. Rev. D}\ }\textbf {\bibinfo {volume}
  {78}},\ \bibinfo {pages} {034007} (\bibinfo {year} {2008})},\ \Eprint
  {https://arxiv.org/abs/0805.3653} {arXiv:0805.3653 [hep-ph]} \BibitemShut
  {NoStop}%
\bibitem [{\citenamefont {Calle~Cord{\'o}n}\ and\ \citenamefont
  {Ruiz~Arriola}(2010)}]{CalleCordon:2009pit}%
  \BibitemOpen
  \bibfield  {author} {\bibinfo {author} {\bibfnamefont {A.}~\bibnamefont
  {Calle~Cord{\'o}n}}\ and\ \bibinfo {author} {\bibfnamefont {E.}~\bibnamefont
  {Ruiz~Arriola}},\ }\bibfield  {title} {\bibinfo {title} {{Renormalization
  versus strong form factors for one-boson-exchange potentials}},\ }\href
  {https://doi.org/10.1103/PhysRevC.81.044002} {\bibfield  {journal} {\bibinfo
  {journal} {Phys. Rev. C}\ }\textbf {\bibinfo {volume} {81}},\ \bibinfo
  {pages} {044002} (\bibinfo {year} {2010})},\ \Eprint
  {https://arxiv.org/abs/0905.4933} {arXiv:0905.4933 [nucl-th]} \BibitemShut
  {NoStop}%
\bibitem [{\citenamefont {Reinert}\ \emph {et~al.}(2018)\citenamefont
  {Reinert}, \citenamefont {Krebs},\ and\ \citenamefont
  {Epelbaum}}]{Reinert:2017usi}%
  \BibitemOpen
  \bibfield  {author} {\bibinfo {author} {\bibfnamefont {P.}~\bibnamefont
  {Reinert}}, \bibinfo {author} {\bibfnamefont {H.}~\bibnamefont {Krebs}},\
  and\ \bibinfo {author} {\bibfnamefont {E.}~\bibnamefont {Epelbaum}},\
  }\bibfield  {title} {\bibinfo {title} {{Semilocal momentum-space regularized
  chiral two-nucleon potentials up to fifth order}},\ }\href
  {https://doi.org/10.1140/epja/i2018-12516-4} {\bibfield  {journal} {\bibinfo
  {journal} {Eur. Phys. J. A}\ }\textbf {\bibinfo {volume} {54}},\ \bibinfo
  {pages} {86} (\bibinfo {year} {2018})},\ \Eprint
  {https://arxiv.org/abs/1711.08821} {arXiv:1711.08821 [nucl-th]} \BibitemShut
  {NoStop}%
\bibitem [{\citenamefont {Ding}\ \emph {et~al.}(2009)\citenamefont {Ding},
  \citenamefont {Liu},\ and\ \citenamefont {Yan}}]{Ding:2009vj}%
  \BibitemOpen
  \bibfield  {author} {\bibinfo {author} {\bibfnamefont {G.-J.}\ \bibnamefont
  {Ding}}, \bibinfo {author} {\bibfnamefont {J.-F.}\ \bibnamefont {Liu}},\ and\
  \bibinfo {author} {\bibfnamefont {M.-L.}\ \bibnamefont {Yan}},\ }\bibfield
  {title} {\bibinfo {title} {{Dynamics of Hadronic Molecule in One-Boson
  Exchange Approach and Possible Heavy Flavor Molecules}},\ }\href
  {https://doi.org/10.1103/PhysRevD.79.054005} {\bibfield  {journal} {\bibinfo
  {journal} {Phys. Rev. D}\ }\textbf {\bibinfo {volume} {79}},\ \bibinfo
  {pages} {054005} (\bibinfo {year} {2009})},\ \Eprint
  {https://arxiv.org/abs/0901.0426} {arXiv:0901.0426 [hep-ph]} \BibitemShut
  {NoStop}%
\bibitem [{\citenamefont {Hao}\ \emph {et~al.}(2022)\citenamefont {Hao},
  \citenamefont {Lu},\ and\ \citenamefont {Zou}}]{Hao:2022vwt}%
  \BibitemOpen
  \bibfield  {author} {\bibinfo {author} {\bibfnamefont {W.}~\bibnamefont
  {Hao}}, \bibinfo {author} {\bibfnamefont {Y.}~\bibnamefont {Lu}},\ and\
  \bibinfo {author} {\bibfnamefont {B.-S.}\ \bibnamefont {Zou}},\ }\bibfield
  {title} {\bibinfo {title} {{Coupled channel effects for the charmed-strange
  mesons}},\ }\href {https://doi.org/10.1103/PhysRevD.106.074014} {\bibfield
  {journal} {\bibinfo  {journal} {Phys. Rev. D}\ }\textbf {\bibinfo {volume}
  {106}},\ \bibinfo {pages} {074014} (\bibinfo {year} {2022})},\ \Eprint
  {https://arxiv.org/abs/2208.10915} {arXiv:2208.10915 [hep-ph]} \BibitemShut
  {NoStop}%
\bibitem [{\citenamefont {R{\"o}nchen}\ \emph {et~al.}(2013)\citenamefont
  {R{\"o}nchen}, \citenamefont {D{\"o}ring}, \citenamefont {Huang},
  \citenamefont {Haberzettl}, \citenamefont {Haidenbauer}, \citenamefont
  {Hanhart}, \citenamefont {Krewald}, \citenamefont {Mei\ss{}ner},\ and\
  \citenamefont {Nakayama}}]{Ronchen:2012eg}%
  \BibitemOpen
  \bibfield  {author} {\bibinfo {author} {\bibfnamefont {D.}~\bibnamefont
  {R{\"o}nchen}}, \bibinfo {author} {\bibfnamefont {M.}~\bibnamefont
  {D{\"o}ring}}, \bibinfo {author} {\bibfnamefont {F.}~\bibnamefont {Huang}},
  \bibinfo {author} {\bibfnamefont {H.}~\bibnamefont {Haberzettl}}, \bibinfo
  {author} {\bibfnamefont {J.}~\bibnamefont {Haidenbauer}}, \bibinfo {author}
  {\bibfnamefont {C.}~\bibnamefont {Hanhart}}, \bibinfo {author} {\bibfnamefont
  {S.}~\bibnamefont {Krewald}}, \bibinfo {author} {\bibfnamefont {U.-G.}\
  \bibnamefont {Mei\ss{}ner}},\ and\ \bibinfo {author} {\bibfnamefont
  {K.}~\bibnamefont {Nakayama}},\ }\bibfield  {title} {\bibinfo {title}
  {{Coupled-channel dynamics in the reactions $\pi N \to \pi N, \eta N,
  K\Lambda, K\Sigma$}},\ }\href {https://doi.org/10.1140/epja/i2013-13044-5}
  {\bibfield  {journal} {\bibinfo  {journal} {Eur. Phys. J. A}\ }\textbf
  {\bibinfo {volume} {49}},\ \bibinfo {pages} {44} (\bibinfo {year} {2013})},\
  \Eprint {https://arxiv.org/abs/1211.6998} {arXiv:1211.6998 [nucl-th]}
  \BibitemShut {NoStop}%
\bibitem [{\citenamefont {Wu}\ \emph {et~al.}(2024)\citenamefont {Wu},
  \citenamefont {Cao}, \citenamefont {Dong},\ and\ \citenamefont
  {Guo}}]{Wu:2023uva}%
  \BibitemOpen
  \bibfield  {author} {\bibinfo {author} {\bibfnamefont {B.}~\bibnamefont
  {Wu}}, \bibinfo {author} {\bibfnamefont {X.-H.}\ \bibnamefont {Cao}},
  \bibinfo {author} {\bibfnamefont {X.-K.}\ \bibnamefont {Dong}},\ and\
  \bibinfo {author} {\bibfnamefont {F.-K.}\ \bibnamefont {Guo}},\ }\bibfield
  {title} {\bibinfo {title} {{\ensuremath{\sigma} exchange in the one-boson
  exchange model involving the ground state octet baryons}},\ }\href
  {https://doi.org/10.1103/PhysRevD.109.034026} {\bibfield  {journal} {\bibinfo
   {journal} {Phys. Rev. D}\ }\textbf {\bibinfo {volume} {109}},\ \bibinfo
  {pages} {034026} (\bibinfo {year} {2024})},\ \Eprint
  {https://arxiv.org/abs/2312.01013} {arXiv:2312.01013 [hep-ph]} \BibitemShut
  {NoStop}%
\bibitem [{\citenamefont {Yan}\ \emph {et~al.}(1992)\citenamefont {Yan},
  \citenamefont {Cheng}, \citenamefont {Cheung}, \citenamefont {Lin},
  \citenamefont {Lin},\ and\ \citenamefont {Yu}}]{Yan:1992gz}%
  \BibitemOpen
  \bibfield  {author} {\bibinfo {author} {\bibfnamefont {T.-M.}\ \bibnamefont
  {Yan}}, \bibinfo {author} {\bibfnamefont {H.-Y.}\ \bibnamefont {Cheng}},
  \bibinfo {author} {\bibfnamefont {C.-Y.}\ \bibnamefont {Cheung}}, \bibinfo
  {author} {\bibfnamefont {G.-L.}\ \bibnamefont {Lin}}, \bibinfo {author}
  {\bibfnamefont {Y.~C.}\ \bibnamefont {Lin}},\ and\ \bibinfo {author}
  {\bibfnamefont {H.-L.}\ \bibnamefont {Yu}},\ }\bibfield  {title} {\bibinfo
  {title} {{Heavy quark symmetry and chiral dynamics}},\ }\href
  {https://doi.org/10.1103/PhysRevD.46.1148} {\bibfield  {journal} {\bibinfo
  {journal} {Phys. Rev. D}\ }\textbf {\bibinfo {volume} {46}},\ \bibinfo
  {pages} {1148} (\bibinfo {year} {1992})},\ \bibinfo {note} {[Erratum:
  Phys.Rev.D 55, 5851 (1997)]}\BibitemShut {NoStop}%
\bibitem [{\citenamefont {Liu}\ and\ \citenamefont {Oka}(2012)}]{Liu:2011xc}%
  \BibitemOpen
  \bibfield  {author} {\bibinfo {author} {\bibfnamefont {Y.-R.}\ \bibnamefont
  {Liu}}\ and\ \bibinfo {author} {\bibfnamefont {M.}~\bibnamefont {Oka}},\
  }\bibfield  {title} {\bibinfo {title} {{$\Lambda_c N$ bound states
  revisited}},\ }\href {https://doi.org/10.1103/PhysRevD.85.014015} {\bibfield
  {journal} {\bibinfo  {journal} {Phys. Rev. D}\ }\textbf {\bibinfo {volume}
  {85}},\ \bibinfo {pages} {014015} (\bibinfo {year} {2012})},\ \Eprint
  {https://arxiv.org/abs/1103.4624} {arXiv:1103.4624 [hep-ph]} \BibitemShut
  {NoStop}%
\bibitem [{\citenamefont {Meng}\ \emph {et~al.}(2018)\citenamefont {Meng},
  \citenamefont {Li},\ and\ \citenamefont {Zhu}}]{Meng:2017udf}%
  \BibitemOpen
  \bibfield  {author} {\bibinfo {author} {\bibfnamefont {L.}~\bibnamefont
  {Meng}}, \bibinfo {author} {\bibfnamefont {N.}~\bibnamefont {Li}},\ and\
  \bibinfo {author} {\bibfnamefont {S.-l.}\ \bibnamefont {Zhu}},\ }\bibfield
  {title} {\bibinfo {title} {{Possible hadronic molecules composed of the
  doubly charmed baryon and nucleon}},\ }\href
  {https://doi.org/10.1140/epja/i2018-12578-2} {\bibfield  {journal} {\bibinfo
  {journal} {Eur. Phys. J. A}\ }\textbf {\bibinfo {volume} {54}},\ \bibinfo
  {pages} {143} (\bibinfo {year} {2018})},\ \Eprint
  {https://arxiv.org/abs/1707.03598} {arXiv:1707.03598 [hep-ph]} \BibitemShut
  {NoStop}%
\bibitem [{\citenamefont {Mei{\ss}ner}(1988)}]{Meissner:1987ge}%
  \BibitemOpen
  \bibfield  {author} {\bibinfo {author} {\bibfnamefont {U.-G.}\ \bibnamefont
  {Mei{\ss}ner}},\ }\bibfield  {title} {\bibinfo {title} {{Low-Energy Hadron
  Physics from Effective Chiral Lagrangians with Vector Mesons}},\ }\href
  {https://doi.org/10.1016/0370-1573(88)90090-7} {\bibfield  {journal}
  {\bibinfo  {journal} {Phys. Rept.}\ }\textbf {\bibinfo {volume} {161}},\
  \bibinfo {pages} {213} (\bibinfo {year} {1988})}\BibitemShut {NoStop}%
\bibitem [{\citenamefont {Yang}\ \emph {et~al.}(2019)\citenamefont {Yang},
  \citenamefont {Meng},\ and\ \citenamefont {Zhu}}]{Yang:2018amd}%
  \BibitemOpen
  \bibfield  {author} {\bibinfo {author} {\bibfnamefont {B.}~\bibnamefont
  {Yang}}, \bibinfo {author} {\bibfnamefont {L.}~\bibnamefont {Meng}},\ and\
  \bibinfo {author} {\bibfnamefont {S.-L.}\ \bibnamefont {Zhu}},\ }\bibfield
  {title} {\bibinfo {title} {{Hadronic molecular states composed of
  spin-$\frac{3}{2}$ singly charmed baryons}},\ }\href
  {https://doi.org/10.1140/epja/i2019-12686-5} {\bibfield  {journal} {\bibinfo
  {journal} {Eur. Phys. J. A}\ }\textbf {\bibinfo {volume} {55}},\ \bibinfo
  {pages} {21} (\bibinfo {year} {2019})},\ \Eprint
  {https://arxiv.org/abs/1810.03332} {arXiv:1810.03332 [hep-ph]} \BibitemShut
  {NoStop}%
\bibitem [{\citenamefont {Erkelenz}\ \emph {et~al.}(1971)\citenamefont
  {Erkelenz}, \citenamefont {Alzetta},\ and\ \citenamefont
  {Holinde}}]{Erkelenz:1971caz}%
  \BibitemOpen
  \bibfield  {author} {\bibinfo {author} {\bibfnamefont {K.}~\bibnamefont
  {Erkelenz}}, \bibinfo {author} {\bibfnamefont {R.}~\bibnamefont {Alzetta}},\
  and\ \bibinfo {author} {\bibfnamefont {K.}~\bibnamefont {Holinde}},\
  }\bibfield  {title} {\bibinfo {title} {{Momentum space calculations and
  helicity formalism in nuclear physics}},\ }\href
  {https://doi.org/10.1016/0375-9474(71)90279-X} {\bibfield  {journal}
  {\bibinfo  {journal} {Nucl. Phys. A}\ }\textbf {\bibinfo {volume} {176}},\
  \bibinfo {pages} {413} (\bibinfo {year} {1971})}\BibitemShut {NoStop}%
\bibitem [{\citenamefont {Erkelenz}(1974)}]{Erkelenz:1974uj}%
  \BibitemOpen
  \bibfield  {author} {\bibinfo {author} {\bibfnamefont {K.}~\bibnamefont
  {Erkelenz}},\ }\bibfield  {title} {\bibinfo {title} {{Current status of the
  relativistic two nucleon one boson exchange potential}},\ }\href
  {https://doi.org/10.1016/0370-1573(74)90008-8} {\bibfield  {journal}
  {\bibinfo  {journal} {Phys. Rept.}\ }\textbf {\bibinfo {volume} {13}},\
  \bibinfo {pages} {191} (\bibinfo {year} {1974})}\BibitemShut {NoStop}%
\bibitem [{\citenamefont {Nagels}\ \emph {et~al.}(1978)\citenamefont {Nagels},
  \citenamefont {Rijken},\ and\ \citenamefont {de~Swart}}]{Nagels:1977ze}%
  \BibitemOpen
  \bibfield  {author} {\bibinfo {author} {\bibfnamefont {M.~M.}\ \bibnamefont
  {Nagels}}, \bibinfo {author} {\bibfnamefont {T.~A.}\ \bibnamefont {Rijken}},\
  and\ \bibinfo {author} {\bibfnamefont {J.~J.}\ \bibnamefont {de~Swart}},\
  }\bibfield  {title} {\bibinfo {title} {{Low-energy nucleon-nucleon potential
  from Regge-pole theory}},\ }\href {https://doi.org/10.1103/PhysRevD.17.768}
  {\bibfield  {journal} {\bibinfo  {journal} {Phys. Rev. D}\ }\textbf {\bibinfo
  {volume} {17}},\ \bibinfo {pages} {768} (\bibinfo {year} {1978})}\BibitemShut
  {NoStop}%
\bibitem [{\citenamefont {Li}\ and\ \citenamefont {Zhu}(2012)}]{Li:2012bt}%
  \BibitemOpen
  \bibfield  {author} {\bibinfo {author} {\bibfnamefont {N.}~\bibnamefont
  {Li}}\ and\ \bibinfo {author} {\bibfnamefont {S.-L.}\ \bibnamefont {Zhu}},\
  }\bibfield  {title} {\bibinfo {title} {{Hadronic Molecular States Composed of
  Heavy Flavor Baryons}},\ }\href {https://doi.org/10.1103/PhysRevD.86.014020}
  {\bibfield  {journal} {\bibinfo  {journal} {Phys. Rev. D}\ }\textbf {\bibinfo
  {volume} {86}},\ \bibinfo {pages} {014020} (\bibinfo {year} {2012})},\
  \Eprint {https://arxiv.org/abs/1204.3364} {arXiv:1204.3364 [hep-ph]}
  \BibitemShut {NoStop}%
\bibitem [{\citenamefont {Zhao}\ \emph {et~al.}(2013)\citenamefont {Zhao},
  \citenamefont {Li}, \citenamefont {Zhu},\ and\ \citenamefont
  {Zou}}]{Zhao:2013ffn}%
  \BibitemOpen
  \bibfield  {author} {\bibinfo {author} {\bibfnamefont {L.}~\bibnamefont
  {Zhao}}, \bibinfo {author} {\bibfnamefont {N.}~\bibnamefont {Li}}, \bibinfo
  {author} {\bibfnamefont {S.-L.}\ \bibnamefont {Zhu}},\ and\ \bibinfo {author}
  {\bibfnamefont {B.-S.}\ \bibnamefont {Zou}},\ }\bibfield  {title} {\bibinfo
  {title} {{Meson-exchange model for the $\Lambda\bar{\Lambda}$ interaction}},\
  }\href {https://doi.org/10.1103/PhysRevD.87.054034} {\bibfield  {journal}
  {\bibinfo  {journal} {Phys. Rev. D}\ }\textbf {\bibinfo {volume} {87}},\
  \bibinfo {pages} {054034} (\bibinfo {year} {2013})},\ \Eprint
  {https://arxiv.org/abs/1302.1770} {arXiv:1302.1770 [hep-ph]} \BibitemShut
  {NoStop}%
\bibitem [{\citenamefont {Liu}\ \emph {et~al.}(2017)\citenamefont {Liu},
  \citenamefont {Jia},\ and\ \citenamefont {Chen}}]{Liu:2017mrh}%
  \BibitemOpen
  \bibfield  {author} {\bibinfo {author} {\bibfnamefont {M.-Z.}\ \bibnamefont
  {Liu}}, \bibinfo {author} {\bibfnamefont {D.-J.}\ \bibnamefont {Jia}},\ and\
  \bibinfo {author} {\bibfnamefont {D.-Y.}\ \bibnamefont {Chen}},\ }\bibfield
  {title} {\bibinfo {title} {{Possible hadronic molecular states composed of
  $S$-wave heavy-light mesons}},\ }\href
  {https://doi.org/10.1088/1674-1137/41/5/053105} {\bibfield  {journal}
  {\bibinfo  {journal} {Chin. Phys. C}\ }\textbf {\bibinfo {volume} {41}},\
  \bibinfo {pages} {053105} (\bibinfo {year} {2017})},\ \Eprint
  {https://arxiv.org/abs/1702.04440} {arXiv:1702.04440 [hep-ph]} \BibitemShut
  {NoStop}%
\bibitem [{\citenamefont {Liu}\ \emph {et~al.}(2018)\citenamefont {Liu},
  \citenamefont {Wu}, \citenamefont {Xie}, \citenamefont {Pavon~Valderrama},\
  and\ \citenamefont {Geng}}]{Liu:2018bkx}%
  \BibitemOpen
  \bibfield  {author} {\bibinfo {author} {\bibfnamefont {M.-Z.}\ \bibnamefont
  {Liu}}, \bibinfo {author} {\bibfnamefont {T.-W.}\ \bibnamefont {Wu}},
  \bibinfo {author} {\bibfnamefont {J.-J.}\ \bibnamefont {Xie}}, \bibinfo
  {author} {\bibfnamefont {M.}~\bibnamefont {Pavon~Valderrama}},\ and\ \bibinfo
  {author} {\bibfnamefont {L.-S.}\ \bibnamefont {Geng}},\ }\bibfield  {title}
  {\bibinfo {title} {{$D \Xi$ and $D^* \Xi$ molecular states from one boson
  exchange}},\ }\href {https://doi.org/10.1103/PhysRevD.98.014014} {\bibfield
  {journal} {\bibinfo  {journal} {Phys. Rev. D}\ }\textbf {\bibinfo {volume}
  {98}},\ \bibinfo {pages} {014014} (\bibinfo {year} {2018})},\ \Eprint
  {https://arxiv.org/abs/1805.08384} {arXiv:1805.08384 [hep-ph]} \BibitemShut
  {NoStop}%
\bibitem [{\citenamefont {Liu}\ \emph {et~al.}(2021)\citenamefont {Liu},
  \citenamefont {Wu}, \citenamefont {S\'anchez~S\'anchez}, \citenamefont
  {Valderrama}, \citenamefont {Geng},\ and\ \citenamefont {Xie}}]{Liu:2019zvb}%
  \BibitemOpen
  \bibfield  {author} {\bibinfo {author} {\bibfnamefont {M.-Z.}\ \bibnamefont
  {Liu}}, \bibinfo {author} {\bibfnamefont {T.-W.}\ \bibnamefont {Wu}},
  \bibinfo {author} {\bibfnamefont {M.}~\bibnamefont {S\'anchez~S\'anchez}},
  \bibinfo {author} {\bibfnamefont {M.~P.}\ \bibnamefont {Valderrama}},
  \bibinfo {author} {\bibfnamefont {L.-S.}\ \bibnamefont {Geng}},\ and\
  \bibinfo {author} {\bibfnamefont {J.-J.}\ \bibnamefont {Xie}},\ }\bibfield
  {title} {\bibinfo {title} {{Spin-parities of the $P_c(4440)$ and $P_c(4457)$
  in the one-boson-exchange model}},\ }\href
  {https://doi.org/10.1103/PhysRevD.103.054004} {\bibfield  {journal} {\bibinfo
   {journal} {Phys. Rev. D}\ }\textbf {\bibinfo {volume} {103}},\ \bibinfo
  {pages} {054004} (\bibinfo {year} {2021})},\ \Eprint
  {https://arxiv.org/abs/1907.06093} {arXiv:1907.06093 [hep-ph]} \BibitemShut
  {NoStop}%
\bibitem [{\citenamefont {Peng}\ \emph {et~al.}(2020)\citenamefont {Peng},
  \citenamefont {Liu}, \citenamefont {S\'anchez~S\'anchez},\ and\ \citenamefont
  {Pavon~Valderrama}}]{Peng:2020xrf}%
  \BibitemOpen
  \bibfield  {author} {\bibinfo {author} {\bibfnamefont {F.-Z.}\ \bibnamefont
  {Peng}}, \bibinfo {author} {\bibfnamefont {M.-Z.}\ \bibnamefont {Liu}},
  \bibinfo {author} {\bibfnamefont {M.}~\bibnamefont {S\'anchez~S\'anchez}},\
  and\ \bibinfo {author} {\bibfnamefont {M.}~\bibnamefont {Pavon~Valderrama}},\
  }\bibfield  {title} {\bibinfo {title} {{Heavy-hadron molecules from
  light-meson-exchange saturation}},\ }\href
  {https://doi.org/10.1103/PhysRevD.102.114020} {\bibfield  {journal} {\bibinfo
   {journal} {Phys. Rev. D}\ }\textbf {\bibinfo {volume} {102}},\ \bibinfo
  {pages} {114020} (\bibinfo {year} {2020})},\ \Eprint
  {https://arxiv.org/abs/2004.05658} {arXiv:2004.05658 [hep-ph]} \BibitemShut
  {NoStop}%
\bibitem [{\citenamefont {Chen}\ \emph {et~al.}(2015)\citenamefont {Chen},
  \citenamefont {Liu}, \citenamefont {Li},\ and\ \citenamefont
  {Zhu}}]{Chen:2015loa}%
  \BibitemOpen
  \bibfield  {author} {\bibinfo {author} {\bibfnamefont {R.}~\bibnamefont
  {Chen}}, \bibinfo {author} {\bibfnamefont {X.}~\bibnamefont {Liu}}, \bibinfo
  {author} {\bibfnamefont {X.-Q.}\ \bibnamefont {Li}},\ and\ \bibinfo {author}
  {\bibfnamefont {S.-L.}\ \bibnamefont {Zhu}},\ }\bibfield  {title} {\bibinfo
  {title} {{Identifying exotic hidden-charm pentaquarks}},\ }\href
  {https://doi.org/10.1103/PhysRevLett.115.132002} {\bibfield  {journal}
  {\bibinfo  {journal} {Phys. Rev. Lett.}\ }\textbf {\bibinfo {volume} {115}},\
  \bibinfo {pages} {132002} (\bibinfo {year} {2015})},\ \Eprint
  {https://arxiv.org/abs/1507.03704} {arXiv:1507.03704 [hep-ph]} \BibitemShut
  {NoStop}%
\bibitem [{\citenamefont {Liu}\ \emph {et~al.}(2008{\natexlab{a}})\citenamefont
  {Liu}, \citenamefont {Liu}, \citenamefont {Deng},\ and\ \citenamefont
  {Zhu}}]{Liu:2007bf}%
  \BibitemOpen
  \bibfield  {author} {\bibinfo {author} {\bibfnamefont {X.}~\bibnamefont
  {Liu}}, \bibinfo {author} {\bibfnamefont {Y.-R.}\ \bibnamefont {Liu}},
  \bibinfo {author} {\bibfnamefont {W.-Z.}\ \bibnamefont {Deng}},\ and\
  \bibinfo {author} {\bibfnamefont {S.-L.}\ \bibnamefont {Zhu}},\ }\bibfield
  {title} {\bibinfo {title} {{Is $Z^+(4430)$ a loosely bound molecular
  state?}},\ }\href {https://doi.org/10.1103/PhysRevD.77.034003} {\bibfield
  {journal} {\bibinfo  {journal} {Phys. Rev. D}\ }\textbf {\bibinfo {volume}
  {77}},\ \bibinfo {pages} {034003} (\bibinfo {year} {2008}{\natexlab{a}})},\
  \Eprint {https://arxiv.org/abs/0711.0494} {arXiv:0711.0494 [hep-ph]}
  \BibitemShut {NoStop}%
\bibitem [{\citenamefont {Liu}\ \emph {et~al.}(2008{\natexlab{b}})\citenamefont
  {Liu}, \citenamefont {Liu}, \citenamefont {Deng},\ and\ \citenamefont
  {Zhu}}]{Liu:2008fh}%
  \BibitemOpen
  \bibfield  {author} {\bibinfo {author} {\bibfnamefont {Y.-R.}\ \bibnamefont
  {Liu}}, \bibinfo {author} {\bibfnamefont {X.}~\bibnamefont {Liu}}, \bibinfo
  {author} {\bibfnamefont {W.-Z.}\ \bibnamefont {Deng}},\ and\ \bibinfo
  {author} {\bibfnamefont {S.-L.}\ \bibnamefont {Zhu}},\ }\bibfield  {title}
  {\bibinfo {title} {{Is $X(3872) $ Really a Molecular State?}},\ }\href
  {https://doi.org/10.1140/epjc/s10052-008-0640-4} {\bibfield  {journal}
  {\bibinfo  {journal} {Eur. Phys. J. C}\ }\textbf {\bibinfo {volume} {56}},\
  \bibinfo {pages} {63} (\bibinfo {year} {2008}{\natexlab{b}})},\ \Eprint
  {https://arxiv.org/abs/0801.3540} {arXiv:0801.3540 [hep-ph]} \BibitemShut
  {NoStop}%
\bibitem [{\citenamefont {Liu}\ \emph {et~al.}(2019{\natexlab{b}})\citenamefont
  {Liu}, \citenamefont {Wu}, \citenamefont {Pavon~Valderrama}, \citenamefont
  {Xie},\ and\ \citenamefont {Geng}}]{Liu:2019stu}%
  \BibitemOpen
  \bibfield  {author} {\bibinfo {author} {\bibfnamefont {M.-Z.}\ \bibnamefont
  {Liu}}, \bibinfo {author} {\bibfnamefont {T.-W.}\ \bibnamefont {Wu}},
  \bibinfo {author} {\bibfnamefont {M.}~\bibnamefont {Pavon~Valderrama}},
  \bibinfo {author} {\bibfnamefont {J.-J.}\ \bibnamefont {Xie}},\ and\ \bibinfo
  {author} {\bibfnamefont {L.-S.}\ \bibnamefont {Geng}},\ }\bibfield  {title}
  {\bibinfo {title} {{Heavy-quark spin and flavor symmetry partners of the
  $X(3872)$ revisited: What can we learn from the one boson exchange model?}},\
  }\href {https://doi.org/10.1103/PhysRevD.99.094018} {\bibfield  {journal}
  {\bibinfo  {journal} {Phys. Rev. D}\ }\textbf {\bibinfo {volume} {99}},\
  \bibinfo {pages} {094018} (\bibinfo {year} {2019}{\natexlab{b}})},\ \Eprint
  {https://arxiv.org/abs/1902.03044} {arXiv:1902.03044 [hep-ph]} \BibitemShut
  {NoStop}%
\bibitem [{\citenamefont {Liu}\ \emph {et~al.}(2020)\citenamefont {Liu},
  \citenamefont {Xie},\ and\ \citenamefont {Geng}}]{Liu:2020nil}%
  \BibitemOpen
  \bibfield  {author} {\bibinfo {author} {\bibfnamefont {M.-Z.}\ \bibnamefont
  {Liu}}, \bibinfo {author} {\bibfnamefont {J.-J.}\ \bibnamefont {Xie}},\ and\
  \bibinfo {author} {\bibfnamefont {L.-S.}\ \bibnamefont {Geng}},\ }\bibfield
  {title} {\bibinfo {title} {{$X_0(2866)$ as a $D^*\bar{K}^*$ molecular
  state}},\ }\href {https://doi.org/10.1103/PhysRevD.102.091502} {\bibfield
  {journal} {\bibinfo  {journal} {Phys. Rev. D}\ }\textbf {\bibinfo {volume}
  {102}},\ \bibinfo {pages} {091502} (\bibinfo {year} {2020})},\ \Eprint
  {https://arxiv.org/abs/2008.07389} {arXiv:2008.07389 [hep-ph]} \BibitemShut
  {NoStop}%
\bibitem [{\citenamefont {Yalikun}\ \emph {et~al.}(2021)\citenamefont
  {Yalikun}, \citenamefont {Lin}, \citenamefont {Guo}, \citenamefont {Kamiya},\
  and\ \citenamefont {Zou}}]{Yalikun:2021bfm}%
  \BibitemOpen
  \bibfield  {author} {\bibinfo {author} {\bibfnamefont {N.}~\bibnamefont
  {Yalikun}}, \bibinfo {author} {\bibfnamefont {Y.-H.}\ \bibnamefont {Lin}},
  \bibinfo {author} {\bibfnamefont {F.-K.}\ \bibnamefont {Guo}}, \bibinfo
  {author} {\bibfnamefont {Y.}~\bibnamefont {Kamiya}},\ and\ \bibinfo {author}
  {\bibfnamefont {B.-S.}\ \bibnamefont {Zou}},\ }\bibfield  {title} {\bibinfo
  {title} {{Coupled-channel effects of the
  $\Sigma_c^{(*)}\bar{D}^{(*)}-\Lambda_c(2595)\bar{D}$ system and molecular
  nature of the $P_c$ pentaquark states from one-boson exchange model}},\
  }\href {https://doi.org/10.1103/PhysRevD.104.094039} {\bibfield  {journal}
  {\bibinfo  {journal} {Phys. Rev. D}\ }\textbf {\bibinfo {volume} {104}},\
  \bibinfo {pages} {094039} (\bibinfo {year} {2021})},\ \Eprint
  {https://arxiv.org/abs/2109.03504} {arXiv:2109.03504 [hep-ph]} \BibitemShut
  {NoStop}%
\bibitem [{\citenamefont {Manohar}\ and\ \citenamefont
  {Wise}(1993)}]{Manohar:1992nd}%
  \BibitemOpen
  \bibfield  {author} {\bibinfo {author} {\bibfnamefont {A.~V.}\ \bibnamefont
  {Manohar}}\ and\ \bibinfo {author} {\bibfnamefont {M.~B.}\ \bibnamefont
  {Wise}},\ }\bibfield  {title} {\bibinfo {title} {{Exotic $QQ\bar{q}\bar{q}$
  States in QCD}},\ }\href {https://doi.org/10.1016/0550-3213(93)90614-U}
  {\bibfield  {journal} {\bibinfo  {journal} {Nucl. Phys. B}\ }\textbf
  {\bibinfo {volume} {399}},\ \bibinfo {pages} {17} (\bibinfo {year} {1993})},\
  \Eprint {https://arxiv.org/abs/hep-ph/9212236} {arXiv:hep-ph/9212236}
  \BibitemShut {NoStop}%
\bibitem [{\citenamefont {Dong}\ \emph {et~al.}(2020)\citenamefont {Dong},
  \citenamefont {Lin},\ and\ \citenamefont {Zou}}]{Dong:2019ofp}%
  \BibitemOpen
  \bibfield  {author} {\bibinfo {author} {\bibfnamefont {X.-K.}\ \bibnamefont
  {Dong}}, \bibinfo {author} {\bibfnamefont {Y.-H.}\ \bibnamefont {Lin}},\ and\
  \bibinfo {author} {\bibfnamefont {B.-S.}\ \bibnamefont {Zou}},\ }\bibfield
  {title} {\bibinfo {title} {{Prediction of an exotic state around 4240 MeV
  with $J^{PC}=1^{-+}$ as $C$-parity partner of $Y(4260)$ in molecular
  picture}},\ }\href {https://doi.org/10.1103/PhysRevD.101.076003} {\bibfield
  {journal} {\bibinfo  {journal} {Phys. Rev. D}\ }\textbf {\bibinfo {volume}
  {101}},\ \bibinfo {pages} {076003} (\bibinfo {year} {2020})},\ \Eprint
  {https://arxiv.org/abs/1910.14455} {arXiv:1910.14455 [hep-ph]} \BibitemShut
  {NoStop}%
\bibitem [{\citenamefont {Wu}\ \emph {et~al.}(2019)\citenamefont {Wu},
  \citenamefont {Lee},\ and\ \citenamefont {Zou}}]{Wu:2019adv}%
  \BibitemOpen
  \bibfield  {author} {\bibinfo {author} {\bibfnamefont {J.-J.}\ \bibnamefont
  {Wu}}, \bibinfo {author} {\bibfnamefont {T.-S.~H.}\ \bibnamefont {Lee}},\
  and\ \bibinfo {author} {\bibfnamefont {B.-S.}\ \bibnamefont {Zou}},\
  }\bibfield  {title} {\bibinfo {title} {{Nucleon resonances with hidden charm
  in $\gamma p$ reactions}},\ }\href
  {https://doi.org/10.1103/PhysRevC.100.035206} {\bibfield  {journal} {\bibinfo
   {journal} {Phys. Rev. C}\ }\textbf {\bibinfo {volume} {100}},\ \bibinfo
  {pages} {035206} (\bibinfo {year} {2019})},\ \Eprint
  {https://arxiv.org/abs/1906.05375} {arXiv:1906.05375 [nucl-th]} \BibitemShut
  {NoStop}%
\bibitem [{\citenamefont {Kim}\ and\ \citenamefont {Nam}(2019)}]{Kim:2019kef}%
  \BibitemOpen
  \bibfield  {author} {\bibinfo {author} {\bibfnamefont {S.-H.}\ \bibnamefont
  {Kim}}\ and\ \bibinfo {author} {\bibfnamefont {S.-i.}\ \bibnamefont {Nam}},\
  }\bibfield  {title} {\bibinfo {title} {{Pomeron, nucleon-resonance, and
  $(0^+,0^-,1^+)$-meson contributions in $\phi$-meson photoproduction}},\
  }\href {https://doi.org/10.1103/PhysRevC.100.065208} {\bibfield  {journal}
  {\bibinfo  {journal} {Phys. Rev. C}\ }\textbf {\bibinfo {volume} {100}},\
  \bibinfo {pages} {065208} (\bibinfo {year} {2019})},\ \Eprint
  {https://arxiv.org/abs/1904.05133} {arXiv:1904.05133 [hep-ph]} \BibitemShut
  {NoStop}%
\bibitem [{\citenamefont {Wang}\ \emph {et~al.}(2015)\citenamefont {Wang},
  \citenamefont {Liu},\ and\ \citenamefont {Zhao}}]{Wang:2015jsa}%
  \BibitemOpen
  \bibfield  {author} {\bibinfo {author} {\bibfnamefont {Q.}~\bibnamefont
  {Wang}}, \bibinfo {author} {\bibfnamefont {X.-H.}\ \bibnamefont {Liu}},\ and\
  \bibinfo {author} {\bibfnamefont {Q.}~\bibnamefont {Zhao}},\ }\bibfield
  {title} {\bibinfo {title} {{Photoproduction of hidden charm pentaquark states
  $P_c^+(4380)$ and $P_c^+(4450)$}},\ }\href
  {https://doi.org/10.1103/PhysRevD.92.034022} {\bibfield  {journal} {\bibinfo
  {journal} {Phys. Rev. D}\ }\textbf {\bibinfo {volume} {92}},\ \bibinfo
  {pages} {034022} (\bibinfo {year} {2015})},\ \Eprint
  {https://arxiv.org/abs/1508.00339} {arXiv:1508.00339 [hep-ph]} \BibitemShut
  {NoStop}%
\bibitem [{\citenamefont {Lee}\ \emph {et~al.}(2011)\citenamefont {Lee},
  \citenamefont {Luo}, \citenamefont {Chen},\ and\ \citenamefont
  {Zhu}}]{Lee:2011rka}%
  \BibitemOpen
  \bibfield  {author} {\bibinfo {author} {\bibfnamefont {N.}~\bibnamefont
  {Lee}}, \bibinfo {author} {\bibfnamefont {Z.-G.}\ \bibnamefont {Luo}},
  \bibinfo {author} {\bibfnamefont {X.-L.}\ \bibnamefont {Chen}},\ and\
  \bibinfo {author} {\bibfnamefont {S.-L.}\ \bibnamefont {Zhu}},\ }\bibfield
  {title} {\bibinfo {title} {{Possible Deuteron-like Molecular States Composed
  of Heavy Baryons}},\ }\href {https://doi.org/10.1103/PhysRevD.84.014031}
  {\bibfield  {journal} {\bibinfo  {journal} {Phys. Rev. D}\ }\textbf {\bibinfo
  {volume} {84}},\ \bibinfo {pages} {014031} (\bibinfo {year} {2011})},\
  \Eprint {https://arxiv.org/abs/1104.4257} {arXiv:1104.4257 [hep-ph]}
  \BibitemShut {NoStop}%
\bibitem [{\citenamefont {Riska}\ and\ \citenamefont
  {Brown}(2001)}]{Riska:2000gd}%
  \BibitemOpen
  \bibfield  {author} {\bibinfo {author} {\bibfnamefont {D.~O.}\ \bibnamefont
  {Riska}}\ and\ \bibinfo {author} {\bibfnamefont {G.~E.}\ \bibnamefont
  {Brown}},\ }\bibfield  {title} {\bibinfo {title} {{Nucleon resonance
  transition couplings to vector mesons}},\ }\href
  {https://doi.org/10.1016/S0375-9474(00)00362-6} {\bibfield  {journal}
  {\bibinfo  {journal} {Nucl. Phys. A}\ }\textbf {\bibinfo {volume} {679}},\
  \bibinfo {pages} {577} (\bibinfo {year} {2001})},\ \Eprint
  {https://arxiv.org/abs/nucl-th/0005049} {arXiv:nucl-th/0005049} \BibitemShut
  {NoStop}%
\bibitem [{\citenamefont {Machleidt}(2001)}]{Machleidt:2000ge}%
  \BibitemOpen
  \bibfield  {author} {\bibinfo {author} {\bibfnamefont {R.}~\bibnamefont
  {Machleidt}},\ }\bibfield  {title} {\bibinfo {title} {{High-precision,
  charge-dependent Bonn nucleon-nucleon potential}},\ }\href
  {https://doi.org/10.1103/PhysRevC.63.024001} {\bibfield  {journal} {\bibinfo
  {journal} {Phys. Rev. C}\ }\textbf {\bibinfo {volume} {63}},\ \bibinfo
  {pages} {024001} (\bibinfo {year} {2001})},\ \Eprint
  {https://arxiv.org/abs/nucl-th/0006014} {arXiv:nucl-th/0006014} \BibitemShut
  {NoStop}%
\bibitem [{\citenamefont {Sun}\ \emph {et~al.}(2011)\citenamefont {Sun},
  \citenamefont {He}, \citenamefont {Liu}, \citenamefont {Luo},\ and\
  \citenamefont {Zhu}}]{Sun:2011uh}%
  \BibitemOpen
  \bibfield  {author} {\bibinfo {author} {\bibfnamefont {Z.-F.}\ \bibnamefont
  {Sun}}, \bibinfo {author} {\bibfnamefont {J.}~\bibnamefont {He}}, \bibinfo
  {author} {\bibfnamefont {X.}~\bibnamefont {Liu}}, \bibinfo {author}
  {\bibfnamefont {Z.-G.}\ \bibnamefont {Luo}},\ and\ \bibinfo {author}
  {\bibfnamefont {S.-L.}\ \bibnamefont {Zhu}},\ }\bibfield  {title} {\bibinfo
  {title} {{$Z_b(10610)^\pm$ and $Z_b(10650)^\pm$ as the $B^*\bar{B}$ and
  $B^*\bar{B}^{*}$ molecular states}},\ }\href
  {https://doi.org/10.1103/PhysRevD.84.054002} {\bibfield  {journal} {\bibinfo
  {journal} {Phys. Rev. D}\ }\textbf {\bibinfo {volume} {84}},\ \bibinfo
  {pages} {054002} (\bibinfo {year} {2011})},\ \Eprint
  {https://arxiv.org/abs/1106.2968} {arXiv:1106.2968 [hep-ph]} \BibitemShut
  {NoStop}%
\end{thebibliography}%

\end{document}